# Reliable density functional and $G_0W_0$ approaches to the calculation of bandgaps in 2D materials.

Musen Li,[1,2] Michael J. Ford,[2] Rika Kobayashi,[3] Roger D. Amos[2] and Jeffrey R. Reimers[1,2]

*1 International Centre for Quantum and Molecular Structures and Department of Physics, Shanghai University, Shanghai 200444, China.*

*2 University of Technology Sydney, School of Mathematical and Physical Sciences, Ultimo, New South Wales 2007, Australia.*

*3 ANU Supercomputer Facility, Leonard Huxley Bldg. 56, Mills Rd, Canberra, ACT, 2601, Australia.*

Email: jeffrey.reimers@uts.edu.au

**ABSTRACT** Optimizing density-functional theory (DFT) and $G_0W_0$ calculations present coupled problems as orbitals from DFT are needed as $G_0W_0$ starting points. Applied to 341 two-dimensional (2D) materials, we demonstrate that CAM-B3LYP provides minimal changes in bandgap (e.g., mean absolute deviation of 0.23 eV) when used to start $G_0W_0$ calculations, compared to traditional functionals such as PBE, PBE0, and HSE06 (1.07 eV, 1.48 eV, and 1.51 eV, respectively). CAM-B3LYP also delivers the smallest changes in orbital representation. These and other results indicate the suitability of CAM-B3LYP as a density-functional approach for modelling 2D materials, as well as for use in optimizing $G_0W_0$ calculations. Our findings parallel well established features of applications to molecules, as well as for spectroscopic applications involving 3D materials.

Two-dimensional (2D) materials sparked widespread research, owing to their attractive electronic, photonic, and straintronic characteristics [1, 2] and distinctive spintronic and optoelectronic quantum features [3, 4]. They show great promise and provide numerous opportunities in related fields such as catalysis, hybrids energy storage solar cells, and ultrathin devices [5]. In addition to properties that differ from conventional bulk or surface materials, the distinctive manipulability of 2D materials by mechanics [6, 7], light [8], electric field [1], and magnetic field [9] have further expanded the prospects for 2D materials applications. At the same time, many 2D materials have been prepared by tailored preparation methods, driving research in areas such as low-dimensional devices [7, 10].

Among the many applications of 2D materials, band-structure engineering is an important one that plays a vital role in optoelectronics [8, 11, 12], catalysis [13], communications [14-16], photoluminescence emitters [7] and sensors [17]. From black phosphorus to monolayer hexagonal boron nitride (h-BN), 2D semiconductors can show highly tunable bandgaps produced, e.g., by alloying [18], doping [19, 20], and adding defects [21].

As the chemical and structural properties of 2D materials can be very different to 3D ones, and as intuition concerning the optical properties of 2D materials may be difficult to estimate from known properties of 3D materials owing to the significant difference in dielectric properties [22], accurate first-principles simulation methods are required to boost 2D material design and data interpretation [10]. Considered herein are applications of density-functional theory (DFT) [23] and *ab initio* Green's function theory implemented at the $G_0W_0$ level [24].

As a most effective first-principles method, DFT has had widespread use for the modelling and design of both molecular systems and bulk materials. Nevertheless, the precise exchange-correlation formation of DFT is unknown, and in all postulated functionals, the exchange component of DFT has a different asymptotic shape compared to the Coulomb term. As a result, these DFT functionals manifest flaws, such as self-interaction errors and errors associated with charge partial occupancy [25]. These problems can lead to a less-than-accurate description of the electronic structure that significantly impacts on properties such as the bandgap, spectral transition energies, exciton binding strengths, and spectral assignment [26-30]. To improve the description of electron exchange, hybrid DFT methods such as B3LYP [31, 32], PBE0 [33], and HSE06 [34] have been developed. For materials, HSE06 proved successful for the description of metals and important semiconductors such as silicon [35-38], with this success owing to the functional-design criteria and the lack of internal charge transfer in these systems. Further, HSE06 is currently perceived as being the best widely-available functional for the evaluation of materials spectroscopic properties [39], giving results generally superior to those from generalized-gradient approximation (GGA) functions such as PBE [40] and its many variants.

Despite significant achievements, the performance of most known functionals, including HSE06 and PBE0, actually remains poor for general 3D materials [24, 29, 30]. For them, mean absolute error compared to experiment for bandgap calculations of 3D materials has been recorded in the range of 0.5 – 0.8 eV [29, 39], making them unsuitable for choosing between multiple possible experimental spectroscopic data assignments. General issues involve the failure of most functionals to properly separate charge, their incorrect representation of the dielectric constant, and its neglect of dispersion forces. These three effects are related, yet usually treated differently during density-functional design. Additional corrections such as D3 [41] are often added to empirically account for dispersion, whereas the charge-separation and dielectric-constant issues have been the subject of numerous recent DFT developments [29, 30, 35-37, 42-44].

Practical solutions to the charge-separation problem alone have been known for two decades, with many approaches now in common use. This work follows advances [45, 46] associated with the development of the CAM-B3LYP density functional [47]. Recently, CAM-B3LYP has become available [29, 30, 43] to materials simulations, driving significant improvement in the simulation of spectroscopic properties of 3D insulators and semiconductors [29].

Methods like CAM-B3LYP are essential for the understanding of the spectroscopy of large aromatic molecules [43, 48, 49], including photosystems [50-52], and the difference in dielectric properties between 2D and 3D materials [22] suggests that its influence on 2D simulations will be profound. For defects in h-BN, CAM-B3LYP has already been shown to be able to usefully evaluate excited-state energies of widely varying types to order spectroscopic transitions and facilitate experimental assignment, applications for which HSE06 was found to be inadequate [29, 53]. It has also been shown to give results similar to *ab initio* approaches to defect spectroscopic simulations [53].

As an alternative to DFT, *ab initio* approaches using the $G_0W_0$ [24] approximation, in general, provide significantly improved accuracy but at much higher computational cost [54]. For periodic materials, this method computes quasiparticle energies from a single $GW$ iteration by disregarding all off-diagonal matrix members of the self-energy [55]. It provides a non-interaction Green's function and screened-Coulomb interaction convolution, expanded around the band energies given by some prior DFT calculation. Unfortunately, the results obtained depend strongly on the quality of the DFT method used to provide the initial wavefunction [56, 57]. If the electronic distribution presented in the prior DFT calculation is good, then $G_0W_0$ will introduce only minor corrections, whereas if the prior DFT calculations are poor, then the $G_0W_0$ corrections will be large and a qualitatively realistic solution may not necessarily be obtained [54].

Finding the most suitable density functional for some application, and optimizing the results of $G_0W_0$ calculations, therefore present as coupled problems [44, 58-61]. CAM-B3lYP has long been recognized as an optimal starting point for $G_0W_0$ calculations on molecules. For ionization energies, this conclusion arises from a study of 34 molecules using 10 density functionals [58], as well as a study of 100 molecules using 55 density functionals [61]. In the larger study, a mean deviation (MD) of 0.05 eV and a mean absolute deviation (MAD) of 0.24 eV were reported for CAM-B3LYP, compared, e.g., to much less useful results for other approaches such as PBE (-0.57 eV and 0.61 eV), SCAN [62] (-0.48 eV and 0.54 eV), HSE06 (-0.30 eV and 0.37 eV) and PBE0 (-0.20 eV and 0.30 eV). Its performance for materials in general, and 2D materials in particular, therefore also needs to be further assessed. Supporting this, at the level of pure DFT, spectroscopic properties evaluated using CAM-B3LYP have already been demonstrated to provide substantial qualitative and quantitative improvements in the simulation of defects in hexagonal boron nitride (h-BN) compared to HSE06 [29, 53, 63-66].

CAM-B3LYP is chosen for study as it provides an example from amongst the numerous class of established density functionals that embody corrections to the asymptotic potential. Indeed, many related methods also appear worthy of similar investigation [29, 61], including those that correct for the dielectric constant that hold even more promise [35-37, 42, 67]. Of these approaches, one that has already been applied to enhance $G_0W_0$ calculations of specifically 3D materials is WOT-SRSH [67]. This functional delivers similar results to CAM-B3LYP in predicting bandgaps [29] and has been shown to provide significant improvements to $G_0W_0$ energies. Its current drawback is that it include internal parameters in the functional for which the values used are material-dependent and not clearly defined [44]. CAM-B3LYP is chosen herein as it provides an old and clearly established functional description that is universally applicable.

Two methods are widely used to postulate and screen 2D materials: (i) construction inspired by the properties of layered bulk materials, and (ii) brute-force construction by applying optimization algorithms based on atomic composition [68]. Many databases listing possible 2D materials have been created using these approaches, including C2DB [69] and MATHUB-2D [70]. In this work, 341 2D materials (see Supplemental Material) were chosen randomly for study from the C2DB database. The selected structures embody 53 elements and cover 26 different space groups, enough to make them suitable for usage as a benchmark data set.

All calculations were performed using the Vienna ab initio simulation package (VASP) [71, 72], using a customized version of VASP.6.3.2 that implements the CAM-B3LYP functional [30]. Self-consistent calculations are performed for the DFT and $G_0W_0$ band gaps using an energy-convergence criterion of $10^{-7}$ eV. The projector augmented wave (PAW) method [73] was used to describe the interactions between the valence electrons and the core electrons, with the contributions of the PAW sphere's non-spherical density gradients are taken into account. Standard "PBE" pseudopotentials [73] were not used as the shortcomings of CAM-B3LYP and $G_0W_0$ calculations are typically less than the shortcomings that these pseudopotentials provide [30]. Hence "GW" pseudopotentials, as specified in VASP POTCAR version 5.4, were used in all calculations. The plane-wave basis set kinetic energy cutoff is set to 500 eV to ensure the accuracy of the calculation; full details of the resulting basis set, plus the $k$-samplings used, are provided in Supplemental Material. The $G_0W_0$ calculations are performed using either CAM-B3LYP, PBE0, HSE06, or PBE starting orbitals; the results are named "$G_0W_0$@CAM-B3LYP", "$G_0W_0$@PBE0", "$G_0W_0$@HSE06", and "$G_0W_0$@PBE", respectively. Individually optimized atomic structures are obtained for each density functional, using a force convergence criterion of $10^{-3}$ eV Å$^{-1}$.

Full results, including optimized coordinates and calculated bandgaps for the 341 2D materials, are provided in Supplemental Material. Figure 1 compares the calculated $G_0W_0$@PBE0, $G_0W_0$@HSE06, and $G_0W_0$@PBE bandgaps $\Delta E$ with those calculated by $G_0W_0$@CAM-B3LYP. Key statistical analyses are listed in Table I, including the mean differences (MD), mean average differences (MAD), the most negative differences, the most positive differences, the maximum absolute differences (MAX), and the maximum absolute percentage differences (MAPD). The bandgaps calculated starting with PBE0 orbitals and CAM-B3LYP orbitals are similar, with a MD of -0.12 eV, MAD of 0.14 eV, and MAX of 0.66 eV. In contrast, the differences to HSE06 results are ca. 3 times larger, with those to PBE results being up to 10 times larger. Supporting Material shows that the differences from $G_0W_0$@HSE06 and $G_0W_0$@PBE results to $G_0W_0$@PBE0 ones are analogous.

A measure of the quality of the different $G_0W_0$ calculations is the quasiparticle weight [54, 74] $Z$ obtained for the HOMO and LUMO orbitals. This measure is indicative of the similarity of the original orbitals to those as modified by the $G_0W_0$ procedure, with values near unity being desired. It has physical significance in that the HOMO band structures from $G_0W_0$ calculations typically display increased curvature compared to DFT band structures, indicating lower effective mass of the electrons in the material [54]. Results are listed in Table II for the minimum and average values of $Z$ predicted for the 341 2D materials, with $G_0W_0$@CAM-B3LYP providing the best results, $G_0W_0$@PBE0 being slightly poorer, and both $G_0W_0$@HSE06 and $G_0W_0$@PBE providing significantly poorer results. Figure 1 shows the correlation between the $G_0W_0$@CAM-B3LYP results and those from the other methods. Of interest, the treatments of the orbitals afforded by CAM-B3LYP and HSE06 appear to be similar, whereas those for PBE0 and PBE appear uncorrelated.

An alternative measure of quality is the change in the bandgap between that from a $G_0W_0$ calculation and that from its initiating DFT calculation, with results listed in Table I. CAM-B3LYP provides excellent results, realizing a MD of -0.19 eV, MAD of 0.23 eV, MAX of 0.53 eV, and MAPD of 7%. In contrast the changes introduced at the $G_0W_0$ level are 4-6 times larger for PBE0 and 6-8 times larger for $G_0W_0$ calculations starting with HSE06 or PBE orbitals.

Consideration of the pure-DFT bandgaps requires some standard to compare to, and in the absence of experimental data we choose $G_0W_0$@CAM-B3LYP values as they are associated with the highest $Z$ values and the smallest bandgap changes from pure DFT. Results from CAM-B3LYP, PBE0, HSE06, and PBE DFT calculations are compared to them in Figure 2 and Table I. The MD, MAD, MAX, and MAPD values range from 4-6 times larger for PBE0 to 6-10 times larger for HSE06 to 8-14 times larger for PBE.

A feature noted previously (Fig. 1) is the similarity of the CAM-B3LYP and HSE06 wavefunctions as indicated by the $Z$ values manifested in the $G_0W_0$ calculations. To investigate this, the overlaps between all occupied HSE06 and CAM-B3LYP DFT orbitals were computed, and indeed the calculated electron densities appear to be very similar. Overall, the results presented in the figures and tables perceive CAM-B3LYP to be highly accurate, PBE0 to be less accurate but viable, HSE06 to be of poor quality, and PBE to be highly unreliable, yet historically HSE06 has been the method of choice over PBE0 in materials spectroscopic simulations. Figure 1 suggests the possibility that it is the reliable qualitative nature of HSE06 orbitals, rather than the lower quantitative accuracy for bandgaps, that is the reason for this preference, as the impact of the orbital description influences most calculated materials properties.

In terms of qualitative features, an essential aspect of any DFT calculation is that it correctly determines the occupied orbitals. This property forms the basis of the empirical "DFT+U" scheme in which orbital forms are qualitatively retained and bandgaps adjusted to fit experiment. The $G_0W_0$ @PBE (etc.) calculations often reveal negative bandgaps, indicating that PBE actually failed to correctly describe the ground-state of the material. The serious underestimation of bandgaps predicted by PBE0, HSE06, and PBE (Fig. 2) leads to scenarios in which the predicted bandgap becomes less that the worst-case bandgap error for the method, and then even less than the average bandgap error for the method. When this happens, there is no guarantee that the method has correctly identified the ground state electronic structure, with results needing to be carefully considered. To illustrate this effect, Fig. 2 shows the cumulative band-gap-probability distributions obtained for the 341 2D materials using each DFT method, with the high-confidence region (bandgap above worst-case error) shaded in grey, the region of serious concern in which the bandgap is less that the average error shaded in pink, and the intermediate region shaded in green. There is a steady progression from CAM-B3LYP, for which only 2 of the 341 materials are in the intermediary region to PBE, for which most materials are in the region of serious concern, with PBE0 and then HSE06 intermediary.

Combined, the results for 341 2D materials indicate the high quality of the CAM-B3LYP DFT calculations and their suitability as starting orbitals for $G_0W_0$ calculations. Similar results have already been observed for the application of CAM-B3LYP to $G_0W_0$ calculations of molecules [58, 61]. Also, the improvements found for the pure DFT results of CAM-B3LYP exceed the significant improvements already found [29, 61] compared to experiment for 3D materials (bandgap errors reduced by factors of: 1.8 for PBE0, 2.5 for HSE06, 3.7 for SCAN and 4-6 for PBE). In conclusion, for 2D materials (i) for CAM-B3LYP, the quality of the orbitals and energies is high, with $G_0W_0$ making minimal changes to both; (ii) for PBE0, the orbitals and energies are poorer, with $G_0W_0$ introducing adequate corrections; (iii) for HSE06, the orbitals are satisfactory but the energies poor and $G_0W_0$ does not result in quantitatively useful results, and (iv) for BPE, both orbitals and energies are very poor and hence the method is inappropriate for use in starting $G_0W_0$ calculations.

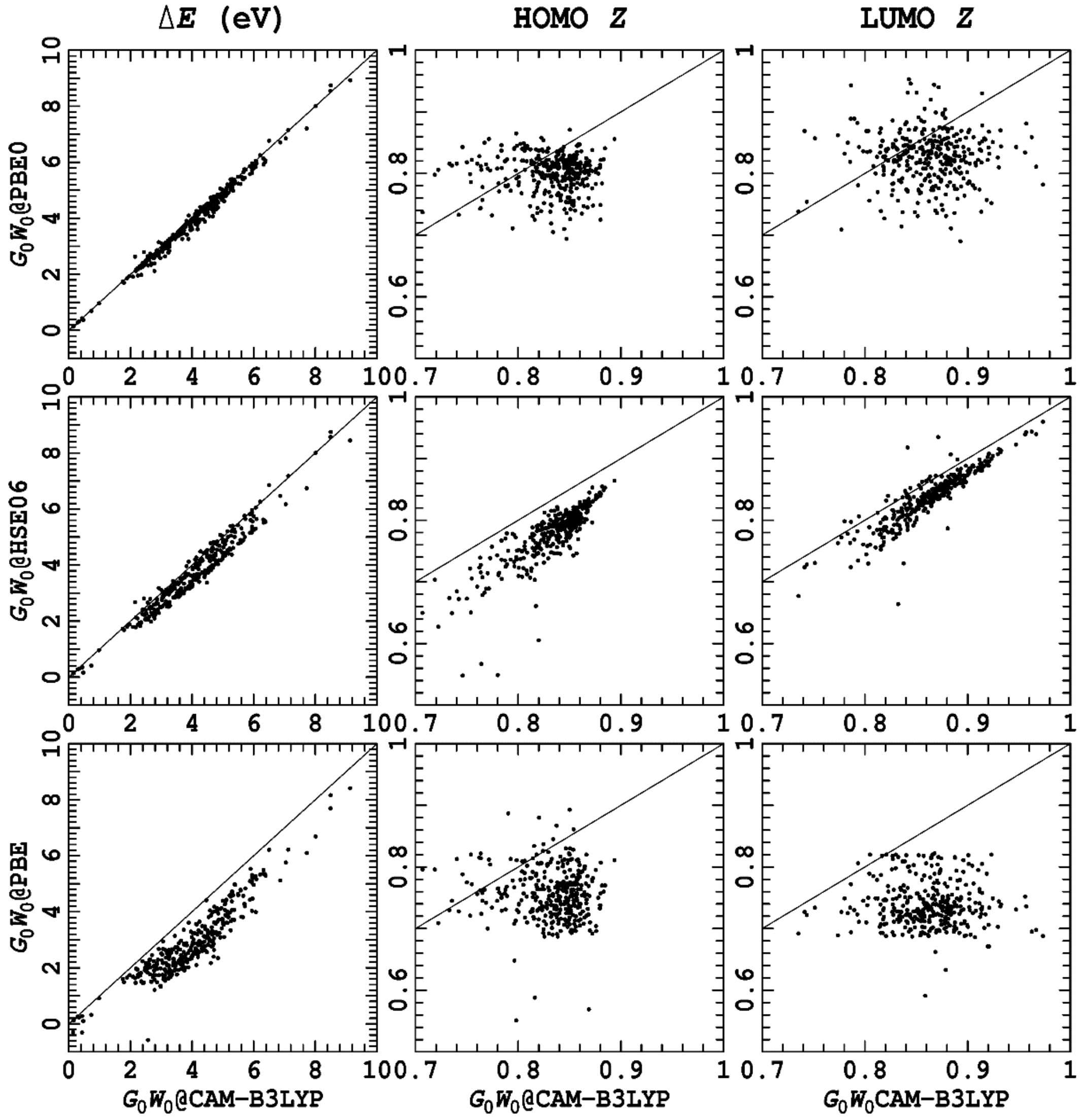

FIG 1. Correlation of the calculated bandgap, $\Delta E$, and the quasiparticle weight $Z$ for the HOMO and LUMO orbitals, for 341 2D materials, as calculated using $G_0W_0$@HSE06, $G_0W_0$@PBE0, and $G_0W_0$@PBE, with those calculated using $G_0W_0$@CAM-B3LYP.

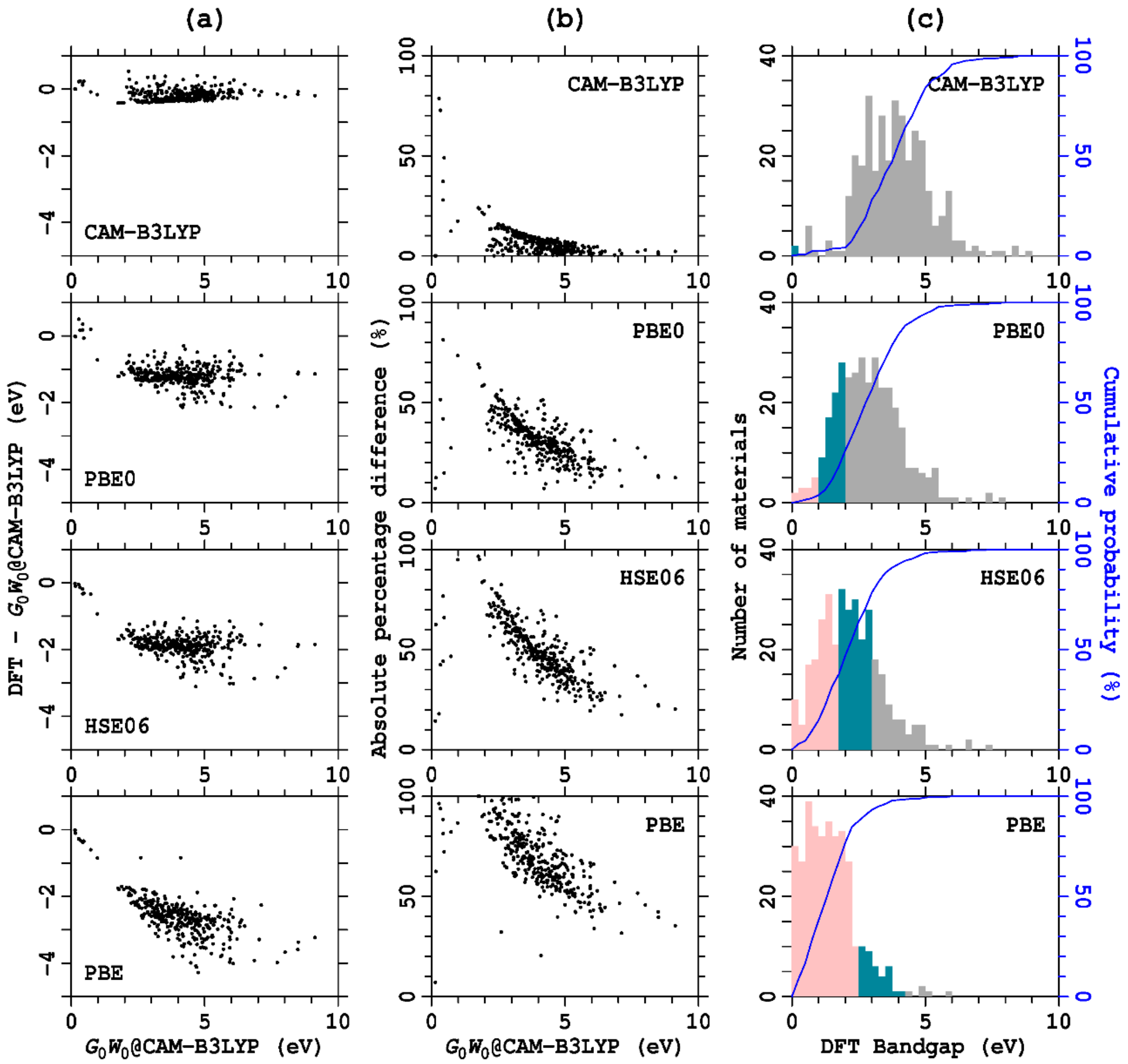

FIG 2. Bandgap analyses for DFT calculations. (a) the difference to $G_0W_0$@CAM-B3LYP versus the $G_0W_0$@CAM-B3LYP values. (b) Analogous percentage absolute deviations. (c) The number of materials predicted with different DFT bandgaps, and their cumulative probability distribution; grey colored regions are ones for which the DFT result can always be taken with confidence, whereas pink colored regions should be taken with caution if significant charge transfer is implicated. Analogous results versus $G_0W_0$@HSE06 are given in SM Fig. S1.

Table I. Comparison of calculated bandgaps of one type of calculation to another, listing the mean difference (MD), mean-absolute difference (MAD), the maximum absolute difference and the most-negative and most-positive differences, as well as the mean absolute percentage difference (MAPD).

| to | of | MD (eV) | MAD (eV) | MAX Abs. (eV) | Most negative (eV) | Most positive (eV) | MAPD (%) |
|---|---|---|---|---|---|---|---|
| $G_0W_0$@CAM-B3LYP | $G_0W_0$@PBE0 | -0.12 | 0.14 | 0.66 | -0.66 | 0.47 | 4 |
| | $G_0W_0$@HSE06 | -0.36 | 0.39 | 1.11 | -1.11 | 0.52 | 11 |
| | $G_0W_0$@PBE | -1.11 | 1.11 | 3.15 | -3.15 | -0.03 | 30 |
| | CAM-B3LYP | -0.19 | 0.23 | 0.53 | -0.42 | 0.53 | 7 |
| | PBE0 | -1.17 | 1.19 | 2.14 | -2.14 | 0.5 | 32 |
| | HSE06 | -1.84 | 1.84 | 3.11 | -3.11 | -0.02 | 49 |
| | PBE | -2.61 | 2.61 | 4.28 | -4.28 | -0.01 | 67 |
| $G_0W_0$@PBE0 | PBE0 | -1.06 | 1.07 | 2.09 | -2.09 | 0.50 | 29 |
| $G_0W_0$@HSE06 | HSE06 | -1.48 | 1.48 | 3.23 | -3.23 | 0.00 | 42 |
| $G_0W_0$@PBE | PBE | -1.50 | 1.51 | 3.38 | -3.38 | 1.07 | 52 |

Table II. The quasiparticle weight (renormalization factor) $Z$ for the initial DFT wavefunction in the final $G_0W_0$ wavefunction for the HOMO and LUMO bands.

| Method | HOMO | | LUMO | |
|---|---|---|---|---|
| | minimum | average | minimum | average |
| $G_0W_0$@CAM-B3LYP | 0.71 | 0.83 | 0.74 | 0.86 |
| $G_0W_0$@PBE0 | 0.69 | 0.83 | 0.69 | 0.80 |
| $G_0W_0$@HSE06 | 0.55 | 0.78 | 0.66 | 0.84 |
| $G_0W_0$@PBE | 0.59 | 0.74 | 0.55 | 0.76 |

| | | | | DFT bandgaps | | | | G0W0 bandgaps | | | | Z for CAM | | | Z for HSE | | | Z for PBE0 | | | Z for PBE | | | | DFT Differences to G0W0@CAM | | | | DFT Differences to G0W0@HSE06 | | | | DFT Differences to G0W0@PBE0 | | | | DFT Differences to G0W0@PBE | | | | G0W0 Diffs to G0W0@CAM | | | G0W0 Diffs to G0W0@PBE0 | | |
|---|---|---|---|---|---|---|---|---|---|---|---|---|---|---|---|---|---|---|---|---|---|---|---|---|---|---|---|---|---|---|---|---|---|---|---|---|---|---|---|---|---|---|---|---|---|
| Material Formula | Space group | Database Index | results file name start<br>add to the end of each either:<br>#CAM-B3LYP.opt<br>#PBE0.opt<br>#HSE06.opt<br>#PBE.opt | CAM | HSE | PBE0 | PBE | CAM | HSE | PBE0 | PBE | Homo | Lumo | Ave | Homo | Lumo | Ave | Humo | Lumo | Ave | Homo | Lumo | Ave | | CAM | HSE | PBE0 | PBE | CAM | HSE | PBE0 | PBE | CAM | HSE | PBE0 | PBE | CAM | HSE | PBE0 | PBE | HSE | PBE0 | PBE | CAM | HSE | PBE |
| | | | | | | | | | | | | | | | | | | | | | | | | average | -0.19 | -1.84 | -1.17 | -2.61 | 0.17 | -1.48 | -0.81 | -2.25 | -0.07 | -1.73 | -1.06 | -2.50 | 0.92 | -0.73 | -0.06 | -1.50 | -0.36 | -0.12 | -1.11 | 0.12 | -0.25 | -1.00 |
| | | | | | | | | | | | | | | | | | | | | | | | | MAD | 0.23 | 1.84 | 1.18 | 2.61 | 0.32 | 1.48 | 0.83 | 2.25 | 0.19 | 1.73 | 1.07 | 2.50 | 0.93 | 0.76 | 0.36 | 1.51 | 0.39 | 0.14 | 1.11 | 0.14 | 0.25 | 1.00 |
| | | | | | | | | | | | | | | | | | | | | | | | | max abs | 0.53 | 3.11 | 2.14 | 4.28 | 0.90 | 3.23 | 1.84 | 4.20 | 0.51 | 3.20 | 2.09 | 4.17 | 3.01 | 2.11 | 2.17 | 3.38 | 1.11 | 0.66 | 3.15 | 0.66 | 0.96 | 2.90 |
| | | | | | | | | | | | | | | | | | | | | | | | | min | -0.42 | -3.11 | -2.14 | -4.28 | -0.40 | -3.23 | -1.84 | -4.20 | -0.40 | -3.20 | -2.09 | -4.17 | -0.27 | -2.11 | -1.39 | -3.38 | -1.11 | -0.66 | -3.15 | -0.47 | -0.96 | -2.90 |
| | | | | | | | | | | | | | | | | | | | | | | | | max | 0.53 | -0.02 | 0.50 | -0.01 | 0.90 | 0.00 | 0.52 | -0.01 | 0.51 | -0.02 | 0.50 | -0.01 | 3.01 | 1.45 | 2.17 | 1.07 | 0.52 | 0.47 | -0.03 | 0.66 | 0.09 | -0.02 |
| | | | | | | | | | | | | | | | | | | | | | | | | number | 341 | 341 | 341 | 341 | 341 | 341 | 341 | 341 | 341 | 341 | 341 | 341 | 341 | 341 | 341 | 341 | 341 | 341 | 341 | 341 | 341 | 341 |
| | | | | | | | | | | | | | | | | | | | | | | | | stdev | 0.18 | 0.41 | 0.36 | 0.63 | 0.31 | 0.49 | 0.44 | 0.65 | 0.21 | 0.43 | 0.37 | 0.63 | 0.48 | 0.46 | 0.47 | 0.58 | 0.29 | 0.15 | 0.49 | 0.15 | 0.21 | 0.47 |
| Ag2Br2 | P1 | 2dm-1949 | Ag2Br2#P1#2dm-1949 | 4.50 | 2.07 | 2.73 | 0.87 | 4.62 | 3.90 | 4.36 | 2.77 | 0.850 | 0.889 | 0.870 | 0.802 | 0.867 | 0.834 | 0.761 | 0.803 | 0.78 | 0.707 | 0.794 | 0.751 | | -0.13 | -2.55 | -1.89 | -3.75 | 0.59 | -1.83 | -1.17 | -3.03 | 0.14 | -2.29 | -1.62 | -3.48 | 1.73 | -0.70 | -0.04 | -1.89 | -0.72 | -0.27 | -1.86 | 0.27 | -0.45 | -1.59 |
| Ag2F2 | Pmm2 | 2dm-1514 | Ag2F2#Pmm2#2dm-1514 | 3.75 | 1.66 | 2.40 | 0.01 | 3.82 | 3.96 | 3.91 | 2.88 | 0.834 | 0.903 | 0.868 | 0.798 | 0.882 | 0.840 | 0.816 | 0.832 | 0.82 | 0.845 | 0.690 | 0.768 | | -0.06 | -2.15 | -1.42 | -3.81 | -0.20 | -2.29 | -1.56 | -3.95 | -0.16 | -2.25 | -1.52 | -3.91 | 0.87 | -1.22 | -0.49 | -2.88 | 0.14 | 0.10 | -0.93 | -0.10 | 0.04 | -1.03 |
| Ag2I2 | Cm | 2dm-3794 | Ag2I2#Cm#2dm-3794 | 4.52 | 3.06 | 3.74 | 2.00 | 4.54 | 4.25 | 4.53 | 3.81 | 0.863 | 0.896 | 0.879 | 0.819 | 0.873 | 0.846 | 0.785 | 0.792 | 0.79 | 0.793 | 0.731 | 0.762 | | -0.02 | -1.48 | -0.79 | -2.54 | 0.27 | -1.19 | -0.51 | -2.26 | -0.02 | -1.48 | -0.79 | -2.54 | 0.70 | -0.76 | -0.07 | -1.82 | -0.28 | 0.00 | -0.72 | 0.00 | -0.28 | -0.72 |
| Ag2I2 | P4/nmm | 2dm-4861 | Ag2I2#P4__nmm#2dm-4861 | 4.59 | 3.12 | 3.81 | 2.05 | 4.72 | 4.30 | 4.57 | 4.01 | 0.867 | 0.901 | 0.884 | 0.819 | 0.877 | 0.848 | 0.805 | 0.839 | 0.82 | 0.755 | 0.805 | 0.780 | | -0.14 | -1.61 | -0.91 | -2.68 | 0.29 | -1.18 | -0.49 | -2.25 | 0.01 | -1.46 | -0.76 | -2.53 | 0.58 | -0.89 | -0.20 | -1.96 | -0.43 | -0.15 | -0.72 | 0.15 | -0.28 | -0.57 |
| Ag2I2 | Pmm2 | 2dm-1519 | Ag2I2#Pmm2#2dm-1519 | 3.41 | 2.00 | 2.65 | 0.97 | 3.78 | 3.60 | 3.70 | 2.48 | 0.857 | 0.885 | 0.871 | 0.804 | 0.862 | 0.833 | 0.835 | 0.874 | 0.85 | 0.782 | 0.755 | 0.769 | | -0.36 | -1.78 | -1.12 | -2.80 | -0.19 | -1.61 | -0.95 | -2.63 | -0.29 | -1.70 | -1.05 | -2.73 | 0.93 | -0.48 | 0.18 | -1.50 | -0.17 | -0.08 | -1.30 | 0.08 | -0.10 | -1.22 |
| Ag2PdO2 | Pm | 2dm-5114 | Ag2PdO2#Pm#2dm-5114 | 3.50 | 1.94 | 2.65 | 0.70 | 3.11 | 3.02 | 3.08 | 1.73 | 0.805 | 0.854 | 0.829 | 0.722 | 0.825 | 0.774 | 0.821 | 0.841 | 0.83 | 0.771 | 0.749 | 0.760 | | 0.38 | -1.18 | -0.46 | -2.41 | 0.48 | -1.08 | -0.37 | -2.31 | 0.41 | -1.15 | -0.43 | -2.38 | 1.77 | 0.21 | 0.92 | -1.02 | -0.10 | -0.03 | -1.38 | 0.03 | -0.07 | -1.36 |
| Ag2S2 | P2/m | 2dm-90 | Ag2S2#P2__m#2dm-90 | 3.53 | 1.91 | 2.58 | 1.14 | 3.78 | 3.14 | 3.69 | 2.37 | 0.814 | 0.821 | 0.817 | 0.743 | 0.783 | 0.763 | 0.797 | 0.822 | 0.81 | 0.814 | 0.746 | 0.780 | | -0.25 | -1.87 | -1.20 | -2.64 | 0.38 | -1.23 | -0.56 | -2.00 | -0.16 | -1.77 | -1.11 | -2.55 | 1.16 | -0.46 | 0.21 | -1.23 | -0.64 | -0.09 | -1.41 | 0.09 | -0.55 | -1.32 |
| Ag2Se2 | P2/m | 2dm-1621 | Ag2Se2#P2__m#2dm-1621 | 0.80 | 0.05 | 0.25 | 0.13 | 0.98 | 0.96 | 0.97 | 0.91 | 0.740 | 0.794 | 0.767 | 0.685 | 0.806 | 0.745 | 0.816 | 0.843 | 0.83 | 0.813 | 0.711 | 0.762 | | -0.17 | -0.93 | -0.72 | -0.85 | -0.16 | -0.91 | -0.71 | -0.83 | -0.17 | -0.92 | -0.72 | -0.84 | -0.11 | -0.86 | -0.66 | -0.78 | -0.02 | 0.00 | -0.06 | 0.00 | -0.01 | -0.06 |
| Ag2Te2 | P2/m | 2dm-70 | Ag2Te2#P2__m#2dm-70 | 2.50 | 0.99 | 1.60 | 0.60 | 2.84 | 2.61 | 2.79 | 2.00 | 0.838 | 0.841 | 0.840 | 0.746 | 0.792 | 0.769 | 0.797 | 0.813 | 0.81 | 0.775 | 0.724 | 0.750 | | -0.34 | -1.85 | -1.24 | -2.24 | -0.11 | -1.62 | -1.01 | -2.01 | -0.29 | -1.80 | -1.19 | -2.19 | 0.50 | -1.01 | -0.40 | -1.40 | -0.23 | -0.05 | -0.84 | 0.05 | -0.18 | -0.79 |
| Ag4S2 | P1 | 2dm-4305 | Ag4S2#P1#2dm-4305 | 4.11 | 2.53 | 3.17 | 1.67 | 4.42 | 3.64 | 4.32 | 3.48 | 0.849 | 0.910 | 0.880 | 0.789 | 0.885 | 0.837 | 0.824 | 0.822 | 0.82 | 0.729 | 0.795 | 0.762 | | -0.31 | -1.89 | -1.25 | -2.75 | 0.46 | -1.12 | -0.48 | -1.97 | -0.21 | -1.79 | -1.15 | -2.65 | 0.63 | -0.95 | -0.31 | -1.81 | -0.77 | -0.10 | -0.94 | 0.10 | -0.67 | -0.84 |
| Ag4Se2 | P1 | 2dm-1423 | Ag4Se2#P1#2dm-1423 | 4.51 | 2.48 | 3.15 | 1.66 | 4.82 | 4.84 | 4.84 | 3.18 | 0.868 | 0.920 | 0.894 | 0.827 | 0.889 | 0.858 | 0.814 | 0.842 | 0.83 | 0.732 | 0.706 | 0.719 | | -0.31 | -2.34 | -1.67 | -3.16 | -0.33 | -2.36 | -1.69 | -3.18 | -0.33 | -2.36 | -1.69 | -3.18 | 1.33 | -0.70 | -0.03 | -1.52 | 0.02 | 0.02 | -1.64 | -0.02 | 0.00 | -1.66 |
| Ag4Se2 | P2_1 | 2dm-4629 | Ag4Se2#P2_1#2dm-4629 | 3.58 | 2.00 | 2.67 | 1.21 | 3.74 | 3.34 | 3.57 | 2.83 | 0.853 | 0.888 | 0.870 | 0.791 | 0.856 | 0.823 | 0.807 | 0.835 | 0.82 | 0.772 | 0.815 | 0.793 | | -0.16 | -1.75 | -1.07 | -2.53 | 0.24 | -1.34 | -0.67 | -2.13 | 0.01 | -1.58 | -0.90 | -2.36 | 0.75 | -0.83 | -0.16 | -1.62 | -0.40 | -0.17 | -0.91 | 0.17 | -0.23 | -0.74 |
| Al2Br2 | P4/nmm | 2dm-2942 | Al2Br2#P4__nmm#2dm-2942 | 4.73 | 2.96 | 3.64 | 2.19 | 4.66 | 4.30 | 4.54 | 3.98 | 0.861 | 0.882 | 0.871 | 0.808 | 0.863 | 0.835 | 0.808 | 0.850 | 0.83 | 0.740 | 0.724 | 0.732 | | 0.07 | -1.70 | -1.01 | -2.47 | 0.42 | -1.35 | -0.66 | -2.12 | 0.19 | -1.58 | -0.90 | -2.35 | 0.75 | -1.02 | -0.34 | -1.80 | -0.35 | -0.12 | -0.68 | 0.12 | -0.23 | -0.56 |
| Al2Cl2 | P4/nmm | 2dm-2973 | Al2Cl2#P4__nmm#2dm-2973 | 5.14 | 3.33 | 4.05 | 2.51 | 5.22 | 4.59 | 5.20 | 4.59 | 0.859 | 0.884 | 0.871 | 0.813 | 0.874 | 0.844 | 0.794 | 0.814 | 0.80 | 0.794 | 0.732 | 0.763 | | -0.08 | -1.89 | -1.18 | -2.72 | 0.55 | -1.25 | -0.54 | -2.08 | -0.06 | -1.87 | -1.16 | -2.70 | 0.55 | -1.26 | -0.54 | -2.08 | -0.64 | -0.02 | -0.63 | 0.02 | -0.62 | -0.61 |
| Al2F2 | P4/nmm | 2dm-2959 | Al2F2#P4__nmm#2dm-2959 | 3.81 | 1.95 | 2.66 | 1.33 | 4.02 | 3.30 | 4.00 | 3.25 | 0.851 | 0.865 | 0.858 | 0.755 | 0.863 | 0.809 | 0.756 | 0.776 | 0.77 | 0.735 | 0.802 | 0.768 | | -0.21 | -2.08 | -1.37 | -2.70 | 0.52 | -1.35 | -0.64 | -1.97 | -0.19 | -2.06 | -1.35 | -2.68 | 0.56 | -1.31 | -0.60 | -1.93 | -0.73 | -0.02 | -0.77 | 0.02 | -0.71 | -0.75 |
| Al2F2 | Pmmn | 2dm-39 | Al2F2#Pmmn#2dm-39 | 3.10 | 1.61 | 2.26 | 0.96 | 3.23 | 2.65 | 3.06 | 2.43 | 0.857 | 0.864 | 0.860 | 0.746 | 0.871 | 0.808 | 0.809 | 0.825 | 0.82 | 0.758 | 0.720 | 0.739 | | -0.13 | -1.62 | -0.97 | -2.26 | 0.45 | -1.04 | -0.39 | -1.69 | 0.03 | -1.46 | -0.81 | -2.10 | 0.67 | -0.83 | -0.18 | -1.47 | -0.58 | -0.16 | -0.79 | 0.16 | -0.42 | -0.63 |
| Al2I2 | P4/nmm | 2dm-814 | Al2I2#P4__nmm#2dm-814 | 4.13 | 2.39 | 3.04 | 1.74 | 4.40 | 3.73 | 4.21 | 2.93 | 0.860 | 0.882 | 0.871 | 0.793 | 0.844 | 0.819 | 0.816 | 0.819 | 0.82 | 0.803 | 0.736 | 0.769 | | -0.27 | -2.02 | -1.36 | -2.66 | 0.40 | -1.34 | -0.69 | -1.99 | -0.08 | -1.82 | -1.16 | -2.47 | 1.20 | -0.55 | 0.11 | -1.19 | -0.67 | -0.19 | -1.47 | 0.19 | -0.48 | -1.28 |
| Al2N2 | P4/nmm | 2dm-1029 | Al2N2#P4__nmm#2dm-1029 | 6.28 | 4.76 | 5.48 | 3.53 | 6.32 | 5.52 | 5.92 | 5.49 | 0.845 | 0.912 | 0.878 | 0.809 | 0.898 | 0.853 | 0.821 | 0.849 | 0.83 | 0.770 | 0.764 | 0.767 | | -0.04 | -1.56 | -0.84 | -2.79 | 0.76 | -0.76 | -0.04 | -2.00 | 0.36 | -1.16 | -0.44 | -2.39 | 0.79 | -0.73 | -0.01 | -1.97 | -0.80 | -0.40 | -0.83 | 0.40 | -0.40 | -0.43 |
| Al2P2 | P4/nmm | 2dm-962 | Al2P2#P4__nmm#2dm-962 | 3.48 | 1.93 | 2.58 | 1.19 | 3.83 | 3.64 | 3.82 | 2.80 | 0.847 | 0.870 | 0.859 | 0.777 | 0.850 | 0.813 | 0.694 | 0.736 | 0.71 | 0.710 | 0.778 | 0.744 | | -0.35 | -1.91 | -1.26 | -2.64 | -0.16 | -1.72 | -1.07 | -2.45 | -0.34 | -1.89 | -1.24 | -2.63 | 0.68 | -0.88 | -0.23 | -1.61 | -0.19 | -0.01 | -1.03 | 0.01 | -0.18 | -1.02 |
| Al2S2 | Cm | 2dm-1791 | Al2S2#Cm#2dm-1791 | 4.90 | 3.04 | 3.75 | 2.23 | 5.09 | 4.38 | 4.80 | 3.87 | 0.854 | 0.875 | 0.864 | 0.817 | 0.853 | 0.835 | 0.756 | 0.854 | 0.80 | 0.746 | 0.744 | 0.745 | | -0.19 | -2.05 | -1.33 | -2.85 | 0.52 | -1.34 | -0.62 | -2.14 | 0.10 | -1.76 | -1.04 | -2.56 | 1.03 | -0.83 | -0.12 | -1.64 | -0.71 | -0.29 | -1.22 | 0.29 | -0.42 | -0.93 |
| Al2S3 | P1 | 2dm-1445 | Al2S3#P1#2dm-1445 | 6.24 | 4.77 | 5.49 | 3.67 | 6.38 | 5.54 | 6.07 | 5.29 | 0.865 | 0.914 | 0.889 | 0.825 | 0.888 | 0.856 | 0.810 | 0.852 | 0.83 | 0.797 | 0.758 | 0.778 | | -0.14 | -1.60 | -0.89 | -2.71 | 0.70 | -0.77 | -0.06 | -1.88 | 0.17 | -1.30 | -0.59 | -2.40 | 0.95 | -0.52 | 0.19 | -1.62 | -0.83 | -0.30 | -1.08 | 0.30 | -0.53 | -0.78 |
| Al2S5 | P1 | 2dm-20 | Al2S5#P1#2dm-20 | 5.87 | 3.11 | 3.83 | 2.07 | 5.96 | 5.56 | 5.75 | 4.47 | 0.858 | 0.890 | 0.874 | 0.810 | 0.871 | 0.840 | 0.845 | 0.864 | 0.85 | 0.764 | 0.749 | 0.757 | | -0.09 | -2.84 | -2.13 | -3.88 | 0.31 | -2.45 | -1.73 | -3.49 | 0.12 | -2.64 | -1.92 | -3.68 | 1.40 | -1.36 | -0.64 | -2.40 | -0.39 | -0.20 | -1.48 | 0.20 | -0.19 | -1.28 |
| Al2Se2 | Cm | 2dm-835 | Al2Se2#Cm#2dm-835 | 4.77 | 2.94 | 3.64 | 2.20 | 5.11 | 4.88 | 5.08 | 3.61 | 0.855 | 0.874 | 0.865 | 0.815 | 0.857 | 0.836 | 0.829 | 0.848 | 0.84 | 0.771 | 0.706 | 0.738 | | -0.35 | -2.17 | -1.47 | -2.91 | -0.11 | -1.94 | -1.23 | -2.67 | -0.31 | -2.14 | -1.44 | -2.88 | 1.16 | -0.67 | 0.04 | -1.40 | -0.24 | -0.03 | -1.51 | 0.03 | -0.21 | -1.48 |
| Al2Se2 | P1 | 2dm-1742 | Al2Se2#P1#2dm-1742 | 4.54 | 2.41 | 3.10 | 1.65 | 4.60 | 4.78 | 4.74 | 3.26 | 0.880 | 0.897 | 0.888 | 0.845 | 0.870 | 0.857 | 0.810 | 0.828 | 0.82 | 0.733 | 0.712 | 0.723 | | -0.07 | -2.19 | -1.51 | -2.95 | -0.25 | -2.37 | -1.69 | -3.13 | -0.21 | -2.33 | -1.65 | -3.09 | 1.27 | -0.85 | -0.17 | -1.61 | 0.18 | 0.14 | -1.34 | -0.14 | 0.04 | -1.48 |
| Al2Se3 | P1 | 2dm-926 | Al2Se3#P1#2dm-926 | 5.36 | 3.90 | 4.58 | 2.91 | 5.50 | 4.76 | 5.24 | 4.60 | 0.868 | 0.902 | 0.885 | 0.815 | 0.881 | 0.848 | 0.781 | 0.785 | 0.78 | 0.732 | 0.757 | 0.744 | | -0.14 | -1.61 | -0.92 | -2.60 | 0.60 | -0.87 | -0.18 | -1.86 | 0.13 | -1.34 | -0.65 | -2.33 | 0.76 | -0.71 | -0.02 | -1.70 | -0.74 | -0.27 | -0.90 | 0.27 | -0.47 | -0.63 |
| Al2Se5 | P1 | 2dm-388 | Al2Se5#P1#2dm-388 | 4.06 | 2.50 | 3.16 | 1.64 | 3.96 | 3.63 | 3.90 | 3.29 | 0.845 | 0.865 | 0.855 | 0.794 | 0.829 | 0.812 | 0.742 | 0.809 | 0.78 | 0.792 | 0.700 | 0.746 | | 0.10 | -1.46 | -0.80 | -2.32 | 0.43 | -1.14 | -0.48 | -1.99 | 0.16 | -1.40 | -0.74 | -2.26 | 0.77 | -0.79 | -0.13 | -1.65 | -0.32 | -0.06 | -0.67 | 0.06 | -0.27 | -0.61 |
| Al2Se5 | P1 | 2dm-1706 | Al2Se5#P1#2dm-1706 | 4.20 | 2.61 | 2.96 | 1.57 | 4.38 | 4.43 | 4.42 | 3.13 | 0.858 | 0.881 | 0.870 | 0.806 | 0.839 | 0.823 | 0.795 | 0.826 | 0.81 | 0.812 | 0.716 | 0.764 | | -0.17 | -1.76 | -1.41 | -2.80 | -0.23 | -1.82 | -1.47 | -2.86 | -0.22 | -1.81 | -1.45 | -2.85 | 1.07 | -0.52 | -0.17 | -1.56 | 0.06 | 0.04 | -1.24 | -0.04 | 0.02 | -1.28 |
| Al2Te2 | Amm2 | 2dm-57 | Al2Te2#Amm2#2dm-57 | 4.28 | 2.50 | 3.18 | 1.86 | 4.56 | 3.90 | 4.08 | 2.96 | 0.857 | 0.874 | 0.865 | 0.804 | 0.851 | 0.827 | 0.776 | 0.806 | 0.79 | 0.750 | 0.743 | 0.746 | | -0.28 | -2.05 | -1.38 | -2.70 | 0.38 | -1.40 | -0.72 | -2.04 | 0.20 | -1.57 | -0.90 | -2.22 | 1.32 | -0.45 | 0.22 | -1.10 | -0.66 | -0.48 | -1.60 | 0.48 | -0.18 | -1.12 |
| Al2Te2 | Cm | 2dm-794 | Al2Te2#Cm#2dm-794 | 4.46 | 2.79 | 3.46 | 2.09 | 4.66 | 4.00 | 4.40 | 3.28 | 0.861 | 0.874 | 0.867 | 0.806 | 0.849 | 0.828 | 0.803 | 0.819 | 0.81 | 0.791 | 0.796 | 0.794 | | -0.19 | -1.87 | -1.20 | -2.56 | 0.47 | -1.21 | -0.54 | -1.90 | 0.06 | -1.61 | -0.94 | -2.31 | 1.19 | -0.49 | 0.18 | -1.18 | -0.66 | -0.26 | -1.38 | 0.26 | -0.40 | -1.12 |
| Al2Te2 | P1 | 2dm-132 | Al2Te2#P1#2dm-132 | 4.30 | 2.21 | 3.87 | 1.58 | 4.27 | 4.49 | 4.40 | 2.83 | 0.876 | 0.887 | 0.882 | 0.840 | 0.859 | 0.850 | 0.807 | 0.822 | 0.81 | 0.696 | 0.704 | 0.700 | | 0.02 | -2.07 | -0.40 | -2.70 | -0.19 | -2.28 | -0.62 | -2.91 | -0.10 | -2.19 | -0.53 | -2.82 | 1.47 | -0.62 | 1.04 | -1.25 | 0.21 | 0.12 | -1.45 | -0.12 | 0.09 | -1.57 |
| AlN | Amm2 | 2dm-3085 | AlN#Amm2#2dm-3085 | 5.53 | 4.09 | 4.80 | 2.95 | 5.57 | 4.89 | 5.23 | 4.79 | 0.839 | 0.920 | 0.879 | 0.810 | 0.903 | 0.856 | 0.815 | 0.822 | 0.82 | 0.753 | 0.671 | 0.712 | | -0.03 | -1.48 | -0.77 | -2.62 | 0.65 | -0.80 | -0.09 | -1.94 | 0.30 | -1.15 | -0.43 | -2.28 | 0.75 | -0.70 | 0.01 | -1.84 | -0.68 | -0.34 | -0.78 | 0.34 | -0.34 | -0.44 |
| As2 | C2/m | 2dm-3312 | As2#C2__m#2dm-3312 | 3.78 | 2.21 | 2.87 | 1.58 | 4.10 | 3.92 | 4.08 | 2.52 | 0.854 | 0.864 | 0.859 | 0.788 | 0.836 | 0.812 | 0.791 | 0.787 | 0.79 | 0.755 | 0.783 | 0.769 | | -0.32 | -1.89 | -1.23 | -2.52 | -0.14 | -1.71 | -1.05 | -2.34 | -0.30 | -1.87 | -1.21 | -2.50 | 1.26 | -0.32 | 0.35 | -0.94 | -0.18 | -0.02 | -1.57 | 0.02 | -0.16 | -1.55 |
| As2N2 | P1 | 2dm-2210 | As2N2#P1#2dm-2210 | 4.31 | 2.58 | 3.29 | 1.63 | 4.64 | 4.55 | 4.63 | 3.45 | 0.844 | 0.881 | 0.863 | 0.805 | 0.857 | 0.831 | 0.818 | 0.858 | 0.84 | 0.761 | 0.753 | 0.757 | | -0.33 | -2.06 | -1.35 | -3.01 | -0.23 | -1.96 | -1.26 | -2.91 | -0.32 | -2.05 | -1.34 | -3.00 | 0.86 | -0.87 | -0.16 | -1.82 | -0.10 | -0.01 | -1.20 | 0.01 | -0.09 | -1.18 |
| As2O3 | P1 | 2dm-2192 | As2O3#P1#2dm-2192 | 3.37 | 1.63 | 2.33 | 0.84 | 3.73 | 3.55 | 3.73 | 3.36 | 0.799 | 0.844 | 0.822 | 0.751 | 0.852 | 0.802 | 0.845 | 0.841 | 0.84 | 0.730 | 0.699 | 0.715 | | -0.36 | -2.10 | -1.40 | -2.88 | -0.18 | -1.92 | -1.22 | -2.70 | -0.36 | -2.10 | -1.40 | -2.88 | 0.01 | -1.73 | -1.03 | -2.52 | -0.18 | 0.00 | -0.37 | 0.00 | -0.18 | -0.37 |
| As2O3 | Pm | 2dm-548 | As2O3#Pm#2dm-548 | 4.64 | 2.93 | 3.63 | 1.94 | 4.82 | 4.10 | 4.58 | 4.02 | 0.825 | 0.864 | 0.844 | 0.786 | 0.841 | 0.813 | 0.854 | 0.920 | 0.89 | 0.837 | 0.735 | 0.786 | | -0.18 | -1.89 | -1.19 | -2.87 | 0.53 | -1.18 | -0.47 | -2.16 | 0.05 | -1.66 | -0.95 | -2.64 | 0.61 | -1.10 | -0.39 | -2.08 | -0.71 | -0.23 | -0.79 | 0.23 | -0.48 | -0.56 |
| As2P2 | P1 | 2dm-2226 | As2P2#P1#2dm-2226 | 3.05 | 1.43 | 2.11 | 0.79 | 3.35 | 3.16 | 3.26 | 1.77 | 0.850 | 0.868 | 0.859 | 0.788 | 0.838 | 0.813 | 0.794 | 0.798 | 0.80 | 0.822 | 0.822 | 0.822 | | -0.30 | -1.92 | -1.24 | -2.56 | -0.12 | -1.74 | -1.06 | -2.38 | -0.21 | -1.83 | -1.15 | -2.47 | 1.28 | -0.34 | 0.34 | -0.98 | -0.18 | -0.09 | -1.58 | 0.09 | -0.10 | -1.49 |
| As2Se3 | Cm | 2dm-753 | As2Se3#Cm#2dm-753 | 2.85 | 1.36 | 1.95 | 0.85 | 2.98 | 2.49 | 2.83 | 1.98 | 0.834 | 0.854 | 0.844 | 0.790 | 0.825 | 0.808 | 0.705 | 0.768 | 0.74 | 0.738 | 0.722 | 0.730 | | -0.13 | -1.63 | -1.03 | -2.13 | 0.36 | -1.14 | -0.54 | -1.64 | 0.02 | -1.47 | -0.87 | -1.97 | 0.87 | -0.63 | -0.03 | -1.13 | -0.49 | -0.16 | -1.00 | 0.16 | -0.33 | -0.84 |
| As4 | Pmna | 2dm-3447 | As4#Pmna#2dm-3447 | 2.78 | 1.24 | 1.89 | 0.69 | 2.98 | 2.52 | 2.84 | 1.33 | 0.856 | 0.867 | 0.862 | 0.788 | 0.836 | 0.812 | 0.824 | 0.861 | 0.84 | 0.696 | 0.754 | 0.725 | | -0.21 | -1.74 | -1.09 | -2.29 | 0.26 | -1.28 | -0.62 | -1.82 | -0.06 | -1.60 | -0.94 | -2.14 | 1.44 | -0.09 | 0.56 | -0.64 | -0.47 | -0.15 | -1.65 | 0.15 | -0.32 | -1.51 |
| AsBr3 | Amm2 | 2dm-2682 | AsBr3#Amm2#2dm-2682 | 2.88 | 1.29 | 1.92 | 0.82 | 3.26 | 2.55 | 2.84 | 1.89 | 0.818 | 0.834 | 0.826 | 0.729 | 0.779 | 0.754 | 0.824 | 0.866 | 0.85 | 0.771 | 0.721 | 0.746 | | -0.38 | -1.97 | -1.34 | -2.45 | 0.33 | -1.25 | -0.62 | -1.73 | 0.04 | -1.55 | -0.92 | -2.02 | 0.99 | -0.59 | 0.04 | -1.07 | -0.72 | -0.42 | -1.38 | 0.42 | -0.29 | -0.95 |
| AsCl3 | Amm2 | 2dm-2488 | AsCl3#Amm2#2dm-2488 | 3.45 | 1.81 | 2.47 | 1.18 | 3.73 | 2.99 | 3.60 | 2.56 | 0.824 | 0.841 | 0.833 | 0.744 | 0.794 | 0.769 | 0.817 | 0.846 | 0.83 | 0.693 | 0.730 | 0.712 | | -0.28 | -1.92 | -1.26 | -2.56 | 0.46 | -1.18 | -0.52 | -1.82 | -0.15 | -1.79 | -1.13 | -2.43 | 0.90 | -0.74 | -0.08 | -1.38 | -0.74 | -0.13 | -1.18 | 0.13 | -0.61 | -1.05 |
| AsF3 | Amm2 | 2dm-2491 | AsF3#Amm2#2dm-2491 | 4.95 | 3.22 | 3.95 | 2.13 | 5.29 | 4.37 | 5.16 | 4.37 | 0.809 | 0.852 | 0.831 | 0.748 | 0.819 | 0.783 | 0.845 | 0.891 | 0.87 | 0.802 | 0.819 | 0.811 | | -0.35 | -2.07 | -1.34 | -3.16 | 0.58 | -1.15 | -0.42 | -2.24 | -0.21 | -1.94 | -1.21 | -3.03 | 0.58 | -1.15 | -0.42 | -2.24 | -0.93 | -0.14 | -0.92 | 0.14 | -0.79 | -0.79 |
| AsI3 | Pm | 2dm-2342 | AsI3#Pm#2dm-2342 | 2.94 | 1.04 | 1.67 | 0.67 | 3.15 | 2.71 | 2.94 | 2.17 | 0.813 | 0.830 | 0.822 | 0.722 | 0.772 | 0.747 | 0.821 | 0.821 | 0.82 | 0.782 | 0.820 | 0.801 | | -0.21 | -2.10 | -1.48 | -2.48 | 0.23 | -1.66 | -1.04 | -2.03 | 0.00 | -1.90 | -1.27 | -2.27 | 0.77 | -1.12 | -0.50 | -1.50 | -0.44 | -0.21 | -0.98 | 0.21 | -0.23 | -0.77 |
| Au2Br2 | Cm | 2dm-654 | Au2Br2#Cm#2dm-654 | 2.47 | 0.94 | 1.60 | 0.44 | 2.18 | 2.13 | 2.17 | 1.97 | 0.828 | 0.862 | 0.845 | 0.788 | 0.837 | 0.812 | 0.797 | 0.798 | 0.80 | 0.742 | 0.711 | 0.726 | | 0.29 | -1.24 | -0.58 | -1.74 | 0.34 | -1.19 | -0.53 | -1.69 | 0.30 | -1.23 | -0.57 | -1.74 | 0.50 | -1.03 | -0.37 | -1.53 | -0.05 | -0.01 | -0.21 | 0.01 | -0.04 | -0.20 |
| Au2Br2 | Cmme | 2dm-4576 | Au2Br2#Cmme#2dm-4576 | 4.49 | 2.99 | 3.67 | 1.90 | 4.78 | 4.05 | 4.45 | 3.61 | 0.842 | 0.856 | 0.849 | 0.788 | 0.862 | 0.825 | 0.824 | 0.843 | 0.83 | 0.739 | 0.703 | 0.721 | | -0.29 | -1.79 | -1.11 | -2.88 | 0.45 | -1.06 | -0.37 | -2.14 | 0.05 | -1.46 | -0.77 | -2.54 | 0.88 | -0.62 | 0.06 | -1.71 | -0.74 | -0.34 | -1.17 | 0.34 | -0.40 | -0.83 |
| Au2Cl2 | P4/nmm | 2dm-509 | Au2Cl2#P4__nmm#2dm-509 | 2.78 | 1.30 | 1.93 | 0.04 | 3.16 | 3.12 | 3.15 | 2.61 | 0.832 | 0.851 | 0.842 | 0.758 | 0.833 | 0.796 | 0.750 | 0.731 | 0.74 | 0.808 | 0.732 | 0.770 | | -0.38 | -1.86 | -1.23 | -3.11 | -0.34 | -1.82 | -1.19 | -3.07 | -0.37 | -1.85 | -1.22 | -3.10 | 0.16 | -1.32 | -0.69 | -2.57 | -0.04 | -0.01 | -0.54 | 0.01 | -0.03 | -0.53 |
| Au2I2 | Cm | 2dm-1766 | Au2I2#Cm#2dm-1766 | 4.46 | 1.97 | 2.65 | 0.99 | 4.19 | 3.90 | 4.15 | 2.39 | 0.846 | 0.877 | 0.861 | 0.797 | 0.841 | 0.819 | 0.819 | 0.874 | 0.85 | 0.744 | 0.708 | 0.726 | | 0.28 | -2.22 | -1.54 | -3.19 | 0.56 | -1.93 | -1.25 | -2.91 | 0.31 | -2.18 | -1.51 | -3.16 | 2.08 | -0.42 | 0.26 | -1.40 | -0.29 | -0.03 | -1.80 | 0.03 | -0.25 | -1.76 |
| Au2S2 | P2/m | 2dm-1477 | Au2S2#P2__m#2dm-1477 | 4.39 | 2.26 | 2.94 | 1.34 | 4.63 | 3.92 | 4.43 | 2.94 | 0.826 | 0.840 | 0.833 | 0.776 | 0.808 | 0.792 | 0.767 | 0.774 | 0.77 | 0.687 | 0.819 | 0.753 | | -0.24 | -2.38 | -1.69 | -3.29 | 0.48 | -1.66 | -0.97 | -2.57 | -0.04 | -2.17 | -1.49 | -3.09 | 1.45 | -0.68 | 0.00 | -1.60 | -0.72 | -0.21 | -1.69 | 0.21 | -0.51 | -1.49 |
| Au4S2 | P1 | 2dm-1444 | Au4S2#P1#2dm-1444 | 3.45 | 1.88 | 2.49 | 1.24 | 3.11 | 3.03 | 3.09 | 1.94 | 0.849 | 0.912 | 0.880 | 0.805 | 0.881 | 0.843 | 0.753 | 0.774 | 0.76 | 0.811 | 0.741 | 0.776 | | 0.33 | -1.24 | -0.63 | -1.87 | 0.42 | -1.15 | -0.55 | -1.79 | 0.36 | -1.21 | -0.61 | -1.85 | 1.51 | -0.06 | 0.55 | -0.70 | -0.08 | -0.02 | -1.17 | 0.02 | -0.06 | -1.15 |
| Au4Se2 | P1 | 2dm-543 | Au4Se2#P1#2dm-543 | 3.39 | 1.82 | 2.43 | 1.25 | 3.75 | 3.55 | 3.70 | 1.91 | 0.854 | 0.908 | 0.881 | 0.809 | 0.880 | 0.844 | 0.802 | 0.813 | 0.81 | 0.743 | 0.729 | 0.736 | | -0.36 | -1.92 | -1.31 | -2.50 | -0.17 | -1.73 | -1.12 | -2.31 | -0.32 | -1.88 | -1.27 | -2.46 | 1.48 | -0.08 | 0.53 | -0.66 | -0.20 | -0.05 | -1.84 | 0.05 | -0.15 | -1.80 |
| AuI | Amm2 | 2dm-242 | AuI#Amm2#2dm-242 | 3.22 | 0.97 | 1.56 | 0.05 | 3.25 | 2.95 | 3.18 | 2.63 | 0.852 | 0.879 | 0.866 | 0.785 | 0.856 | 0.820 | 0.750 | 0.731 | 0.74 | 0.813 | 0.734 | 0.774 | | -0.03 | -2.29 | -1.69 | -3.20 | 0.27 | -1.99 | -1.39 | -2.90 | 0.04 | -2.21 | -1.62 | -3.12 | 0.59 | -1.66 | -1.07 | -2.58 | -0.30 | -0.08 | -0.62 | 0.08 | -0.23 | -0.55 |
| B2F2 | P1 | 2dm-622 | B2F2#P1#2dm-622 | 2.31 | 0.45 | 1.16 | 0.05 | 2.25 | 1.87 | 2.19 | 1.75 | 0.835 | 0.843 | 0.839 | 0.746 | 0.820 | 0.783 | 0.812 | 0.871 | 0.84 | 0.732 | 0.695 | 0.713 | | 0.06 | -1.81 | -1.09 | -2.20 | 0.44 | -1.42 | -0.71 | -1.81 | 0.12 | -1.74 | -1.03 | -2.13 | 0.56 | -1.30 | -0.59 | -1.70 | -0.39 | -0.07 | -0.50 | 0.07 | -0.32 | -0.44 |
| B2I2 | P4/nmm | 2dm-2981 | B2I2#P4__nmm#2dm-2981 | 2.82 | 1.30 | 1.89 | 0.66 | 3.18 | 2.60 | 2.97 | 2.09 | 0.818 | 0.827 | 0.823 | 0.745 | 0.803 | 0.774 | 0.772 | 0.773 | 0.77 | 0.820 | 0.739 | 0.779 | | -0.36 | -1.88 | -1.29 | -2.52 | 0.22 | -1.30 | -0.71 | -1.93 | -0.15 | -1.67 | -1.08 | -2.31 | 0.73 | -0.79 | -0.20 | -1.43 | -0.59 | -0.21 | -1.09 | 0.21 | -0.37 | -0.88 |
| B2N2 | P4/nmm | 2dm-1004 | B2N2#P4__nmm#2dm-1004 | 8.42 | 6.65 | 7.41 | 5.13 | 8.50 | 8.75 | 8.75 | 8.16 | 0.861 | 0.884 | 0.872 | 0.811 | 0.907 | 0.859 | 0.818 | 0.824 | 0.82 | 0.715 | 0.722 | 0.718 | | -0.08 | -1.85 | -1.09 | -3.37 | -0.33 | -2.10 | -1.34 | -3.62 | -0.33 | -2.10 | -1.33 | -3.61 | 0.26 | -1.51 | -0.75 | -3.03 | 0.25 | 0.24 | -0.34 | -0.24 | 0.00 | -0.58 |
| B2O3 | P1 | 2dm-2805 | B2O3#P1#2dm-2805 | 2.43 | 0.87 | 1.59 | 0.49 | 2.57 | 2.06 | 2.32 | -0.58 | 0.795 | 0.884 | 0.839 | 0.778 | 0.865 | 0.821 | 0.711 | 0.711 | 0.71 | 0.778 | 0.726 | 0.752 | | -0.14 | -1.70 | -0.97 | -2.07 | 0.37 | -1.19 | -0.47 | -1.57 | 0.11 | -1.45 | -0.73 | -1.83 | 3.01 | 1.45 | 2.17 | 1.07 | -0.50 | -0.25 | -3.15 | 0.25 | -0.26 | -2.90 |
| B2S2 | Cm | 2dm-1472 | B2S2#Cm#2dm-1472 | 5.99 | 4.01 | 4.75 | 2.92 | 6.00 | 5.74 | 5.90 | 4.96 | 0.863 | 0.878 | 0.870 | 0.822 | 0.861 | 0.841 | 0.790 | 0.799 | 0.79 | 0.728 | 0.773 | 0.750 | | -0.01 | -2.00 | -1.26 | -3.08 | 0.26 | -1.73 | -0.99 | -2.81 | 0.10 | -1.89 | -1.15 | -2.98 | 1.04 | -0.95 | -0.21 | -2.03 | -0.27 | -0.11 | -1.05 | 0.11 | -0.16 | -0.94 |
| B2S2 | P1 | 2dm-1084 | B2S2#P1#2dm-1084 | 5.60 | 3.18 | 3.92 | 2.12 | 5.70 | 5.76 | 5.75 | 4.28 | 0.872 | 0.883 | 0.878 | 0.853 | 0.866 | 0.860 | 0.804 | 0.801 | 0.80 | 0.692 | 0.728 | 0.710 | | -0.10 | -2.51 | -1.78 | -3.58 | -0.16 | -2.57 | -1.84 | -3.64 | -0.15 | -2.57 | -1.83 | -3.63 | 1.31 | -1.10 | -0.37 | -2.17 | 0.06 | 0.05 | -1.41 | -0.05 | 0.01 | -1.47 |
| B2S3 | P1 | 2dm-3796 | B2S3#P1#2dm-3796 | 4.64 | 2.88 | 3.57 | 1.85 | 4.58 | 4.15 | 4.42 | 4.05 | 0.847 | 0.871 | 0.859 | 0.795 | 0.838 | 0.817 | 0.809 | 0.841 | 0.82 | 0.725 | 0.724 | 0.725 | | 0.06 | -1.70 | -1.01 | -2.73 | 0.49 | -1.27 | -0.58 | -2.30 | 0.21 | -1.55 | -0.85 | -2.58 | 0.58 | -1.18 | -0.48 | -2.21 | -0.43 | -0.16 | -0.52 | 0.16 | -0.27 | -0.37 |
| B2S5 | P1 | 2dm-657 | B2S5#P1#2dm-657 | 3.08 | 1.39 | 2.04 | 0.69 | 3.45 | 2.85 | 3.26 | 1.96 | 0.845 | 0.856 | 0.851 | 0.788 | 0.824 | 0.806 | 0.795 | 0.839 | 0.82 | 0.812 | 0.694 | 0.753 | | -0.37 | -2.06 | -1.41 | -2.76 | 0.22 | -1.47 | -0.81 | -2.17 | -0.18 | -1.87 | -1.22 | -2.57 | 1.12 | -0.57 | 0.09 | -1.27 | -0.59 | -0.19 | -1.49 | 0.19 | -0.41 | -1.30 |
| B2S5 | P1 | 2dm-1236 | B2S5#P1#2dm-1236 | 2.96 | 1.31 | 1.96 | 0.68 | 3.27 | 2.68 | 3.07 | 2.41 | 0.826 | 0.832 | 0.829 | 0.764 | 0.800 | 0.782 | 0.814 | 0.839 | 0.83 | 0.735 | 0.715 | 0.725 | | -0.31 | -1.96 | -1.31 | -2.59 | 0.28 | -1.37 | -0.71 | -2.00 | -0.11 | -1.76 | -1.10 | -2.39 | 0.55 | -1.10 | -0.44 | -1.72 | -0.59 | -0.20 | -0.87 | 0.20 | -0.39 | -0.66 |
| B2Se2 | Cm | 2dm-471 | B2Se2#Cm#2dm-471 | 5.61 | 3.93 | 4.63 | 2.95 | 5.88 | 5.82 | 5.86 | 4.43 | 0.870 | 0.876 | 0.873 | 0.824 | 0.854 | 0.839 | 0.806 | 0.803 | 0.80 | 0.710 | 0.747 | 0.728 | | -0.27 | -1.95 | -1.25 | -2.93 | -0.21 | -1.89 | -1.19 | -2.87 | -0.26 | -1.93 | -1.24 | -2.92 | 1.17 | -0.50 | 0.20 | -1.49 | -0.06 | -0.01 | -1.44 | 0.01 | -0.05 | -1.43 |
| B2Se2 | P1 | 2dm-1937 | B2Se2#P1#2dm-1937 | 5.21 | 2.87 | 3.59 | 1.89 | 5.31 | 5.40 | 5.40 | 3.72 | 0.869 | 0.873 | 0.871 | 0.845 | 0.854 | 0.850 | 0.820 | 0.840 | 0.83 | 0.753 | 0.793 | 0.773 | | -0.09 | -2.44 | -1.72 | -3.42 | -0.18 | -2.52 | -1.80 | -3.50 | -0.19 | -2.53 | -1.81 | -3.51 | 1.49 | -0.85 | -0.13 | -1.83 | 0.09 | 0.09 | -1.59 | -0.09 | 0.00 | -1.68 |
| B2Se3 | P1 | 2dm-1703 | B2Se3#P1#2dm-1703 | 4.18 | 2.46 | 3.13 | 1.57 | 4.52 | 3.73 | 4.39 | 3.20 | 0.850 | 0.858 | 0.854 | 0.791 | 0.836 | 0.814 | 0.871 | 0.905 | 0.89 | 0.830 | 0.692 | 0.761 | | -0.35 | -2.06 | -1.40 | -2.95 | 0.45 | -1.27 | -0.61 | -2.16 | -0.21 | -1.93 | -1.26 | -2.82 | 0.98 | -0.74 | -0.07 | -1.63 | -0.79 | -0.13 | -1.32 | 0.13 | -0.66 | -1.19 |
| B2Se5 | P1 | 2dm-609 | B2Se5#P1#2dm-609 | 3.09 | 1.47 | 2.12 | 0.84 | 3.39 | 2.84 | 3.38 | 2.08 | 0.845 | 0.870 | 0.857 | 0.794 | 0.842 | 0.818 | 0.791 | 0.852 | 0.82 | 0.761 | 0.697 | 0.729 | | -0.30 | -1.92 | -1.26 | -2.55 | 0.25 | -1.37 | -0.72 | -2.00 | -0.29 | -1.92 | -1.26 | -2.55 | 1.01 | -0.61 | 0.05 | -1.24 | -0.55 | -0.01 | -1.31 | 0.01 | -0.54 | -1.30 |
| B2Se5 | P1 | 2dm-1255 | B2Se5#P1#2dm-1255 | 3.01 | 1.46 | 2.10 | 0.86 | 3.39 | 3.17 | 3.34 | 2.20 | 0.850 | 0.850 | 0.850 | 0.772 | 0.819 | 0.796 | 0.803 | 0.817 | 0.81 | 0.820 | 0.716 | 0.768 | | -0.38 | -1.93 | -1.29 | -2.53 | -0.15 | -1.71 | -1.07 | -2.31 | -0.33 | -1.88 | -1.25 | -2.49 | 0.81 | -0.74 | -0.10 | -1.34 | -0.22 | -0.05 | -1.19 | 0.05 | -0.18 | -1.14 |
| B2Te2 | Amm2 | 2dm-374 | B2Te2#Amm2#2dm-374 | 3.83 | 2.17 | 2.81 | 1.52 | 3.98 | 3.41 | 3.61 | 2.40 | 0.869 | 0.872 | 0.870 | 0.824 | 0.849 | 0.836 | 0.798 | 0.940 | 0.87 | 0.569 | 0.738 | 0.654 | | -0.15 | -1.82 | -1.17 | -2.47 | 0.42 | -1.25 | -0.60 | -1.89 | 0.23 | -1.44 | -0.80 | -2.09 | 1.43 | -0.23 | 0.41 | -0.88 | -0.57 | -0.38 | -1.59 | 0.38 | -0.19 | -1.21 |
| B2Te2 | Cm | 2dm-749 | B2Te2#Cm#2dm-749 | 4.18 | 2.49 | 3.14 | 1.83 | 4.36 | 3.74 | 3.96 | 2.72 | 0.865 | 0.872 | 0.868 | 0.822 | 0.849 | 0.835 | 0.801 | 0.794 | 0.80 | 0.753 | 0.816 | 0.784 | | -0.18 | -1.87 | -1.22 | -2.53 | 0.44 | -1.25 | -0.60 | -1.91 | 0.23 | -1.47 | -0.82 | -2.13 | 1.46 | -0.23 | 0.42 | -0.89 | -0.62 | -0.40 | -1.64 | 0.40 | -0.22 | -1.24 |
| B2Te5 | P1 | 2dm-1677 | B2Te5#P1#2dm-1677 | 2.10 | 0.57 | 1.17 | 0.22 | 2.46 | 2.22 | 2.34 | 1.92 | 0.857 | 0.863 | 0.860 | 0.794 | 0.830 | 0.812 | 0.785 | 0.796 | 0.79 | 0.746 | 0.711 | 0.729 | | -0.36 | -1.89 | -1.29 | -2.24 | -0.12 | -1.66 | -1.05 | -2.00 | -0.24 | -1.78 | -1.18 | -2.12 | 0.17 | -1.36 | -0.76 | -1.70 | -0.24 | -0.12 | -0.54 | 0.12 | -0.12 | -0.42 |
| B2Te5 | P1 | 2dm-1717 | B2Te5#P1#2dm-1717 | 2.57 | 0.68 | 1.29 | 0.28 | 2.64 | 2.79 | 2.74 | 2.30 | 0.872 | 0.874 | 0.873 | 0.830 | 0.841 | 0.836 | 0.838 | 0.832 | 0.84 | 0.777 | 0.713 | 0.745 | | -0.07 | -1.96 | -1.35 | -2.35 | -0.22 | -2.11 | -1.49 | -2.50 | -0.18 | -2.07 | -1.45 | -2.46 | 0.27 | -1.63 | -1.01 | -2.02 | 0.15 | 0.10 | -0.34 | -0.10 | 0.04 | -0.44 |
| Ba2TiO4 | Pmm2 | 2dm-3897 | Ba2TiO4#Pmm2#2dm-3897 | 4.57 | 3.37 | 3.94 | 1.99 | 4.71 | 4.77 | 4.77 | 3.06 | 0.815 | 0.973 | 0.894 | 0.767 | 0.960 | 0.864 | 0.763 | 0.782 | 0.77 | 0.719 | 0.688 | 0.703 | | -0.15 | -1.35 | -0.78 | -2.72 | -0.20 | -1.40 | -0.83 | -2.77 | -0.20 | -1.40 | -0.83 | -2.78 | 1.50 | 0.30 | 0.87 | -1.07 | 0.05 | 0.06 | -1.65 | -0.06 | 0.00 | -1.71 |
| BaO2 | P4/mmm | 2dm-772 | BaO2#P4__mmm#2dm-772 | 5.93 | 4.56 | 5.15 | 2.93 | 6.21 | 6.26 | 6.25 | 5.33 | 0.823 | 0.947 | 0.885 | 0.796 | 0.923 | 0.859 | 0.817 | 0.843 | 0.83 | 0.764 | 0.731 | 0.747 | | -0.28 | -1.64 | -1.06 | -3.28 | -0.33 | -1.70 | -1.12 | -3.33 | -0.32 | -1.68 | -1.10 | -3.32 | 0.59 | -0.77 | -0.19 | -2.41 | 0.05 | 0.04 | -0.87 | -0.04 | 0.02 | -0.91 |
| BeBr2 | C2/m | 2dm-1112 | BeBr2#C2__m#2dm-1112 | 7.77 | 5.47 | 6.18 | 4.34 | 8.01 | 8.01 | 8.01 | 6.69 | 0.882 | 0.900 | 0.891 | 0.855 | 0.880 | 0.867 | 0.784 | 0.784 | 0.78 | 0.770 | 0.703 | 0.736 | | -0.24 | -2.55 | -1.83 | -3.67 | -0.24 | -2.55 | -1.83 | -3.67 | -0.24 | -2.54 | -1.83 | -3.67 | 1.08 | -1.22 | -0.51 | -2.35 | 0.00 | 0.00 | -1.32 | 0.00 | 0.00 | -1.32 |
| BeCl2 | C2/m | 2dm-91 | BeCl2#C2__m#2dm-91 | 8.93 | 7.28 | 7.99 | 5.90 | 9.13 | 8.45 | 8.93 | 8.41 | 0.880 | 0.967 | 0.923 | 0.844 | 0.940 | 0.892 | 0.814 | 0.811 | 0.81 | 0.738 | 0.698 | 0.718 | | -0.20 | -1.85 | -1.14 | -3.23 | 0.48 | -1.17 | -0.46 | -2.55 | 0.00 | -1.65 | -0.95 | -3.03 | 0.52 | -1.13 | -0.42 | -2.51 | -0.68 | -0.20 | -0.72 | 0.20 | -0.48 | -0.52 |
| BeI2 | C2/m | 2dm-817 | BeI2#C2__m#2dm-817 | 5.09 | 3.45 | 4.10 | 2.61 | 5.37 | 5.27 | 5.33 | 4.07 | 0.869 | 0.875 | 0.872 | 0.834 | 0.871 | 0.853 | 0.801 | 0.749 | 0.77 | 0.769 | 0.721 | 0.745 | | -0.28 | -1.92 | -1.27 | -2.76 | -0.18 | -1.82 | -1.17 | -2.66 | -0.23 | -1.87 | -1.22 | -2.72 | 1.02 | -0.62 | 0.03 | -1.46 | -0.10 | -0.04 | -1.30 | 0.04 | -0.06 | -1.26 |
| BeI2 | P-4m2 | 2dm-760 | BeI2#P-4m2#2dm-760 | 5.88 | 4.17 | 4.87 | 3.14 | 5.90 | 5.61 | 5.85 | 5.53 | 0.880 | 0.894 | 0.887 | 0.840 | 0.883 | 0.862 | 0.749 | 0.821 | 0.78 | 0.757 | 0.751 | 0.754 | | -0.03 | -1.74 | -1.03 | -2.76 | 0.26 | -1.45 | -0.74 | -2.47 | 0.02 | -1.69 | -0.98 | -2.71 | 0.35 | -1.36 | -0.65 | -2.38 | -0.29 | -0.05 | -0.38 | 0.05 | -0.24 | -0.33 |
| Bi2 | C2/m | 2dm-3737 | Bi2#C2__m#2dm-3737 | 2.36 | 1.00 | 1.56 | 0.57 | 2.76 | 2.59 | 2.69 | 2.06 | 0.860 | 0.879 | 0.869 | 0.786 | 0.850 | 0.818 | 0.800 | 0.829 | 0.81 | 0.705 | 0.711 | 0.708 | | -0.39 | -1.75 | -1.19 | -2.19 | -0.23 | -1.59 | -1.03 | -2.03 | -0.33 | -1.69 | -1.13 | -2.13 | 0.30 | -1.06 | -0.50 | -1.49 | -0.16 | -0.06 | -0.70 | 0.06 | -0.10 | -0.64 |
| Bi2As2 | P1 | 2dm-2067 | Bi2As2#P1#2dm-2067 | 2.11 | 0.77 | 1.32 | 0.48 | 2.30 | 1.95 | 2.29 | 1.80 | 0.859 | 0.869 | 0.864 | 0.797 | 0.845 | 0.821 | 0.812 | 0.833 | 0.82 | 0.808 | 0.694 | 0.751 | | -0.19 | -1.53 | -0.98 | -1.82 | 0.16 | -1.18 | -0.63 | -1.47 | -0.18 | -1.52 | -0.96 | -1.80 | 0.31 | -1.03 | -0.48 | -1.31 | -0.35 | -0.01 | -0.50 | 0.01 | -0.33 | -0.49 |
| Bi2N2 | P1 | 2dm-2525 | Bi2N2#P1#2dm-2525 | 3.72 | 2.21 | 2.87 | 1.49 | 3.50 | 3.37 | 3.42 | 2.17 | 0.814 | 0.887 | 0.851 | 0.774 | 0.865 | 0.820 | 0.731 | 0.736 | 0.73 | 0.792 | 0.742 | 0.767 | | 0.22 | -1.28 | -0.63 | -2.00 | 0.35 | -1.16 | -0.50 | -1.87 | 0.30 | -1.20 | -0.55 | -1.92 | 1.55 | 0.05 | 0.70 | -0.67 | -0.13 | -0.08 | -1.33 | 0.08 | -0.05 | -1.25 |
| Bi2O2F2 | C2/m | 2dm-3810 | Bi2O2F2#C2__m#2dm-3810 | 6.37 | 4.76 | 5.47 | 4.02 | 6.09 | 5.64 | 5.96 | 5.16 | 0.814 | 0.905 | 0.859 | 0.788 | 0.888 | 0.838 | 0.816 | 0.832 | 0.82 | 0.732 | 0.802 | 0.767 | | 0.27 | -1.33 | -0.63 | -2.07 | 0.73 | -0.88 | -0.17 | -1.62 | 0.40 | -1.20 | -0.49 | -1.94 | 1.21 | -0.40 | 0.31 | -1.14 | -0.46 | -0.13 | -0.94 | 0.13 | -0.32 | -0.81 |
| Bi2O3 | P1 | 2dm-3968 | Bi2O3#P1#2dm-3968 | 5.08 | 3.46 | 4.17 | 2.52 | 5.25 | 4.55 | 5.03 | 4.54 | 0.807 | 0.892 | 0.849 | 0.765 | 0.875 | 0.820 | 0.806 | 0.810 | 0.81 | 0.718 | 0.724 | 0.721 | | -0.17 | -1.79 | -1.09 | -2.74 | 0.53 | -1.09 | -0.38 | -2.03 | 0.05 | -1.56 | -0.86 | -2.51 | 0.54 | -1.08 | -0.37 | -2.02 | -0.70 | -0.23 | -0.72 | 0.23 | -0.48 | -0.49 |
| Bi2P2 | P1 | 2dm-2112 | Bi2P2#P1#2dm-2112 | 2.35 | 0.91 | 1.53 | 0.52 | 2.34 | 1.96 | 2.31 | 1.82 | 0.849 | 0.866 | 0.858 | 0.790 | 0.842 | 0.816 | 0.792 | 0.843 | 0.82 | 0.712 | 0.766 | 0.739 | | 0.00 | -1.43 | -0.82 | -1.83 | 0.39 | -1.04 | -0.43 | -1.44 | 0.04 | -1.40 | -0.78 | -1.79 | 0.53 | -0.90 | -0.29 | -1.30 | -0.39 | -0.04 | -0.53 | 0.04 | -0.35 | -0.49 |
| Bi2Sb2 | P1 | 2dm-2083 | Bi2Sb2#P1#2dm-2083 | 1.93 | 0.64 | 1.19 | 0.44 | 2.23 | 2.10 | 2.22 | 1.94 | 0.864 | 0.865 | 0.864 | 0.797 | 0.838 | 0.818 | 0.780 | 0.773 | 0.78 | 0.759 | 0.816 | 0.787 | | -0.30 | -1.59 | -1.04 | -1.79 | -0.16 | -1.46 | -0.91 | -1.66 | -0.28 | -1.58 | -1.03 | -1.78 | -0.01 | -1.30 | -0.75 | -1.50 | -0.13 | -0.01 | -0.29 | 0.01 | -0.12 | -0.28 |
| BiBr3 | Amm2 | 2dm-2094 | BiBr3#Amm2#2dm-2094 | 4.63 | 2.03 | 2.71 | 1.43 | 4.70 | 4.34 | 4.69 | 2.86 | 0.840 | 0.869 | 0.854 | 0.789 | 0.836 | 0.812 | 0.810 | 0.740 | 0.77 | 0.725 | 0.662 | 0.694 | | -0.07 | -2.67 | -1.99 | -3.28 | 0.30 | -2.30 | -1.63 | -2.91 | -0.06 | -2.66 | -1.98 | -3.26 | 1.77 | -0.83 | -0.15 | -1.44 | -0.37 | -0.01 | -1.84 | 0.01 | -0.35 | -1.83 |
| BiCl3 | Amm2 | 2dm-2057 | BiCl3#Amm2#2dm-2057 | 4.73 | 2.45 | 3.15 | 1.69 | 4.98 | 4.22 | 4.65 | 3.75 | 0.841 | 0.881 | 0.861 | 0.788 | 0.851 | 0.819 | 0.822 | 0.841 | 0.83 | 0.743 | 0.733 | 0.738 | | -0.25 | -2.53 | -1.83 | -3.29 | 0.51 | -1.78 | -1.07 | -2.53 | 0.08 | -2.20 | -1.50 | -2.96 | 0.98 | -1.30 | -0.60 | -2.05 | -0.76 | -0.33 | -1.23 | 0.33 | -0.43 | -0.90 |
| BiI3 | Pm | 2dm-2497 | BiI3#Pm#2dm-2497 | 3.19 | 1.53 | 2.19 | 1.07 | 3.50 | 2.90 | 3.38 | 2.55 | 0.838 | 0.859 | 0.848 | 0.788 | 0.819 | 0.803 | 0.812 | 0.835 | 0.82 | 0.691 | 0.761 | 0.726 | | -0.31 | -1.97 | -1.31 | -2.43 | 0.30 | -1.36 | -0.71 | -1.83 | -0.19 | -1.85 | -1.20 | -2.32 | 0.64 | -1.02 | -0.36 | -1.48 | -0.61 | -0.12 | -0.95 | 0.12 | -0.49 | -0.84 |
| BiTeBr | Cm | 2dm-4356 | BiTeBr#Cm#2dm-4356 | 3.80 | 2.30 | 2.92 | 1.69 | 4.14 | 4.05 | 4.12 | 2.60 | 0.846 | 0.876 | 0.861 | 0.794 | 0.847 | 0.821 | 0.774 | 0.873 | 0.82 | 0.741 | 0.729 | 0.735 | | -0.34 | -1.84 | -1.22 | -2.45 | -0.25 | -1.75 | -1.13 | -2.35 | -0.32 | -1.82 | -1.20 | -2.43 | 1.20 | -0.30 | 0.32 | -0.91 | -0.10 | -0.02 | -1.54 | 0.02 | -0.08 | -1.52 |
| BiTeCl | P1 | 2dm-3732 | BiTeCl#P1#2dm-3732 | 3.94 | 2.46 | 3.08 | 1.80 | 4.20 | 3.63 | 3.90 | 2.81 | 0.850 | 0.876 | 0.863 | 0.796 | 0.851 | 0.823 | 0.803 | 0.898 | 0.85 | 0.771 | 0.735 | 0.753 | | -0.27 | -1.74 | -1.12 | -2.40 | 0.31 | -1.16 | -0.54 | -1.83 | 0.03 | -1.44 | -0.82 | -2.10 | 1.13 | -0.34 | 0.28 | -1.01 | -0.58 | -0.30 | -1.40 | 0.30 | -0.28 | -1.10 |
| BiTeI | Cm | 2dm-3590 | BiTeI#Cm#2dm-3590 | 3.66 | 2.18 | 2.79 | 1.61 | 4.02 | 3.85 | 3.97 | 2.43 | 0.844 | 0.868 | 0.856 | 0.790 | 0.842 | 0.816 | 0.830 | 0.853 | 0.84 | 0.776 | 0.747 | 0.762 | | -0.36 | -1.84 | -1.23 | -2.41 | -0.19 | -1.68 | -1.06 | -2.24 | -0.31 | -1.80 | -1.19 | -2.36 | 1.23 | -0.25 | 0.36 | -0.82 | -0.17 | -0.05 | -1.59 | 0.05 | -0.12 | -1.54 |
| BN | Amm2 | 2dm-4991 | BN#Amm2#2dm-4991 | 7.05 | 5.89 | 6.53 | 4.87 | 7.12 | 7.18 | 7.15 | 6.22 | 0.854 | 0.962 | 0.908 | 0.804 | 0.944 | 0.874 | 0.782 | 0.859 | 0.82 | 0.740 | 0.695 | 0.718 | | -0.08 | -1.24 | -0.59 | -2.26 | -0.13 | -1.29 | -0.65 | -2.31 | -0.10 | -1.27 | -0.62 | -2.29 | 0.83 | -0.34 | 0.31 | -1.36 | 0.05 | 0.03 | -0.90 | -0.03 | 0.03 | -0.93 |
| C2 | Cmmm | 2dm-3993 | C2#Cmmm#2dm-3993 | 3.89 | 2.47 | 3.05 | 2.04 | 4.05 | 3.50 | 3.89 | 3.01 | 0.841 | 0.842 | 0.841 | 0.746 | 0.918 | 0.832 | 0.762 | 0.771 | 0.77 | 0.731 | 0.766 | 0.749 | | -0.17 | -1.58 | -1.00 | -2.02 | 0.38 | -1.04 | -0.46 | -1.47 | 0.00 | -1.42 | -0.84 | -1.85 | 0.87 | -0.54 | 0.03 | -0.98 | -0.55 | -0.16 | -1.04 | 0.16 | -0.38 | -0.87 |
| C2N4 | P-42_1m | 2dm-3079 | C2N4#P-42_1m#2dm-3079 | 8.33 | 6.60 | 7.35 | 4.90 | 8.49 | 8.57 | 8.55 | 7.69 | 0.860 | 0.906 | 0.883 | 0.829 | 0.887 | 0.858 | 0.847 | 0.859 | 0.85 | 0.784 | 0.773 | 0.778 | | -0.16 | -1.90 | -1.14 | -3.59 | -0.25 | -1.98 | -1.22 | -3.67 | -0.23 | -1.96 | -1.20 | -3.65 | 0.64 | -1.09 | -0.34 | -2.79 | 0.08 | 0.06 | -0.80 | -0.06 | 0.02 | -0.87 |
| C2O2 | P4/nmm | 2dm-1090 | C2O2#P4__nmm#2dm-1090 | 2.13 | 0.72 | 1.32 | 0.28 | 2.24 | 1.87 | 1.96 | 1.68 | 0.762 | 0.787 | 0.774 | 0.723 | 0.797 | 0.760 | 0.828 | 0.943 | 0.89 | 0.815 | 0.694 | 0.755 | | -0.11 | -1.52 | -0.92 | -1.96 | 0.27 | -1.15 | -0.55 | -1.59 | 0.17 | -1.24 | -0.64 | -1.68 | 0.45 | -0.96 | -0.36 | -1.40 | -0.37 | -0.28 | -0.56 | 0.28 | -0.09 | -0.28 |
| C2S2 | Pmn2_1 | 2dm-1207 | C2S2#Pmn2_1#2dm-1207 | 3.68 | 1.91 | 2.61 | 1.02 | 4.02 | 3.86 | 3.90 | 2.47 | 0.851 | 0.864 | 0.858 | 0.782 | 0.842 | 0.812 | 0.742 | 0.854 | 0.80 | 0.695 | 0.731 | 0.713 | | -0.34 | -2.11 | -1.41 | -3.00 | -0.18 | -1.95 | -1.25 | -2.84 | -0.22 | -1.99 | -1.29 | -2.88 | 1.21 | -0.56 | 0.15 | -1.45 | -0.16 | -0.13 | -1.56 | 0.13 | -0.03 | -1.43 |
| Ca2O2 | Aem2 | 2dm-3363 | Ca2O2#Aem2#2dm-3363 | 5.00 | 3.67 | 4.31 | 2.49 | 5.15 | 4.53 | 4.90 | 4.00 | 0.818 | 0.956 | 0.887 | 0.790 | 0.939 | 0.865 | 0.804 | 0.834 | 0.82 | 0.768 | 0.752 | 0.760 | | -0.15 | -1.48 | -0.84 | -2.66 | 0.47 | -0.86 | -0.22 | -2.04 | 0.10 | -1.23 | -0.59 | -2.41 | 1.00 | -0.34 | 0.30 | -1.51 | -0.62 | -0.25 | -1.14 | 0.25 | -0.37 | -0.89 |
| Ca3SiBr2 | P2/m | 2dm-4523 | Ca3SiBr2#P2__m#2dm-4523 | 2.67 | 1.28 | 1.91 | 0.89 | 2.68 | 2.30 | 2.39 | 1.50 | 0.860 | 0.917 | 0.889 | 0.826 | 0.896 | 0.861 | 0.810 | 0.855 | 0.83 | 0.735 | 0.723 | 0.729 | | -0.01 | -1.40 | -0.77 | -1.79 | 0.37 | -1.02 | -0.39 | -1.41 | 0.29 | -1.10 | -0.47 | -1.49 | 1.17 | -0.22 | 0.41 | -0.61 | -0.38 | -0.30 | -1.18 | 0.30 | -0.08 | -0.88 |
| CBr2 | Cm | 2dm-156 | CBr2#Cm#2dm-156 | 3.89 | 2.38 | 3.04 | 1.52 | 4.14 | 3.46 | 3.89 | 2.66 | 0.816 | 0.818 | 0.817 | 0.750 | 0.783 | 0.767 | 0.769 | 0.798 | 0.78 | 0.757 | 0.746 | 0.752 | | -0.25 | -1.76 | -1.10 | -2.62 | 0.44 | -1.08 | -0.42 | -1.94 | 0.00 | -1.51 | -0.85 | -2.37 | 1.23 | -0.28 | 0.38 | -1.14 | -0.68 | -0.25 | -1.48 | 0.25 | -0.43 | -1.23 |
| CCl2 | Amm2 | 2dm-2194 | CCl2#Amm2#2dm-2194 | 4.35 | 2.75 | 3.45 | 1.84 | 4.67 | 3.90 | 4.07 | 2.98 | 0.793 | 0.798 | 0.796 | 0.736 | 0.761 | 0.748 | 0.772 | 0.836 | 0.80 | 0.813 | 0.745 | 0.779 | | -0.32 | -1.92 | -1.23 | -2.83 | 0.45 | -1.15 | -0.45 | -2.06 | 0.28 | -1.32 | -0.63 | -2.23 | 1.38 | -0.23 | 0.47 | -1.14 | -0.77 | -0.60 | -1.69 | 0.60 | -0.17 | -1.09 |
| CCl2 | Cm | 2dm-1237 | CCl2#Cm#2dm-1237 | 4.35 | 2.81 | 3.50 | 1.81 | 4.66 | 3.98 | 4.31 | 3.18 | 0.807 | 0.815 | 0.811 | 0.752 | 0.777 | 0.765 | 0.845 | 0.866 | 0.86 | 0.780 | 0.703 | 0.741 | | -0.31 | -1.85 | -1.17 | -2.85 | 0.37 | -1.17 | -0.49 | -2.17 | 0.05 | -1.50 | -0.81 | -2.50 | 1.18 | -0.36 | 0.32 | -1.37 | -0.68 | -0.35 | -1.49 | 0.35 | -0.33 | -1.13 |
| Cd2S2 | P4/nmm | 2dm-2392 | Cd2S2#P4__nmm#2dm-2392 | 5.47 | 3.26 | 3.97 | 2.28 | 5.71 | 5.00 | 5.45 | 4.20 | 0.845 | 0.888 | 0.866 | 0.797 | 0.873 | 0.835 | 0.710 | 0.775 | 0.74 | 0.740 | 0.710 | 0.725 | | -0.23 | -2.45 | -1.73 | -3.43 | 0.47 | -1.75 | -1.03 | -2.73 | 0.02 | -2.20 | -1.48 | -3.18 | 1.27 | -0.94 | -0.23 | -1.92 | -0.70 | -0.25 | -1.51 | 0.25 | -0.45 | -1.25 |
| Cd2Se2 | P4/nmm | 2dm-2156 | Cd2Se2#P4__nmm#2dm-2156 | 5.34 | 2.92 | 3.61 | 2.04 | 5.26 | 5.05 | 5.25 | 3.67 | 0.855 | 0.889 | 0.872 | 0.798 | 0.869 | 0.833 | 0.830 | 0.829 | 0.83 | 0.708 | 0.738 | 0.723 | | 0.09 | -2.34 | -1.65 | -3.21 | 0.29 | -2.14 | -1.45 | -3.01 | 0.09 | -2.34 | -1.65 | -3.21 | 1.67 | -0.75 | -0.06 | -1.63 | -0.20 | 0.00 | -1.59 | 0.00 | -0.20 | -1.58 |
| CdI2 | C2/m | 2dm-4402 | CdI2#C2__m#2dm-4402 | 4.93 | 3.45 | 4.14 | 2.46 | 5.08 | 4.36 | 4.95 | 4.33 | 0.859 | 0.894 | 0.876 | 0.821 | 0.872 | 0.847 | 0.760 | 0.779 | 0.77 | 0.713 | 0.708 | 0.711 | | -0.15 | -1.63 | -0.94 | -2.62 | 0.57 | -0.91 | -0.22 | -1.90 | -0.02 | -1.50 | -0.81 | -2.49 | 0.60 | -0.88 | -0.19 | -1.87 | -0.72 | -0.13 | -0.75 | 0.13 | -0.59 | -0.62 |
| CdS | Cm | 2dm-3242 | CdS#Cm#2dm-3242 | 4.03 | 2.70 | 3.37 | 1.64 | 4.15 | 3.56 | 4.08 | 3.54 | 0.851 | 0.887 | 0.869 | 0.789 | 0.870 | 0.830 | 0.822 | 0.843 | 0.83 | 0.727 | 0.743 | 0.735 | | -0.13 | -1.46 | -0.78 | -2.52 | 0.47 | -0.86 | -0.19 | -1.92 | -0.05 | -1.38 | -0.71 | -2.44 | 0.49 | -0.84 | -0.16 | -1.90 | -0.59 | -0.07 | -0.62 | 0.07 | -0.52 | -0.54 |
| CeSe2 | C2/m | 2dm-3334 | CeSe2#C2__m#2dm-3334 | 3.70 | 2.01 | 2.74 | 1.01 | 3.95 | 3.32 | 3.58 | 2.65 | 0.746 | 0.790 | 0.768 | 0.548 | 0.769 | 0.658 | 0.843 | 0.889 | 0.87 | 0.788 | 0.730 | 0.759 | | -0.26 | -1.94 | -1.21 | -2.95 | 0.38 | -1.30 | -0.58 | -2.31 | 0.11 | -1.57 | -0.84 | -2.58 | 1.04 | -0.64 | 0.09 | -1.65 | -0.64 | -0.37 | -1.30 | 0.37 | -0.27 | -0.93 |
| CeSe2 | P2/m | 2dm-3173 | CeSe2#P2__m#2dm-3173 | 3.70 | 1.98 | 2.70 | 0.64 | 3.67 | 3.52 | 3.57 | 2.31 | 0.780 | 0.881 | 0.830 | 0.549 | 0.787 | 0.668 | 0.827 | 0.868 | 0.85 | 0.765 | 0.709 | 0.737 | | 0.03 | -1.69 | -0.97 | -3.03 | 0.19 | -1.54 | -0.81 | -2.87 | 0.13 | -1.59 | -0.87 | -2.93 | 1.39 | -0.33 | 0.39 | -1.67 | -0.16 | -0.10 | -1.36 | 0.10 | -0.05 | -1.26 |
| CF2 | Cm | 2dm-1080 | CF2#Cm#2dm-1080 | 4.97 | 3.36 | 4.11 | 1.80 | 5.07 | 4.74 | 4.99 | 4.06 | 0.773 | 0.812 | 0.792 | 0.738 | 0.784 | 0.761 | 0.805 | 0.827 | 0.82 | 0.804 | 0.779 | 0.791 | | -0.11 | -1.71 | -0.96 | -3.27 | 0.23 | -1.38 | -0.63 | -2.94 | -0.03 | -1.64 | -0.89 | -3.19 | 0.91 | -0.70 | 0.05 | -2.25 | -0.33 | -0.08 | -1.02 | 0.08 | -0.25 | -0.94 |
| CF2 | P1 | 2dm-2834 | CF2#P1#2dm-2834 | 5.76 | 4.18 | 4.93 | 2.66 | 5.95 | 5.87 | 5.95 | 5.11 | 0.764 | 0.804 | 0.784 | 0.715 | 0.789 | 0.752 | 0.744 | 0.756 | 0.75 | 0.822 | 0.789 | 0.806 | | -0.19 | -1.77 | -1.02 | -3.29 | -0.11 | -1.69 | -0.94 | -3.21 | -0.18 | -1.77 | -1.02 | -3.29 | 0.65 | -0.93 | -0.18 | -2.45 | -0.08 | 0.00 | -0.84 | 0.00 | -0.08 | -0.84 |
| CI2 | Amm2 | 2dm-2906 | CI2#Amm2#2dm-2906 | 2.43 | 0.84 | 1.49 | 0.34 | 2.48 | 2.06 | 2.46 | 2.01 | 0.798 | 0.827 | 0.813 | 0.736 | 0.780 | 0.758 | 0.792 | 0.848 | 0.82 | 0.811 | 0.748 | 0.779 | | -0.04 | -1.64 | -0.99 | -2.14 | 0.37 | -1.22 | -0.58 | -1.72 | -0.02 | -1.62 | -0.97 | -2.12 | 0.42 | -1.17 | -0.53 | -1.67 | -0.41 | -0.02 | -0.46 | 0.02 | -0.39 | -0.44 |
| CI2 | Cm | 2dm-1640 | CI2#Cm#2dm-1640 | 3.55 | 2.11 | 2.73 | 1.45 | 3.83 | 3.15 | 3.42 | 2.30 | 0.814 | 0.831 | 0.823 | 0.755 | 0.793 | 0.774 | 0.803 | 0.819 | 0.81 | 0.758 | 0.728 | 0.743 | | -0.28 | -1.72 | -1.10 | -2.38 | 0.39 | -1.04 | -0.42 | -1.71 | 0.13 | -1.31 | -0.69 | -1.97 | 1.24 | -0.19 | 0.43 | -0.86 | -0.67 | -0.41 | -1.52 | 0.41 | -0.26 | -1.11 |
| CrO3 | Pm | 2dm-1240 | CrO3#Pm#2dm-1240 | 4.87 | 2.70 | 3.44 | 1.53 | 4.91 | 5.08 | 5.06 | 4.10 | 0.801 | 0.818 | 0.810 | 0.752 | 0.765 | 0.758 | 0.786 | 0.810 | 0.80 | 0.740 | 0.719 | 0.730 | | -0.04 | -2.21 | -1.47 | -3.38 | -0.21 | -2.38 | -1.64 | -3.55 | -0.19 | -2.37 | -1.62 | -3.53 | 0.78 | -1.40 | -0.65 | -2.56 | 0.17 | 0.15 | -0.81 | -0.15 | 0.02 | -0.97 |
| CS | Cm | 2dm-1143 | CS#Cm#2dm-1143 | 4.24 | 2.37 | 3.11 | 1.34 | 4.57 | 3.87 | 4.43 | 3.01 | 0.818 | 0.852 | 0.835 | 0.756 | 0.827 | 0.792 | 0.778 | 0.774 | 0.78 | 0.764 | 0.725 | 0.745 | | -0.33 | -2.20 | -1.46 | -3.23 | 0.37 | -1.50 | -0.76 | -2.53 | -0.19 | -2.06 | -1.32 | -3.09 | 1.23 | -0.64 | 0.10 | -1.67 | -0.70 | -0.14 | -1.56 | 0.14 | -0.56 | -1.42 |
| Cu2Br2 | Cm | 2dm-1454 | Cu2Br2#Cm#2dm-1454 | 5.20 | 3.16 | 3.88 | 1.54 | 5.25 | 5.39 | 5.39 | 3.34 | 0.821 | 0.912 | 0.867 | 0.784 | 0.891 | 0.837 | 0.823 | 0.839 | 0.83 | 0.733 | 0.719 | 0.726 | | -0.05 | -2.09 | -1.37 | -3.71 | -0.19 | -2.22 | -1.51 | -3.85 | -0.19 | -2.22 | -1.51 | -3.85 | 1.86 | -0.17 | 0.54 | -1.80 | 0.14 | 0.14 | -1.91 | -0.14 | 0.00 | -2.05 |
| Cu2Br2 | Pmm2 | 2dm-1054 | Cu2Br2#Pmm2#2dm-1054 | 3.90 | 2.12 | 3.27 | 0.65 | 4.14 | 4.27 | 4.26 | 2.10 | 0.812 | 0.896 | 0.854 | 0.772 | 0.855 | 0.813 | 0.839 | 0.843 | 0.84 | 0.727 | 0.792 | 0.759 | | -0.24 | -2.01 | -0.86 | -3.48 | -0.38 | -2.15 | -1.00 | -3.62 | -0.37 | -2.14 | -0.99 | -3.61 | 1.80 | 0.02 | 1.17 | -1.44 | 0.14 | 0.13 | -2.04 | -0.13 | 0.01 | -2.17 |
| Cu2Cl2 | Pmm2 | 2dm-1748 | Cu2Cl2#Pmm2#2dm-1748 | 4.72 | 2.29 | 2.97 | 0.59 | 4.70 | 4.20 | 4.21 | 2.40 | 0.772 | 0.873 | 0.822 | 0.689 | 0.856 | 0.773 | 0.856 | 0.875 | 0.87 | 0.812 | 0.790 | 0.801 | | 0.02 | -2.41 | -1.74 | -4.12 | 0.52 | -1.90 | -1.23 | -3.61 | 0.51 | -1.92 | -1.25 | -3.62 | 2.32 | -0.10 | 0.57 | -1.81 | -0.51 | -0.49 | -2.31 | 0.49 | -0.01 | -1.82 |
| Cu2Se2 | P2/m | 2dm-1729 | Cu2Se2#P2__m#2dm-1729 | 2.94 | 1.01 | 1.68 | 0.42 | 3.00 | 3.11 | 3.11 | 2.51 | 0.832 | 0.828 | 0.830 | 0.790 | 0.770 | 0.780 | 0.763 | 0.784 | 0.77 | 0.777 | 0.722 | 0.749 | | -0.07 | -2.00 | -1.32 | -2.59 | -0.18 | -2.11 | -1.43 | -2.70 | -0.17 | -2.10 | -1.43 | -2.69 | 0.43 | -1.50 | -0.83 | -2.09 | 0.11 | 0.11 | -0.50 | -0.11 | 0.01 | -0.60 |
| Cu4S2 | P1 | 2dm-1510 | Cu4S2#P1#2dm-1510 | 3.93 | 1.73 | 2.35 | 0.80 | 4.23 | 3.40 | 3.89 | 2.95 | 0.801 | 0.892 | 0.846 | 0.723 | 0.861 | 0.792 | 0.795 | 0.828 | 0.81 | 0.739 | 0.693 | 0.716 | | -0.29 | -2.50 | -1.88 | -3.42 | 0.54 | -1.67 | -1.04 | -2.59 | 0.04 | -2.16 | -1.54 | -3.09 | 0.98 | -1.22 | -0.60 | -2.15 | -0.83 | -0.33 | -1.28 | 0.33 | -0.50 | -0.94 |
| Cu4Se2 | P1 | 2dm-1061 | Cu4Se2#P1#2dm-1061 | 2.93 | 1.36 | 1.99 | 0.64 | 3.30 | 3.05 | 3.25 | 2.54 | 0.818 | 0.878 | 0.848 | 0.756 | 0.843 | 0.800 | 0.803 | 0.856 | 0.83 | 0.789 | 0.700 | 0.744 | | -0.37 | -1.94 | -1.31 | -2.66 | -0.12 | -1.70 | -1.06 | -2.42 | -0.32 | -1.89 | -1.26 | -2.61 | 0.39 | -1.19 | -0.56 | -1.91 | -0.25 | -0.05 | -0.76 | 0.05 | -0.20 | -0.71 |
| CuBr | Amm2 | 2dm-3405 | CuBr#Amm2#2dm-3405 | 4.09 | 2.75 | 3.42 | 1.02 | 4.40 | 3.57 | 4.26 | 3.11 | 0.796 | 0.898 | 0.847 | 0.733 | 0.888 | 0.810 | 0.840 | 0.858 | 0.85 | 0.708 | 0.689 | 0.698 | | -0.30 | -1.64 | -0.97 | -3.38 | 0.52 | -0.82 | -0.15 | -2.56 | -0.17 | -1.51 | -0.84 | -3.25 | 0.99 | -0.36 | 0.32 | -2.09 | -0.82 | -0.13 | -1.29 | 0.13 | -0.69 | -1.16 |
| CuHO2 | P2/m | 2dm-5004 | CuHO2#P2__m#2dm-5004 | 4.02 | 2.71 | 3.40 | 0.75 | 4.06 | 3.65 | 4.03 | 2.55 | 0.750 | 0.751 | 0.751 | 0.673 | 0.731 | 0.702 | 0.818 | 0.857 | 0.84 | 0.775 | 0.734 | 0.755 | | -0.04 | -1.36 | -0.67 | -3.31 | 0.37 | -0.95 | -0.26 | -2.90 | -0.01 | -1.33 | -0.64 | -3.28 | 1.47 | 0.16 | 0.85 | -1.79 | -0.41 | -0.03 | -1.52 | 0.03 | -0.38 | -1.49 |
| CuI | Amm2 | 2dm-1652 | CuI#Amm2#2dm-1652 | 4.82 | 2.93 | 3.62 | 1.43 | 4.96 | 5.05 | 5.04 | 3.41 | 0.830 | 0.905 | 0.868 | 0.794 | 0.889 | 0.841 | 0.770 | 0.781 | 0.78 | 0.753 | 0.725 | 0.739 | | -0.14 | -2.03 | -1.34 | -3.53 | -0.23 | -2.12 | -1.43 | -3.63 | -0.22 | -2.11 | -1.42 | -3.62 | 1.41 | -0.48 | 0.21 | -1.99 | 0.09 | 0.08 | -1.55 | -0.08 | 0.01 | -1.63 |
| Ga2Br2 | P4/nmm | 2dm-1343 | Ga2Br2#P4__nmm#2dm-1343 | 5.18 | 3.49 | 4.17 | 2.71 | 5.00 | 4.91 | 4.99 | 4.39 | 0.866 | 0.912 | 0.889 | 0.819 | 0.885 | 0.852 | 0.825 | 0.839 | 0.83 | 0.741 | 0.760 | 0.750 | | 0.19 | -1.51 | -0.83 | -2.29 | 0.27 | -1.43 | -0.74 | -2.20 | 0.20 | -1.50 | -0.82 | -2.28 | 0.79 | -0.91 | -0.23 | -1.69 | -0.09 | -0.01 | -0.60 | 0.01 | -0.08 | -0.59 |
| Ga2Cl2 | P4/nmm | 2dm-958 | Ga2Cl2#P4__nmm#2dm-958 | 5.75 | 4.02 | 4.73 | 3.17 | 5.88 | 5.15 | 5.58 | 5.10 | 0.866 | 0.921 | 0.893 | 0.827 | 0.895 | 0.861 | 0.818 | 0.832 | 0.82 | 0.773 | 0.734 | 0.754 | | -0.13 | -1.85 | -1.15 | -2.71 | 0.60 | -1.12 | -0.42 | -1.98 | 0.17 | -1.56 | -0.86 | -2.41 | 0.65 | -1.07 | -0.37 | -1.93 | -0.73 | -0.29 | -0.78 | 0.29 | -0.43 | -0.48 |
| Ga2I2 | P4/nmm | 2dm-1033 | Ga2I2#P4__nmm#2dm-1033 | 4.72 | 3.03 | 3.69 | 2.36 | 5.00 | 4.29 | 4.70 | 3.70 | 0.867 | 0.897 | 0.882 | 0.814 | 0.867 | 0.841 | 0.745 | 0.764 | 0.75 | 0.785 | 0.688 | 0.737 | | -0.28 | -1.97 | -1.30 | -2.63 | 0.42 | -1.26 | -0.60 | -1.93 | 0.02 | -1.67 | -1.00 | -2.33 | 1.02 | -0.67 | 0.00 | -1.33 | -0.71 | -0.30 | -1.30 | 0.30 | -0.40 | -1.00 |
| Ga2O3 | P1 | 2dm-2585 | Ga2O3#P1#2dm-2585 | 2.91 | 1.32 | 2.01 | 0.69 | 3.27 | 2.53 | 3.26 | 1.95 | 0.820 | 0.827 | 0.823 | 0.605 | 0.780 | 0.693 | 0.786 | 0.783 | 0.78 | 0.715 | 0.727 | 0.721 | | -0.36 | -1.95 | -1.26 | -2.58 | 0.38 | -1.21 | -0.52 | -1.84 | -0.35 | -1.94 | -1.25 | -2.57 | 0.95 | -0.63 | 0.05 | -1.26 | -0.74 | -0.01 | -1.32 | 0.01 | -0.73 | -1.30 |
| Ga2S2 | Amm2 | 2dm-3608 | Ga2S2#Amm2#2dm-3608 | 4.96 | 3.32 | 4.03 | 2.45 | 5.23 | 4.54 | 5.14 | 3.98 | 0.856 | 0.880 | 0.868 | 0.821 | 0.860 | 0.840 | 0.827 | 0.847 | 0.84 | 0.720 | 0.745 | 0.732 | | -0.27 | -1.91 | -1.20 | -2.78 | 0.42 | -1.23 | -0.52 | -2.09 | -0.18 | -1.82 | -1.11 | -2.69 | 0.98 | -0.66 | 0.04 | -1.53 | -0.68 | -0.09 | -1.25 | 0.09 | -0.59 | -1.15 |
| Ga2S2 | C2/m | 2dm-4063 | Ga2S2#C2__m#2dm-4063 | 4.78 | 3.29 | 3.99 | 2.27 | 4.81 | 4.55 | 4.74 | 4.12 | 0.857 | 0.879 | 0.868 | 0.820 | 0.860 | 0.840 | 0.815 | 0.857 | 0.84 | 0.745 | 0.694 | 0.720 | | -0.03 | -1.52 | -0.82 | -2.53 | 0.23 | -1.26 | -0.56 | -2.27 | 0.03 | -1.45 | -0.76 | -2.47 | 0.65 | -0.83 | -0.13 | -1.85 | -0.26 | -0.06 | -0.68 | 0.06 | -0.20 | -0.62 |
| Ga2S2 | Pm | 2dm-2968 | Ga2S2#Pm#2dm-2968 | 2.70 | 1.14 | 1.74 | 0.97 | 3.07 | 2.87 | 3.02 | 1.59 | 0.820 | 0.831 | 0.825 | 0.759 | 0.777 | 0.768 | 0.837 | 0.841 | 0.84 | 0.764 | 0.749 | 0.756 | | -0.37 | -1.93 | -1.33 | -2.10 | -0.17 | -1.73 | -1.13 | -1.90 | -0.32 | -1.88 | -1.28 | -2.05 | 1.12 | -0.45 | 0.15 | -0.62 | -0.20 | -0.05 | -1.49 | 0.05 | -0.15 | -1.44 |
| Ga2S5 | P1 | 2dm-1200 | Ga2S5#P1#2dm-1200 | 5.82 | 3.06 | 4.76 | 2.06 | 6.07 | 6.05 | 6.06 | 3.99 | 0.878 | 0.916 | 0.897 | 0.839 | 0.888 | 0.863 | 0.800 | 0.857 | 0.83 | 0.750 | 0.701 | 0.726 | | -0.26 | -3.02 | -1.31 | -4.01 | -0.23 | -2.99 | -1.29 | -3.99 | -0.25 | -3.01 | -1.30 | -4.00 | 1.82 | -0.93 | 0.77 | -1.93 | -0.02 | -0.01 | -2.08 | 0.01 | -0.01 | -2.07 |
| Ga2Se2 | Amm2 | 2dm-3530 | Ga2Se2#Amm2#2dm-3530 | 4.19 | 2.77 | 3.44 | 1.87 | 4.33 | 3.74 | 4.25 | 3.49 | 0.858 | 0.879 | 0.869 | 0.815 | 0.858 | 0.836 | 0.800 | 0.819 | 0.81 | 0.772 | 0.726 | 0.749 | | -0.14 | -1.57 | -0.90 | -2.46 | 0.45 | -0.97 | -0.30 | -1.87 | -0.06 | -1.48 | -0.81 | -2.38 | 0.70 | -0.73 | -0.06 | -1.63 | -0.59 | -0.08 | -0.84 | 0.08 | -0.51 | -0.76 |
| Ga2Se2 | Cm | 2dm-1630 | Ga2Se2#Cm#2dm-1630 | 4.94 | 2.51 | 3.17 | 1.63 | 4.89 | 4.69 | 4.85 | 3.24 | 0.858 | 0.878 | 0.868 | 0.814 | 0.858 | 0.836 | 0.847 | 0.890 | 0.87 | 0.781 | 0.743 | 0.762 | | 0.05 | -2.38 | -1.72 | -3.26 | 0.25 | -2.18 | -1.52 | -3.06 | 0.09 | -2.34 | -1.67 | -3.22 | 1.70 | -0.73 | -0.06 | -1.61 | -0.20 | -0.05 | -1.65 | 0.05 | -0.16 | -1.61 |
| Ga2Se2 | P1 | 2dm-1592 | Ga2Se2#P1#2dm-1592 | 4.35 | 2.44 | 3.12 | 1.68 | 4.49 | 4.53 | 4.52 | 3.35 | 0.880 | 0.899 | 0.890 | 0.842 | 0.873 | 0.858 | 0.725 | 0.802 | 0.76 | 0.738 | 0.706 | 0.722 | | -0.14 | -2.05 | -1.37 | -2.81 | -0.18 | -2.09 | -1.41 | -2.85 | -0.17 | -2.08 | -1.41 | -2.84 | 1.00 | -0.90 | -0.23 | -1.67 | 0.04 | 0.03 | -1.14 | -0.03 | 0.01 | -1.18 |
| Ga2Se3 | P1 | 2dm-1123 | Ga2Se3#P1#2dm-1123 | 5.77 | 2.92 | 4.59 | 1.97 | 5.94 | 5.96 | 5.96 | 4.03 | 0.883 | 0.914 | 0.899 | 0.849 | 0.881 | 0.865 | 0.823 | 0.878 | 0.85 | 0.760 | 0.735 | 0.747 | | -0.17 | -3.02 | -1.34 | -3.97 | -0.18 | -3.04 | -1.36 | -3.99 | -0.18 | -3.04 | -1.36 | -3.99 | 1.75 | -1.11 | 0.57 | -2.06 | 0.02 | 0.02 | -1.91 | -0.02 | 0.00 | -1.93 |
| Ga2Se5 | P1 | 2dm-572 | Ga2Se5#P1#2dm-572 | 4.22 | 2.62 | 3.31 | 1.72 | 4.43 | 3.68 | 4.39 | 3.38 | 0.852 | 0.862 | 0.857 | 0.795 | 0.867 | 0.831 | 0.754 | 0.755 | 0.75 | 0.785 | 0.734 | 0.759 | | -0.21 | -1.81 | -1.12 | -2.71 | 0.54 | -1.06 | -0.36 | -1.95 | -0.17 | -1.77 | -1.08 | -2.67 | 0.84 | -0.76 | -0.07 | -1.65 | -0.75 | -0.04 | -1.05 | 0.04 | -0.71 | -1.01 |
| Ga2Te2 | Cm | 2dm-1639 | Ga2Te2#Cm#2dm-1639 | 3.97 | 2.15 | 2.81 | 1.45 | 4.34 | 4.16 | 4.32 | 2.61 | 0.861 | 0.879 | 0.870 | 0.796 | 0.851 | 0.824 | 0.816 | 0.847 | 0.83 | 0.764 | 0.702 | 0.733 | | -0.37 | -2.19 | -1.53 | -2.89 | -0.19 | -2.00 | -1.35 | -2.71 | -0.35 | -2.17 | -1.51 | -2.87 | 1.37 | -0.45 | 0.21 | -1.16 | -0.18 | -0.02 | -1.73 | 0.02 | -0.16 | -1.72 |
| Ga2Te5 | P1 | 2dm-1545 | Ga2Te5#P1#2dm-1545 | 3.05 | 1.54 | 2.20 | 1.04 | 3.42 | 3.28 | 3.39 | 2.19 | 0.854 | 0.879 | 0.866 | 0.793 | 0.853 | 0.823 | 0.799 | 0.766 | 0.78 | 0.774 | 0.719 | 0.747 | | -0.37 | -1.87 | -1.22 | -2.38 | -0.22 | -1.73 | -1.08 | -2.23 | -0.34 | -1.85 | -1.20 | -2.35 | 0.87 | -0.64 | 0.01 | -1.14 | -0.14 | -0.02 | -1.23 | 0.02 | -0.12 | -1.21 |
| GaN | P1 | 2dm-2992 | GaN#P1#2dm-2992 | 4.91 | 3.49 | 4.20 | 2.21 | 5.06 | 4.44 | 4.86 | 4.26 | 0.823 | 0.893 | 0.858 | 0.790 | 0.879 | 0.834 | 0.839 | 0.855 | 0.85 | 0.737 | 0.779 | 0.758 | | -0.14 | -1.57 | -0.86 | -2.85 | 0.48 | -0.95 | -0.23 | -2.23 | 0.06 | -1.37 | -0.65 | -2.65 | 0.65 | -0.77 | -0.06 | -2.05 | -0.62 | -0.20 | -0.80 | 0.20 | -0.42 | -0.59 |
| Ge2 | P-1 | 2dm-3095 | Ge2#P-1#2dm-3095 | 1.34 | 0.06 | 0.55 | 0.00 | 1.75 | 1.73 | 1.74 | 1.60 | 0.839 | 0.841 | 0.840 | 0.775 | 0.843 | 0.809 | 0.808 | 0.883 | 0.85 | 0.708 | 0.712 | 0.710 | | -0.42 | -1.69 | -1.21 | -1.75 | -0.40 | -1.67 | -1.18 | -1.73 | -0.40 | -1.68 | -1.19 | -1.74 | -0.27 | -1.54 | -1.06 | -1.60 | -0.02 | -0.01 | -0.15 | 0.01 | -0.01 | -0.14 |
| Ge2H2 | C2/m | 2dm-124 | Ge2H2#C2__m#2dm-124 | 3.98 | 1.80 | 2.43 | 1.05 | 4.17 | 3.64 | 3.76 | 2.49 | 0.872 | 0.881 | 0.877 | 0.796 | 0.871 | 0.834 | 0.776 | 0.872 | 0.82 | 0.780 | 0.749 | 0.765 | | -0.19 | -2.37 | -1.74 | -3.13 | 0.33 | -1.84 | -1.21 | -2.60 | 0.22 | -1.96 | -1.33 | -2.72 | 1.49 | -0.68 | -0.06 | -1.44 | -0.53 | -0.41 | -1.69 | 0.41 | -0.12 | -1.28 |
| Ge2O2 | P4/nmm | 2dm-831 | Ge2O2#P4__nmm#2dm-831 | 5.00 | 3.31 | 4.05 | 2.22 | 5.11 | 4.43 | 5.00 | 4.42 | 0.847 | 0.902 | 0.874 | 0.817 | 0.884 | 0.851 | 0.815 | 0.816 | 0.82 | 0.712 | 0.730 | 0.721 | | -0.11 | -1.80 | -1.06 | -2.89 | 0.57 | -1.12 | -0.38 | -2.21 | 0.00 | -1.69 | -0.95 | -2.78 | 0.57 | -1.11 | -0.37 | -2.20 | -0.68 | -0.11 | -0.69 | 0.11 | -0.57 | -0.58 |
| Ge2Se2 | Pmm2 | 2dm-805 | Ge2Se2#Pmm2#2dm-805 | 3.87 | 2.28 | 2.93 | 1.68 | 4.21 | 4.02 | 4.14 | 2.91 | 0.851 | 0.876 | 0.864 | 0.796 | 0.855 | 0.825 | 0.807 | 0.895 | 0.85 | 0.736 | 0.736 | 0.736 | | -0.33 | -1.92 | -1.27 | -2.53 | -0.14 | -1.73 | -1.08 | -2.34 | -0.27 | -1.86 | -1.21 | -2.46 | 0.96 | -0.63 | 0.02 | -1.23 | -0.19 | -0.07 | -1.30 | 0.07 | -0.12 | -1.23 |
| Ge2Se2 | Pmn2_1 | 2dm-4478 | Ge2Se2#Pmn2_1#2dm-4478 | 3.37 | 1.80 | 2.43 | 1.29 | 3.69 | 3.52 | 3.67 | 2.39 | 0.858 | 0.878 | 0.868 | 0.798 | 0.848 | 0.823 | 0.797 | 0.827 | 0.81 | 0.734 | 0.743 | 0.738 | | -0.31 | -1.88 | -1.25 | -2.40 | -0.14 | -1.72 | -1.08 | -2.23 | -0.30 | -1.87 | -1.24 | -2.38 | 0.99 | -0.58 | 0.05 | -1.10 | -0.17 | -0.02 | -1.30 | 0.02 | -0.15 | -1.28 |
| Ge2Te2 | Pm | 2dm-3330 | Ge2Te2#Pm#2dm-3330 | 2.75 | 1.26 | 1.82 | 0.95 | 3.08 | 2.93 | 3.01 | 1.70 | 0.865 | 0.866 | 0.866 | 0.802 | 0.837 | 0.820 | 0.813 | 0.846 | 0.83 | 0.801 | 0.740 | 0.770 | | -0.33 | -1.83 | -1.27 | -2.14 | -0.17 | -1.67 | -1.11 | -1.98 | -0.26 | -1.75 | -1.20 | -2.07 | 1.06 | -0.44 | 0.12 | -0.75 | -0.16 | -0.07 | -1.39 | 0.07 | -0.08 | -1.32 |
| Ge2Te2 | Pmm2 | 2dm-781 | Ge2Te2#Pmm2#2dm-781 | 3.06 | 1.57 | 2.16 | 1.15 | 3.43 | 3.30 | 3.31 | 1.90 | 0.857 | 0.857 | 0.857 | 0.792 | 0.842 | 0.817 | 0.816 | 0.834 | 0.83 | 0.777 | 0.818 | 0.797 | | -0.36 | -1.86 | -1.27 | -2.28 | -0.24 | -1.73 | -1.14 | -2.15 | -0.24 | -1.74 | -1.15 | -2.15 | 1.16 | -0.34 | 0.25 | -0.75 | -0.13 | -0.12 | -1.52 | 0.12 | 0.00 | -1.40 |
| GeBr2 | P1 | 2dm-66 | GeBr2#P1#2dm-66 | 4.85 | 3.22 | 3.90 | 2.39 | 5.08 | 4.34 | 4.94 | 3.88 | 0.849 | 0.874 | 0.861 | 0.806 | 0.849 | 0.827 | 0.811 | 0.850 | 0.83 | 0.730 | 0.734 | 0.732 | | -0.23 | -1.85 | -1.18 | -2.69 | 0.51 | -1.12 | -0.44 | -1.95 | -0.09 | -1.71 | -1.04 | -2.55 | 0.97 | -0.65 | 0.02 | -1.49 | -0.74 | -0.14 | -1.20 | 0.14 | -0.60 | -1.06 |
| GeC | P1 | 2dm-2803 | GeC#P1#2dm-2803 | 4.66 | 3.14 | 3.83 | 2.37 | 5.00 | 4.82 | 4.94 | 3.63 | 0.826 | 0.860 | 0.843 | 0.776 | 0.864 | 0.820 | 0.772 | 0.823 | 0.80 | 0.721 | 0.806 | 0.763 | | -0.35 | -1.86 | -1.17 | -2.64 | -0.16 | -1.68 | -0.99 | -2.45 | -0.28 | -1.80 | -1.11 | -2.57 | 1.03 | -0.49 | 0.20 | -1.26 | -0.18 | -0.07 | -1.38 | 0.07 | -0.12 | -1.31 |
| GeCl2 | Cm | 2dm-497 | GeCl2#Cm#2dm-497 | 5.37 | 3.71 | 4.41 | 2.77 | 5.66 | 4.88 | 5.64 | 4.57 | 0.846 | 0.880 | 0.863 | 0.813 | 0.857 | 0.835 | 0.821 | 0.826 | 0.82 | 0.732 | 0.731 | 0.732 | | -0.29 | -1.95 | -1.25 | -2.89 | 0.49 | -1.17 | -0.47 | -2.11 | -0.26 | -1.93 | -1.23 | -2.87 | 0.80 | -0.86 | -0.16 | -1.80 | -0.78 | -0.02 | -1.09 | 0.02 | -0.76 | -1.07 |
| GeF2 | Amm2 | 2dm-2019 | GeF2#Amm2#2dm-2019 | 6.83 | 3.99 | 4.72 | 2.94 | 6.86 | 6.47 | 6.71 | 5.12 | 0.824 | 0.869 | 0.847 | 0.768 | 0.872 | 0.820 | 0.836 | 0.899 | 0.87 | 0.826 | 0.762 | 0.794 | | -0.03 | -2.87 | -2.14 | -3.92 | 0.36 | -2.48 | -1.75 | -3.53 | 0.12 | -2.72 | -1.99 | -3.77 | 1.71 | -1.13 | -0.40 | -2.18 | -0.39 | -0.15 | -1.74 | 0.15 | -0.24 | -1.59 |
| GeI2 | P-1 | 2dm-3455 | GeI2#P-1#2dm-3455 | 4.48 | 2.86 | 3.52 | 2.16 | 4.68 | 4.00 | 4.34 | 3.30 | 0.853 | 0.869 | 0.861 | 0.808 | 0.843 | 0.826 | 0.818 | 0.831 | 0.82 | 0.753 | 0.723 | 0.738 | | -0.20 | -1.83 | -1.16 | -2.52 | 0.48 | -1.15 | -0.48 | -1.84 | 0.14 | -1.48 | -0.82 | -2.18 | 1.18 | -0.44 | 0.22 | -1.14 | -0.68 | -0.34 | -1.38 | 0.34 | -0.34 | -1.04 |

| | | | | | | | | | | | | | | | | | | | | | | | | | | | | | | | | | | | | | | | | | | | | | |
|---|---|---|---|---|---|---|---|---|---|---|---|---|---|---|---|---|---|---|---|---|---|---|---|---|---|---|---|---|---|---|---|---|---|---|---|---|---|---|---|---|---|---|---|---|---|
| GeS2 | C2/m | 2dm-705 | GeS2#C2__m#2dm-705 | 3.05 | 1.50 | 2.17 | 0.73 | 3.44 | 3.29 | 3.38 | 2.19 | 0.828 | 0.858 | 0.843 | 0.788 | 0.836 | 0.812 | 0.815 | 0.856 | 0.84 | 0.804 | 0.712 | 0.758 | -0.39 | -1.95 | -1.27 | -2.71 | -0.25 | -1.80 | -1.12 | -2.57 | -0.33 | -1.89 | -1.21 | -2.65 | 0.86 | -0.69 | -0.02 | -1.46 | -0.15 | -0.06 | -1.25 | 0.06 | -0.09 | -1.19 |
| GeS2 | P-4m2 | 2dm-3352 | GeS2#P-4m2#2dm-3352 | 4.08 | 2.54 | 3.27 | 1.37 | 4.24 | 3.66 | 3.97 | 3.65 | 0.841 | 0.879 | 0.860 | 0.794 | 0.859 | 0.827 | 0.837 | 0.873 | 0.85 | 0.773 | 0.710 | 0.742 | -0.16 | -1.71 | -0.97 | -2.87 | 0.42 | -1.12 | -0.39 | -2.29 | 0.11 | -1.43 | -0.70 | -2.60 | 0.43 | -1.12 | -0.38 | -2.29 | -0.59 | -0.27 | -0.59 | 0.27 | -0.31 | -0.31 |
| GeTe | Cm | 2dm-3977 | GeTe#Cm#2dm-3977 | 3.88 | 2.37 | 3.01 | 1.75 | 4.11 | 3.52 | 3.80 | 2.69 | 0.852 | 0.862 | 0.857 | 0.791 | 0.840 | 0.815 | 0.792 | 0.796 | 0.79 | 0.693 | 0.730 | 0.712 | -0.23 | -1.74 | -1.10 | -2.36 | 0.35 | -1.16 | -0.52 | -1.77 | 0.08 | -1.43 | -0.79 | -2.05 | 1.18 | -0.33 | 0.31 | -0.94 | -0.59 | -0.31 | -1.42 | 0.31 | -0.28 | -1.11 |
| H2C2 | C2/m | 2dm-3228 | H2C2#C2__m#2dm-3228 | 5.82 | 4.38 | 5.05 | 3.41 | 5.94 | 5.27 | 5.62 | 5.24 | 0.885 | 0.957 | 0.921 | 0.853 | 0.943 | 0.898 | 0.841 | 0.881 | 0.86 | 0.770 | 0.736 | 0.753 | -0.12 | -1.56 | -0.89 | -2.53 | 0.55 | -0.89 | -0.22 | -1.86 | 0.20 | -1.24 | -0.56 | -2.21 | 0.58 | -0.85 | -0.18 | -1.82 | -0.67 | -0.32 | -0.71 | 0.32 | -0.35 | -0.38 |
| H2C2S | C2/m | 2dm-3158 | H2C2S#C2__m#2dm-3158 | 2.96 | 1.16 | 1.87 | 0.31 | 3.12 | 2.61 | 2.64 | 1.54 | 0.852 | 0.871 | 0.862 | 0.812 | 0.935 | 0.873 | 0.825 | 0.828 | 0.83 | 0.763 | 0.774 | 0.768 | -0.16 | -1.96 | -1.24 | -2.81 | 0.36 | -1.45 | -0.73 | -2.29 | 0.32 | -1.49 | -0.77 | -2.33 | 1.42 | -0.38 | 0.34 | -1.22 | -0.51 | -0.47 | -1.58 | 0.47 | -0.04 | -1.11 |
| H2Pb2 | P-1 | 2dm-284 | H2Pb2#P-1#2dm-284 | 1.37 | 0.09 | 0.58 | 0.00 | 1.79 | 1.68 | 1.70 | 1.49 | 0.847 | 0.867 | 0.857 | 0.791 | 0.830 | 0.811 | 0.780 | 0.871 | 0.83 | 0.783 | 0.721 | 0.752 | -0.42 | -1.70 | -1.21 | -1.79 | -0.31 | -1.59 | -1.10 | -1.68 | -0.34 | -1.61 | -1.12 | -1.70 | -0.12 | -1.40 | -0.91 | -1.49 | -0.11 | -0.08 | -0.30 | 0.08 | -0.02 | -0.22 |
| Hf2Br4 | P2_1/m | 2dm-17 | Hf2Br4#P2_1__m#2dm-17 | 2.91 | 1.32 | 1.98 | 0.80 | 3.25 | 2.66 | 3.08 | 2.09 | 0.837 | 0.840 | 0.839 | 0.786 | 0.819 | 0.803 | 0.822 | 0.833 | 0.83 | 0.731 | 0.722 | 0.727 | -0.34 | -1.93 | -1.27 | -2.44 | 0.25 | -1.34 | -0.69 | -1.86 | -0.17 | -1.76 | -1.10 | -2.27 | 0.82 | -0.77 | -0.11 | -1.28 | -0.59 | -0.17 | -1.16 | 0.17 | -0.42 | -0.99 |
| Hf2Br4 | Pm | 2dm-652 | Hf2Br4#Pm#2dm-652 | 3.15 | 1.55 | 2.21 | 1.03 | 3.46 | 2.88 | 3.38 | 2.08 | 0.839 | 0.843 | 0.841 | 0.788 | 0.828 | 0.808 | 0.809 | 0.827 | 0.82 | 0.798 | 0.722 | 0.760 | -0.31 | -1.91 | -1.25 | -2.43 | 0.27 | -1.33 | -0.68 | -1.85 | -0.22 | -1.83 | -1.17 | -2.35 | 1.08 | -0.53 | 0.13 | -1.05 | -0.58 | -0.08 | -1.38 | 0.08 | -0.50 | -1.30 |
| Hf2Cl4 | P2_1/m | 2dm-19 | Hf2Cl4#P2_1__m#2dm-19 | 3.21 | 1.61 | 2.27 | 1.00 | 3.56 | 3.33 | 3.48 | 2.20 | 0.842 | 0.842 | 0.842 | 0.784 | 0.819 | 0.801 | 0.765 | 0.772 | 0.77 | 0.716 | 0.800 | 0.758 | -0.35 | -1.95 | -1.28 | -2.56 | -0.12 | -1.72 | -1.05 | -2.33 | -0.27 | -1.87 | -1.20 | -2.48 | 1.01 | -0.60 | 0.07 | -1.21 | -0.23 | -0.08 | -1.35 | 0.08 | -0.15 | -1.27 |
| Hf2Cl4 | Pm | 2dm-337 | Hf2Cl4#Pm#2dm-337 | 3.25 | 1.61 | 2.28 | 1.00 | 3.60 | 2.97 | 3.48 | 2.20 | 0.834 | 0.846 | 0.840 | 0.781 | 0.818 | 0.799 | 0.796 | 0.826 | 0.81 | 0.711 | 0.687 | 0.699 | -0.35 | -1.99 | -1.32 | -2.60 | 0.29 | -1.35 | -0.68 | -1.97 | -0.23 | -1.87 | -1.20 | -2.48 | 1.05 | -0.59 | 0.08 | -1.20 | -0.64 | -0.12 | -1.40 | 0.12 | -0.51 | -1.28 |
| Hf2I4 | Pm | 2dm-1614 | Hf2I4#Pm#2dm-1614 | 2.69 | 1.10 | 1.75 | 0.71 | 3.05 | 2.87 | 2.87 | 2.17 | 0.840 | 0.844 | 0.842 | 0.789 | 0.833 | 0.811 | 0.810 | 0.808 | 0.81 | 0.747 | 0.720 | 0.734 | -0.36 | -1.95 | -1.31 | -2.34 | -0.18 | -1.76 | -1.12 | -2.15 | -0.18 | -1.77 | -1.13 | -2.16 | 0.52 | -1.07 | -0.43 | -1.46 | -0.19 | -0.18 | -0.88 | 0.18 | -0.01 | -0.70 |
| HfI4 | P4/mmm | 2dm-2401 | HfI4#P4__mmm#2dm-2401 | 4.08 | 1.49 | 2.15 | 0.71 | 4.21 | 3.85 | 3.85 | 2.59 | 0.845 | 0.854 | 0.850 | 0.798 | 0.825 | 0.811 | 0.789 | 0.818 | 0.80 | 0.715 | 0.689 | 0.702 | -0.14 | -2.73 | -2.06 | -3.50 | 0.23 | -2.36 | -1.70 | -3.14 | 0.23 | -2.36 | -1.70 | -3.14 | 1.49 | -1.10 | -0.43 | -1.87 | -0.36 | -0.36 | -1.63 | 0.36 | 0.00 | -1.26 |
| HfS2 | C2/m | 2dm-3834 | HfS2#C2__m#2dm-3834 | 3.77 | 2.17 | 2.84 | 1.33 | 4.04 | 3.90 | 3.98 | 2.87 | 0.847 | 0.856 | 0.851 | 0.795 | 0.834 | 0.815 | 0.810 | 0.867 | 0.84 | 0.766 | 0.779 | 0.772 | -0.27 | -1.87 | -1.20 | -2.71 | -0.13 | -1.73 | -1.06 | -2.57 | -0.21 | -1.81 | -1.14 | -2.65 | 0.90 | -0.70 | -0.04 | -1.54 | -0.14 | -0.06 | -1.17 | 0.06 | -0.08 | -1.11 |
| HfSe2 | C2/m | 2dm-4424 | HfSe2#C2__m#2dm-4424 | 2.96 | 1.38 | 2.03 | 0.68 | 3.33 | 3.14 | 3.24 | 1.88 | 0.848 | 0.856 | 0.852 | 0.799 | 0.830 | 0.815 | 0.802 | 0.864 | 0.83 | 0.745 | 0.717 | 0.731 | -0.37 | -1.94 | -1.30 | -2.65 | -0.18 | -1.76 | -1.11 | -2.46 | -0.28 | -1.86 | -1.21 | -2.57 | 1.08 | -0.50 | 0.15 | -1.21 | -0.19 | -0.09 | -1.45 | 0.09 | -0.10 | -1.36 |
| HfTaNO3 | Pmm2 | 2dm-3034 | HfTaNO3#Pmm2#2dm-3034 | 4.73 | 3.17 | 3.85 | 2.13 | 4.85 | 4.16 | 4.56 | 3.94 | 0.830 | 0.854 | 0.842 | 0.789 | 0.831 | 0.810 | 0.805 | 0.822 | 0.81 | 0.762 | 0.721 | 0.742 | -0.12 | -1.68 | -1.00 | -2.72 | 0.58 | -0.99 | -0.31 | -2.03 | 0.17 | -1.39 | -0.71 | -2.43 | 0.80 | -0.77 | -0.09 | -1.81 | -0.69 | -0.29 | -0.91 | 0.29 | -0.40 | -0.62 |
| HfTeSe4 | P2/m | 2dm-4723 | HfTeSe4#P2__m#2dm-4723 | 2.95 | 1.48 | 2.14 | 1.09 | 3.24 | 3.11 | 3.20 | 2.07 | 0.848 | 0.858 | 0.853 | 0.792 | 0.836 | 0.814 | 0.812 | 0.884 | 0.85 | 0.773 | 0.722 | 0.748 | -0.29 | -1.76 | -1.10 | -2.15 | -0.15 | -1.63 | -0.97 | -2.02 | -0.25 | -1.72 | -1.06 | -2.11 | 0.88 | -0.59 | 0.07 | -0.98 | -0.13 | -0.04 | -1.17 | 0.04 | -0.09 | -1.13 |
| Hg | P2/m | 2dm-4442 | Hg#P2__m#2dm-4442 | 4.18 | 2.58 | 3.21 | 2.22 | 4.51 | 3.77 | 4.20 | 3.24 | 0.854 | 0.898 | 0.876 | 0.743 | 0.862 | 0.802 | 0.800 | 0.818 | 0.81 | 0.828 | 0.771 | 0.799 | -0.33 | -1.94 | -1.30 | -2.30 | 0.42 | -1.19 | -0.56 | -1.55 | -0.02 | -1.62 | -0.99 | -1.99 | 0.94 | -0.67 | -0.03 | -1.03 | -0.75 | -0.31 | -1.27 | 0.31 | -0.43 | -0.96 |
| Hg2S2 | P4/nmm | 2dm-2440 | Hg2S2#P4__nmm#2dm-2440 | 3.72 | 2.08 | 2.77 | 1.07 | 3.96 | 3.40 | 3.74 | 2.73 | 0.828 | 0.887 | 0.858 | 0.758 | 0.863 | 0.810 | 0.811 | 0.820 | 0.82 | 0.739 | 0.698 | 0.719 | -0.24 | -1.88 | -1.18 | -2.89 | 0.32 | -1.32 | -0.62 | -2.33 | -0.02 | -1.66 | -0.97 | -2.68 | 0.99 | -0.65 | 0.05 | -1.66 | -0.56 | -0.21 | -1.23 | 0.21 | -0.34 | -1.01 |
| HgBi2S4 | P-1 | 2dm-4703 | HgBi2S4#P-1#2dm-4703 | 4.11 | 2.59 | 3.27 | 1.79 | 4.48 | 4.30 | 4.47 | 3.06 | 0.841 | 0.879 | 0.860 | 0.791 | 0.851 | 0.821 | 0.756 | 0.815 | 0.79 | 0.696 | 0.728 | 0.712 | -0.37 | -1.89 | -1.21 | -2.69 | -0.18 | -1.71 | -1.03 | -2.51 | -0.36 | -1.88 | -1.20 | -2.68 | 1.06 | -0.47 | 0.21 | -1.27 | -0.18 | -0.01 | -1.42 | 0.01 | -0.17 | -1.41 |
| HgCl2 | C2/m | 2dm-704 | HgCl2#C2__m#2dm-704 | 5.16 | 3.72 | 4.41 | 2.47 | 5.33 | 4.61 | 4.98 | 4.59 | 0.834 | 0.887 | 0.860 | 0.794 | 0.859 | 0.827 | 0.779 | 0.780 | 0.78 | 0.756 | 0.736 | 0.746 | -0.17 | -1.61 | -0.91 | -2.86 | 0.54 | -0.90 | -0.20 | -2.15 | 0.18 | -1.27 | -0.57 | -2.51 | 0.56 | -0.88 | -0.18 | -2.12 | -0.71 | -0.34 | -0.73 | 0.34 | -0.37 | -0.39 |
| HgCl2 | Pmm2 | 2dm-1588 | HgCl2#Pmm2#2dm-1588 | 5.73 | 3.69 | 4.42 | 2.39 | 5.94 | 5.94 | 5.94 | 5.31 | 0.865 | 0.914 | 0.889 | 0.831 | 0.884 | 0.857 | 0.804 | 0.930 | 0.87 | 0.718 | 0.805 | 0.761 | -0.21 | -2.24 | -1.51 | -3.54 | -0.21 | -2.25 | -1.52 | -3.55 | -0.21 | -2.25 | -1.52 | -3.55 | 0.42 | -1.62 | -0.89 | -2.91 | 0.01 | 0.00 | -0.63 | 0.00 | 0.00 | -0.63 |
| HgI2 | Cm | 2dm-3557 | HgI2#Cm#2dm-3557 | 4.45 | 2.89 | 3.61 | 1.94 | 4.61 | 3.95 | 4.44 | 3.87 | 0.862 | 0.874 | 0.868 | 0.804 | 0.859 | 0.832 | 0.735 | 0.787 | 0.76 | 0.754 | 0.737 | 0.745 | -0.16 | -1.72 | -1.01 | -2.68 | 0.51 | -1.05 | -0.34 | -2.01 | 0.02 | -1.54 | -0.83 | -2.50 | 0.58 | -0.98 | -0.27 | -1.94 | -0.67 | -0.18 | -0.74 | 0.18 | -0.49 | -0.56 |
| I4 | Pbam | 2dm-5502 | I4#Pbam#2dm-5502 | 3.28 | 1.76 | 2.40 | 1.08 | 3.40 | 2.85 | 3.38 | 2.32 | 0.843 | 0.850 | 0.847 | 0.765 | 0.804 | 0.785 | 0.795 | 0.835 | 0.82 | 0.765 | 0.729 | 0.747 | -0.13 | -1.64 | -1.00 | -2.33 | 0.42 | -1.09 | -0.46 | -1.78 | -0.10 | -1.62 | -0.98 | -2.30 | 0.95 | -0.56 | 0.08 | -1.25 | -0.55 | -0.02 | -1.08 | 0.02 | -0.53 | -1.06 |
| In2Cl2 | P4/nmm | 2dm-966 | In2Cl2#P4__nmm#2dm-966 | 5.49 | 3.79 | 4.47 | 3.05 | 5.73 | 5.64 | 5.73 | 4.83 | 0.873 | 0.919 | 0.896 | 0.831 | 0.893 | 0.862 | 0.748 | 0.757 | 0.75 | 0.782 | 0.789 | 0.785 | -0.24 | -1.94 | -1.27 | -2.69 | -0.15 | -1.85 | -1.17 | -2.59 | -0.24 | -1.93 | -1.26 | -2.68 | 0.66 | -1.03 | -0.36 | -1.78 | -0.10 | -0.01 | -0.91 | 0.01 | -0.09 | -0.90 |
| In2N2 | P-1 | 2dm-1839 | In2N2#P-1#2dm-1839 | 3.14 | 1.68 | 2.35 | 0.87 | 3.46 | 3.30 | 3.41 | 2.73 | 0.805 | 0.867 | 0.836 | 0.739 | 0.851 | 0.795 | 0.849 | 0.944 | 0.90 | 0.757 | 0.774 | 0.765 | -0.32 | -1.78 | -1.11 | -2.59 | -0.16 | -1.62 | -0.95 | -2.43 | -0.27 | -1.72 | -1.05 | -2.54 | 0.40 | -1.05 | -0.38 | -1.86 | -0.16 | -0.05 | -0.73 | 0.05 | -0.11 | -0.67 |
| In2S2 | Cm | 2dm-1655 | In2S2#Cm#2dm-1655 | 4.01 | 2.54 | 3.23 | 1.68 | 4.30 | 4.14 | 4.27 | 3.26 | 0.845 | 0.876 | 0.860 | 0.808 | 0.849 | 0.828 | 0.809 | 0.815 | 0.81 | 0.767 | 0.752 | 0.759 | -0.29 | -1.76 | -1.08 | -2.62 | -0.13 | -1.60 | -0.91 | -2.46 | -0.25 | -1.72 | -1.04 | -2.58 | 0.75 | -0.72 | -0.03 | -1.58 | -0.16 | -0.04 | -1.04 | 0.04 | -0.12 | -1.00 |
| In2S2 | P1 | 2dm-1715 | In2S2#P1#2dm-1715 | 4.99 | 2.96 | 3.65 | 2.20 | 5.15 | 5.21 | 5.21 | 3.60 | 0.876 | 0.908 | 0.892 | 0.836 | 0.879 | 0.858 | 0.783 | 0.817 | 0.80 | 0.795 | 0.742 | 0.768 | -0.16 | -2.18 | -1.50 | -2.95 | -0.22 | -2.25 | -1.56 | -3.01 | -0.22 | -2.24 | -1.55 | -3.01 | 1.38 | -0.64 | 0.05 | -1.40 | 0.06 | 0.06 | -1.54 | -0.06 | 0.00 | -1.60 |
| In2S2 | Pm | 2dm-2941 | In2S2#Pm#2dm-2941 | 2.51 | 1.15 | 1.68 | 0.96 | 2.75 | 2.31 | 2.58 | 2.24 | 0.833 | 0.835 | 0.834 | 0.748 | 0.785 | 0.767 | 0.818 | 0.872 | 0.84 | 0.723 | 0.766 | 0.744 | -0.24 | -1.60 | -1.07 | -1.79 | 0.21 | -1.16 | -0.63 | -1.34 | -0.07 | -1.43 | -0.90 | -1.62 | 0.27 | -1.09 | -0.56 | -1.28 | -0.44 | -0.17 | -0.51 | 0.17 | -0.27 | -0.34 |
| In2S5 | P1 | 2dm-1413 | In2S5#P1#2dm-1413 | 4.56 | 2.60 | 3.29 | 1.67 | 4.71 | 4.83 | 4.77 | 3.62 | 0.876 | 0.913 | 0.894 | 0.835 | 0.880 | 0.858 | 0.802 | 0.837 | 0.82 | 0.704 | 0.732 | 0.718 | -0.15 | -2.10 | -1.41 | -3.03 | -0.27 | -2.22 | -1.53 | -3.16 | -0.21 | -2.17 | -1.48 | -3.10 | 0.94 | -1.01 | -0.32 | -1.95 | 0.12 | 0.06 | -1.09 | -0.06 | 0.06 | -1.15 |
| In2Se2 | Cm | 2dm-211 | In2Se2#Cm#2dm-211 | 4.61 | 2.17 | 2.84 | 1.41 | 4.52 | 4.33 | 4.43 | 2.82 | 0.854 | 0.880 | 0.867 | 0.813 | 0.856 | 0.834 | 0.754 | 0.780 | 0.77 | 0.697 | 0.746 | 0.721 | 0.09 | -2.35 | -1.69 | -3.11 | 0.29 | -2.16 | -1.49 | -2.92 | 0.19 | -2.25 | -1.59 | -3.02 | 1.79 | -0.65 | 0.02 | -1.41 | -0.19 | -0.10 | -1.70 | 0.10 | -0.10 | -1.61 |
| In2Se5 | P1 | 2dm-1500 | In2Se5#P1#2dm-1500 | 4.86 | 2.37 | 3.06 | 1.49 | 5.04 | 4.33 | 4.61 | 3.10 | 0.853 | 0.897 | 0.875 | 0.791 | 0.865 | 0.828 | 0.815 | 0.835 | 0.83 | 0.722 | 0.709 | 0.715 | -0.18 | -2.67 | -1.98 | -3.55 | 0.53 | -1.96 | -1.27 | -2.84 | 0.26 | -2.24 | -1.55 | -3.12 | 1.76 | -0.73 | -0.04 | -1.61 | -0.71 | -0.43 | -1.94 | 0.43 | -0.28 | -1.51 |
| In2Te2 | Cm | 2dm-145 | In2Te2#Cm#2dm-145 | 3.37 | 1.99 | 2.59 | 1.32 | 3.64 | 3.48 | 3.60 | 2.45 | 0.857 | 0.878 | 0.867 | 0.800 | 0.849 | 0.825 | 0.831 | 0.904 | 0.87 | 0.755 | 0.716 | 0.735 | -0.26 | -1.65 | -1.05 | -2.32 | -0.11 | -1.49 | -0.89 | -2.16 | -0.22 | -1.61 | -1.01 | -2.28 | 0.92 | -0.46 | 0.14 | -1.13 | -0.16 | -0.04 | -1.19 | 0.04 | -0.12 | -1.15 |
| In2Te2 | P1 | 2dm-1769 | In2Te2#P1#2dm-1769 | 3.99 | 2.09 | 2.76 | 1.46 | 4.10 | 4.20 | 4.15 | 2.58 | 0.880 | 0.896 | 0.888 | 0.843 | 0.861 | 0.852 | 0.822 | 0.822 | 0.82 | 0.807 | 0.811 | 0.809 | -0.11 | -2.01 | -1.34 | -2.64 | -0.21 | -2.11 | -1.44 | -2.74 | -0.16 | -2.06 | -1.40 | -2.70 | 1.42 | -0.49 | 0.18 | -1.12 | 0.10 | 0.05 | -1.52 | -0.05 | 0.05 | -1.58 |
| In2Te5 | P1 | 2dm-4306 | In2Te5#P1#2dm-4306 | 3.32 | 1.79 | 2.46 | 1.18 | 3.25 | 2.98 | 3.22 | 2.32 | 0.852 | 0.881 | 0.866 | 0.789 | 0.856 | 0.822 | 0.770 | 0.840 | 0.80 | 0.754 | 0.736 | 0.745 | 0.07 | -1.46 | -0.80 | -2.07 | 0.34 | -1.19 | -0.53 | -1.80 | 0.10 | -1.43 | -0.77 | -2.04 | 1.01 | -0.52 | 0.14 | -1.13 | -0.27 | -0.03 | -0.94 | 0.03 | -0.24 | -0.91 |
| Ir2C4 | P-1 | 2dm-3031 | Ir2C4#P-1#2dm-3031 | 0.70 | 0.16 | 0.40 | 0.13 | 0.47 | 0.16 | 0.37 | 0.10 | 0.798 | 0.818 | 0.808 | 0.772 | 0.802 | 0.787 | 0.805 | 0.812 | 0.81 | 0.713 | 0.715 | 0.714 | 0.23 | -0.31 | -0.07 | -0.34 | 0.54 | 0.00 | 0.24 | -0.04 | 0.33 | -0.21 | 0.03 | -0.24 | 0.60 | 0.06 | 0.30 | 0.03 | -0.31 | -0.10 | -0.37 | 0.10 | -0.21 | -0.27 |
| LaHBr2 | Amm2 | 2dm-4199 | LaHBr2#Amm2#2dm-4199 | 6.85 | 5.17 | 5.89 | 3.75 | 7.04 | 6.17 | 6.85 | 5.76 | 0.864 | 0.879 | 0.871 | 0.825 | 0.864 | 0.844 | 0.747 | 0.807 | 0.78 | 0.708 | 0.633 | 0.671 | -0.19 | -1.87 | -1.15 | -3.29 | 0.69 | -1.00 | -0.28 | -2.42 | 0.00 | -1.69 | -0.96 | -3.11 | 1.09 | -0.60 | 0.13 | -2.02 | -0.88 | -0.19 | -1.28 | 0.19 | -0.69 | -1.09 |
| Li2CeP2 | C2/m | 2dm-4916 | Li2CeP2#C2__m#2dm-4916 | 3.37 | 1.66 | 2.35 | 0.85 | 3.21 | 3.02 | 3.20 | 2.39 | 0.764 | 0.780 | 0.772 | 0.567 | 0.798 | 0.683 | 0.806 | 0.832 | 0.82 | 0.724 | 0.749 | 0.737 | 0.16 | -1.55 | -0.86 | -2.36 | 0.36 | -1.35 | -0.67 | -2.17 | 0.17 | -1.54 | -0.85 | -2.36 | 0.98 | -0.73 | -0.04 | -1.54 | -0.19 | 0.00 | -0.82 | 0.00 | -0.19 | -0.82 |
| Mg3C | Pmm2 | 2dm-3115 | Mg3C#Pmm2#2dm-3115 | 2.14 | 0.68 | 1.26 | 0.35 | 2.08 | 1.79 | 1.92 | 1.68 | 0.808 | 0.827 | 0.818 | 0.719 | 0.780 | 0.749 | 0.820 | 0.843 | 0.83 | 0.752 | 0.813 | 0.782 | 0.06 | -1.40 | -0.82 | -1.73 | 0.35 | -1.11 | -0.53 | -1.44 | 0.22 | -1.24 | -0.66 | -1.58 | 0.46 | -1.00 | -0.42 | -1.33 | -0.29 | -0.16 | -0.40 | 0.16 | -0.14 | -0.24 |
| MgI2 | P-4m2 | 2dm-1587 | MgI2#P-4m2#2dm-1587 | 6.64 | 4.65 | 5.35 | 3.61 | 6.50 | 6.85 | 6.77 | 6.21 | 0.894 | 0.932 | 0.913 | 0.864 | 0.910 | 0.887 | 0.856 | 0.870 | 0.86 | 0.811 | 0.736 | 0.773 | 0.14 | -1.85 | -1.15 | -2.89 | -0.22 | -2.21 | -1.50 | -3.25 | -0.13 | -2.12 | -1.41 | -3.16 | 0.43 | -1.56 | -0.85 | -2.60 | 0.35 | 0.27 | -0.29 | -0.27 | 0.09 | -0.56 |
| Mo2Se4 | Pm | 2dm-139 | Mo2Se4#Pm#2dm-139 | 2.40 | 0.60 | 1.24 | 0.22 | 2.56 | 2.66 | 2.65 | 2.00 | 0.854 | 0.833 | 0.843 | 0.821 | 0.804 | 0.812 | 0.753 | 0.789 | 0.77 | 0.786 | 0.711 | 0.748 | -0.16 | -1.96 | -1.32 | -2.34 | -0.26 | -2.06 | -1.43 | -2.45 | -0.25 | -2.05 | -1.41 | -2.43 | 0.41 | -1.39 | -0.76 | -1.78 | 0.10 | 0.09 | -0.56 | -0.09 | 0.01 | -0.65 |
| MoO3 | Pm | 2dm-3784 | MoO3#Pm#2dm-3784 | 3.48 | 1.82 | 2.55 | 0.77 | 3.57 | 3.00 | 3.38 | 2.88 | 0.774 | 0.835 | 0.805 | 0.757 | 0.814 | 0.785 | 0.732 | 0.786 | 0.76 | 0.734 | 0.763 | 0.748 | -0.08 | -1.74 | -1.02 | -2.79 | 0.48 | -1.17 | -0.45 | -2.23 | 0.11 | -1.55 | -0.83 | -2.61 | 0.60 | -1.06 | -0.34 | -2.11 | -0.57 | -0.19 | -0.68 | 0.19 | -0.38 | -0.49 |
| MoS2 | Amm2 | 2dm-3150 | MoS2#Amm2#2dm-3150 | 4.05 | 2.37 | 3.06 | 1.81 | 4.32 | 4.22 | 4.29 | 2.69 | 0.820 | 0.828 | 0.824 | 0.792 | 0.815 | 0.803 | 0.829 | 0.845 | 0.84 | 0.778 | 0.763 | 0.771 | -0.27 | -1.95 | -1.26 | -2.51 | -0.17 | -1.85 | -1.16 | -2.41 | -0.24 | -1.93 | -1.23 | -2.48 | 1.36 | -0.32 | 0.37 | -0.88 | -0.10 | -0.02 | -1.63 | 0.02 | -0.08 | -1.60 |
| MoSe2 | Amm2 | 2dm-3409 | MoSe2#Amm2#2dm-3409 | 3.79 | 2.16 | 2.82 | 1.67 | 3.93 | 3.40 | 3.70 | 2.31 | 0.823 | 0.827 | 0.825 | 0.785 | 0.817 | 0.801 | 0.794 | 0.799 | 0.80 | 0.730 | 0.818 | 0.774 | -0.14 | -1.77 | -1.11 | -2.26 | 0.39 | -1.24 | -0.58 | -1.73 | 0.08 | -1.54 | -0.88 | -2.03 | 1.48 | -0.14 | 0.52 | -0.64 | -0.53 | -0.22 | -1.62 | 0.22 | -0.30 | -1.40 |
| N2O3 | P1 | 2dm-1886 | N2O3#P1#2dm-1886 | 3.88 | 1.26 | 2.01 | 3.26 | 4.10 | 3.48 | 4.10 | 3.23 | 0.707 | 0.735 | 0.721 | 0.650 | 0.677 | 0.663 | 0.737 | 0.738 | 0.74 | 0.796 | 0.692 | 0.744 | -0.22 | -2.84 | -2.09 | -0.84 | 0.39 | -2.23 | -1.47 | -0.23 | -0.22 | -2.84 | -2.09 | -0.84 | 0.64 | -1.98 | -1.22 | 0.02 | -0.62 | 0.00 | -0.87 | 0.00 | -0.61 | -0.86 |
| NF3 | Amm2 | 2dm-2435 | NF3#Amm2#2dm-2435 | 3.33 | 1.21 | 1.95 | 0.47 | 3.57 | 2.86 | 3.51 | 2.25 | 0.722 | 0.832 | 0.777 | 0.627 | 0.664 | 0.646 | 0.806 | 0.828 | 0.82 | 0.709 | 0.816 | 0.762 | -0.23 | -2.35 | -1.62 | -3.10 | 0.47 | -1.65 | -0.91 | -2.39 | -0.18 | -2.30 | -1.57 | -3.05 | 1.08 | -1.04 | -0.30 | -1.78 | -0.71 | -0.05 | -1.32 | 0.05 | -0.65 | -1.26 |
| Ni2N4 | P1 | 2dm-1274 | Ni2N4#P1#2dm-1274 | 2.69 | 0.71 | 1.37 | 0.18 | 2.15 | 2.67 | 2.63 | 1.63 | 0.817 | 0.803 | 0.810 | 0.661 | 0.730 | 0.695 | 0.839 | 0.869 | 0.85 | 0.786 | 0.750 | 0.768 | 0.53 | -1.44 | -0.78 | -1.97 | 0.02 | -1.96 | -1.30 | -2.49 | 0.06 | -1.91 | -1.26 | -2.45 | 1.05 | -0.92 | -0.27 | -1.46 | 0.52 | 0.47 | -0.52 | -0.47 | 0.05 | -0.99 |
| Ni2Se4 | P2_1/c | 2dm-1115 | Ni2Se4#P2_1__c#2dm-1115 | 4.33 | 2.30 | 3.02 | 1.03 | 4.46 | 4.50 | 4.48 | 4.41 | 0.806 | 0.842 | 0.824 | 0.758 | 0.790 | 0.774 | 0.837 | 0.931 | 0.88 | 0.810 | 0.745 | 0.777 | -0.13 | -2.15 | -1.43 | -3.42 | -0.18 | -2.20 | -1.48 | -3.47 | -0.16 | -2.18 | -1.46 | -3.45 | -0.09 | -2.11 | -1.39 | -3.38 | 0.05 | 0.03 | -0.04 | -0.03 | 0.02 | -0.07 |
| Ni2Te4 | P1 | 2dm-1388 | Ni2Te4#P1#2dm-1388 | 3.91 | 1.93 | 2.63 | 0.95 | 3.89 | 4.12 | 4.10 | 3.61 | 0.847 | 0.847 | 0.847 | 0.766 | 0.792 | 0.779 | 0.753 | 0.796 | 0.77 | 0.759 | 0.728 | 0.744 | 0.02 | -1.96 | -1.26 | -2.94 | -0.21 | -2.19 | -1.49 | -3.17 | -0.19 | -2.17 | -1.47 | -3.15 | 0.29 | -1.68 | -0.99 | -2.66 | 0.23 | 0.21 | -0.28 | -0.21 | 0.02 | -0.48 |
| NiO2 | C2/m | 2dm-4739 | NiO2#C2__m#2dm-4739 | 4.58 | 3.21 | 3.81 | 1.32 | 4.84 | 4.03 | 4.39 | 2.35 | 0.742 | 0.744 | 0.743 | 0.672 | 0.728 | 0.700 | 0.733 | 0.754 | 0.74 | 0.728 | 0.722 | 0.725 | -0.26 | -1.64 | -1.03 | -3.53 | 0.55 | -0.83 | -0.22 | -2.72 | 0.20 | -1.18 | -0.58 | -3.07 | 2.23 | 0.86 | 1.46 | -1.03 | -0.81 | -0.46 | -2.50 | 0.46 | -0.35 | -2.04 |
| NiSe2 | C2/m | 2dm-421 | NiSe2#C2__m#2dm-421 | 2.16 | 0.71 | 1.32 | 0.22 | 2.50 | 2.29 | 2.40 | 1.88 | 0.781 | 0.826 | 0.804 | 0.715 | 0.784 | 0.749 | 0.771 | 0.765 | 0.77 | 0.759 | 0.796 | 0.778 | -0.34 | -1.79 | -1.18 | -2.28 | -0.13 | -1.58 | -0.97 | -2.07 | -0.24 | -1.69 | -1.08 | -2.18 | 0.28 | -1.17 | -0.56 | -1.66 | -0.21 | -0.10 | -0.62 | 0.10 | -0.12 | -0.52 |
| OsBr2 | C2/m | 2dm-637 | OsBr2#C2__m#2dm-637 | 4.50 | 2.81 | 3.54 | 1.27 | 4.71 | 4.28 | 4.31 | 2.64 | 0.823 | 0.850 | 0.836 | 0.768 | 0.816 | 0.792 | 0.858 | 0.874 | 0.87 | 0.806 | 0.789 | 0.798 | -0.21 | -1.90 | -1.18 | -3.44 | 0.23 | -1.47 | -0.74 | -3.01 | 0.19 | -1.50 | -0.78 | -3.04 | 1.86 | 0.17 | 0.90 | -1.37 | -0.43 | -0.40 | -2.07 | 0.40 | -0.03 | -1.67 |
| OsCl2 | C2/m | 2dm-486 | OsCl2#C2__m#2dm-486 | 4.77 | 3.06 | 3.79 | 1.41 | 4.92 | 4.52 | 4.85 | 3.08 | 0.829 | 0.850 | 0.840 | 0.781 | 0.821 | 0.801 | 0.841 | 0.931 | 0.89 | 0.824 | 0.757 | 0.791 | -0.15 | -1.87 | -1.13 | -3.52 | 0.25 | -1.47 | -0.73 | -3.12 | -0.08 | -1.80 | -1.06 | -3.45 | 1.70 | -0.02 | 0.71 | -1.67 | -0.40 | -0.07 | -1.85 | 0.07 | -0.33 | -1.78 |
| OsI2 | C2/m | 2dm-262 | OsI2#C2__m#2dm-262 | 4.28 | 2.58 | 3.28 | 1.36 | 4.61 | 3.81 | 4.53 | 2.17 | 0.828 | 0.853 | 0.841 | 0.761 | 0.824 | 0.792 | 0.837 | 0.852 | 0.84 | 0.733 | 0.752 | 0.743 | -0.33 | -2.03 | -1.33 | -3.25 | 0.47 | -1.23 | -0.53 | -2.46 | -0.25 | -1.95 | -1.25 | -3.17 | 2.10 | 0.41 | 1.11 | -0.82 | -0.80 | -0.08 | -2.44 | 0.08 | -0.72 | -2.36 |
| P2 | C2/m | 2dm-3043 | P2#C2__m#2dm-3043 | 4.55 | 2.78 | 3.51 | 1.95 | 4.78 | 4.10 | 4.32 | 3.27 | 0.845 | 0.857 | 0.851 | 0.785 | 0.829 | 0.807 | 0.812 | 0.808 | 0.81 | 0.706 | 0.726 | 0.716 | -0.23 | -2.01 | -1.27 | -2.83 | 0.45 | -1.33 | -0.59 | -2.15 | 0.23 | -1.55 | -0.81 | -2.37 | 1.28 | -0.50 | 0.24 | -1.32 | -0.68 | -0.46 | -1.51 | 0.46 | -0.22 | -1.05 |
| P2Au2S2 | Pc | 2dm-3793 | P2Au2S2#Pc#2dm-3793 | 3.29 | 1.68 | 2.34 | 1.13 | 3.63 | 3.41 | 3.61 | 2.34 | 0.840 | 0.857 | 0.849 | 0.764 | 0.815 | 0.790 | 0.781 | 0.832 | 0.81 | 0.770 | 0.742 | 0.756 | -0.34 | -1.96 | -1.29 | -2.51 | -0.12 | -1.73 | -1.07 | -2.29 | -0.32 | -1.94 | -1.27 | -2.49 | 0.96 | -0.66 | 0.01 | -1.21 | -0.22 | -0.02 | -1.30 | 0.02 | -0.20 | -1.27 |
| P2N2 | P1 | 2dm-2022 | P2N2#P1#2dm-2022 | 5.72 | 2.87 | 3.61 | 1.80 | 5.60 | 5.31 | 5.48 | 3.82 | 0.855 | 0.873 | 0.864 | 0.808 | 0.854 | 0.831 | 0.823 | 0.860 | 0.84 | 0.754 | 0.696 | 0.725 | 0.12 | -2.73 | -1.99 | -3.80 | 0.41 | -2.44 | -1.70 | -3.51 | 0.24 | -2.61 | -1.87 | -3.68 | 1.90 | -0.95 | -0.21 | -2.02 | -0.29 | -0.12 | -1.78 | 0.12 | -0.17 | -1.66 |
| P2O3 | P1 | 2dm-2448 | P2O3#P1#2dm-2448 | 3.59 | 1.89 | 2.61 | 0.94 | 3.92 | 3.17 | 3.88 | 2.63 | 0.733 | 0.805 | 0.769 | 0.674 | 0.760 | 0.717 | 0.816 | 0.836 | 0.83 | 0.730 | 0.820 | 0.775 | -0.33 | -2.03 | -1.31 | -2.98 | 0.42 | -1.28 | -0.56 | -2.23 | -0.29 | -1.99 | -1.27 | -2.94 | 0.96 | -0.74 | -0.02 | -1.69 | -0.75 | -0.04 | -1.29 | 0.04 | -0.71 | -1.25 |
| P2Se3 | Cm | 2dm-416 | P2Se3#Cm#2dm-416 | 2.70 | 1.19 | 1.77 | 0.71 | 2.88 | 2.41 | 2.67 | 1.97 | 0.838 | 0.846 | 0.842 | 0.794 | 0.815 | 0.804 | 0.817 | 0.841 | 0.83 | 0.735 | 0.716 | 0.726 | -0.17 | -1.69 | -1.10 | -2.17 | 0.29 | -1.23 | -0.64 | -1.70 | 0.04 | -1.48 | -0.89 | -1.96 | 0.73 | -0.78 | -0.20 | -1.26 | -0.46 | -0.21 | -0.91 | 0.21 | -0.26 | -0.70 |
| P4 | P2_1/m | 2dm-5959 | P4#P2_1__m#2dm-5959 | 2.21 | 0.62 | 1.25 | 0.15 | 2.32 | 1.83 | 2.22 | 1.73 | 0.847 | 0.848 | 0.847 | 0.771 | 0.815 | 0.793 | 0.800 | 0.842 | 0.82 | 0.750 | 0.803 | 0.777 | -0.11 | -1.70 | -1.07 | -2.17 | 0.38 | -1.21 | -0.58 | -1.68 | -0.01 | -1.59 | -0.97 | -2.06 | 0.48 | -1.11 | -0.48 | -1.58 | -0.49 | -0.10 | -0.59 | 0.10 | -0.38 | -0.48 |
| P4 | Pmna | 2dm-3404 | P4#Pmna#2dm-3404 | 3.18 | 1.53 | 2.21 | 0.84 | 3.55 | 3.33 | 3.39 | 1.92 | 0.850 | 0.865 | 0.857 | 0.794 | 0.836 | 0.815 | 0.820 | 0.854 | 0.84 | 0.795 | 0.731 | 0.763 | -0.37 | -2.02 | -1.34 | -2.71 | -0.15 | -1.80 | -1.12 | -2.49 | -0.21 | -1.86 | -1.18 | -2.55 | 1.26 | -0.39 | 0.29 | -1.08 | -0.22 | -0.15 | -1.63 | 0.15 | -0.06 | -1.48 |
| P4Pd2 | P1 | 2dm-1457 | P4Pd2#P1#2dm-1457 | 0.15 | 0.06 | 0.14 | 0.06 | 0.16 | 0.14 | 0.15 | 0.13 | 0.780 | 0.808 | 0.794 | 0.743 | 0.779 | 0.761 | 0.808 | 0.812 | 0.81 | 0.745 | 0.785 | 0.765 | 0.00 | -0.10 | -0.02 | -0.10 | 0.01 | -0.09 | 0.00 | -0.08 | 0.00 | -0.10 | -0.01 | -0.09 | 0.02 | -0.08 | 0.01 | -0.07 | -0.01 | 0.00 | -0.03 | 0.00 | -0.01 | -0.02 |
| P4W2 | P1 | 2dm-2531 | P4W2#P1#2dm-2531 | 0.57 | 0.18 | 0.49 | 0.02 | 0.33 | 0.30 | 0.32 | 0.23 | 0.830 | 0.841 | 0.835 | 0.808 | 0.837 | 0.823 | 0.811 | 0.805 | 0.81 | 0.748 | 0.742 | 0.745 | 0.24 | -0.14 | 0.17 | -0.31 | 0.27 | -0.12 | 0.19 | -0.28 | 0.25 | -0.13 | 0.18 | -0.30 | 0.34 | -0.05 | 0.26 | -0.21 | -0.03 | -0.01 | -0.10 | 0.01 | -0.02 | -0.09 |
| Pb2O2 | P4/nmm | 2dm-3561 | Pb2O2#P4__nmm#2dm-3561 | 5.05 | 3.52 | 4.20 | 2.61 | 5.29 | 5.20 | 5.25 | 4.06 | 0.822 | 0.923 | 0.873 | 0.753 | 0.899 | 0.826 | 0.739 | 0.752 | 0.75 | 0.807 | 0.820 | 0.813 | -0.24 | -1.77 | -1.09 | -2.68 | -0.16 | -1.69 | -1.01 | -2.59 | -0.20 | -1.73 | -1.05 | -2.64 | 0.98 | -0.55 | 0.13 | -1.45 | -0.08 | -0.04 | -1.22 | 0.04 | -0.05 | -1.19 |
| Pb2S2 | Pmn2_1 | 2dm-3278 | Pb2S2#Pmn2_1#2dm-3278 | 4.33 | 2.81 | 3.48 | 2.09 | 4.64 | 4.48 | 4.57 | 3.30 | 0.843 | 0.901 | 0.872 | 0.800 | 0.875 | 0.837 | 0.788 | 0.867 | 0.83 | 0.810 | 0.819 | 0.815 | -0.31 | -1.84 | -1.16 | -2.55 | -0.15 | -1.67 | -1.00 | -2.39 | -0.24 | -1.77 | -1.09 | -2.48 | 1.03 | -0.50 | 0.18 | -1.21 | -0.17 | -0.07 | -1.34 | 0.07 | -0.09 | -1.27 |
| Pb2S2 | Pmn2_1 | 2dm-3045 | Pb2S2#Pmn2_1#2dm-3045 | 3.94 | 2.41 | 3.04 | 1.74 | 3.96 | 3.40 | 3.94 | 2.64 | 0.851 | 0.902 | 0.876 | 0.802 | 0.877 | 0.840 | 0.772 | 0.787 | 0.78 | 0.752 | 0.735 | 0.744 | -0.03 | -1.56 | -0.93 | -2.22 | 0.53 | -1.00 | -0.37 | -1.66 | -0.01 | -1.53 | -0.91 | -2.20 | 1.30 | -0.23 | 0.40 | -0.90 | -0.56 | -0.02 | -1.32 | 0.02 | -0.54 | -1.30 |
| Pb2Se2 | Pmm2 | 2dm-150 | Pb2Se2#Pmm2#2dm-150 | 4.07 | 2.63 | 3.27 | 1.96 | 4.34 | 3.73 | 4.25 | 2.89 | 0.849 | 0.892 | 0.871 | 0.793 | 0.866 | 0.830 | 0.804 | 0.800 | 0.80 | 0.746 | 0.746 | 0.746 | -0.28 | -1.71 | -1.08 | -2.38 | 0.34 | -1.10 | -0.47 | -1.77 | -0.18 | -1.62 | -0.98 | -2.29 | 1.18 | -0.26 | 0.38 | -0.93 | -0.61 | -0.09 | -1.46 | 0.09 | -0.52 | -1.36 |
| Pb2Se2 | Pmmn | 2dm-3073 | Pb2Se2#Pmmn#2dm-3073 | 3.50 | 1.98 | 2.59 | 1.39 | 3.87 | 3.69 | 3.85 | 2.52 | 0.855 | 0.893 | 0.874 | 0.802 | 0.867 | 0.835 | 0.740 | 0.690 | 0.72 | 0.713 | 0.736 | 0.724 | -0.36 | -1.89 | -1.27 | -2.47 | -0.19 | -1.71 | -1.10 | -2.30 | -0.34 | -1.87 | -1.25 | -2.45 | 0.99 | -0.54 | 0.08 | -1.13 | -0.18 | -0.02 | -1.35 | 0.02 | -0.15 | -1.33 |
| PbBr2 | Pm | 2dm-2251 | PbBr2#Pm#2dm-2251 | 5.51 | 3.89 | 4.58 | 3.09 | 5.58 | 4.80 | 5.48 | 4.76 | 0.849 | 0.887 | 0.868 | 0.810 | 0.874 | 0.842 | 0.810 | 0.861 | 0.84 | 0.737 | 0.695 | 0.716 | -0.07 | -1.69 | -1.00 | -2.49 | 0.71 | -0.90 | -0.22 | -1.71 | 0.03 | -1.59 | -0.90 | -2.39 | 0.75 | -0.86 | -0.18 | -1.67 | -0.79 | -0.10 | -0.83 | 0.10 | -0.68 | -0.72 |
| PbCl2 | Cm | 2dm-1488 | PbCl2#Cm#2dm-1488 | 5.73 | 4.08 | 4.77 | 3.10 | 5.38 | 5.01 | 5.25 | 4.98 | 0.859 | 0.902 | 0.880 | 0.821 | 0.883 | 0.852 | 0.792 | 0.871 | 0.83 | 0.788 | 0.748 | 0.768 | 0.35 | -1.30 | -0.61 | -2.28 | 0.72 | -0.93 | -0.24 | -1.91 | 0.48 | -1.17 | -0.48 | -2.16 | 0.76 | -0.90 | -0.21 | -1.88 | -0.37 | -0.13 | -0.40 | 0.13 | -0.24 | -0.28 |
| PbCl2 | Pm | 2dm-2389 | PbCl2#Pm#2dm-2389 | 5.97 | 4.33 | 5.03 | 3.44 | 6.06 | 5.30 | 5.83 | 5.27 | 0.848 | 0.890 | 0.869 | 0.811 | 0.882 | 0.847 | 0.761 | 0.759 | 0.76 | 0.806 | 0.778 | 0.792 | -0.08 | -1.73 | -1.02 | -2.62 | 0.68 | -0.97 | -0.26 | -1.86 | 0.15 | -1.50 | -0.79 | -2.39 | 0.70 | -0.94 | -0.24 | -1.84 | -0.76 | -0.23 | -0.78 | 0.23 | -0.53 | -0.55 |
| PbF2 | Amm2 | 2dm-2158 | PbF2#Amm2#2dm-2158 | 7.57 | 4.89 | 5.60 | 3.74 | 7.72 | 6.74 | 7.21 | 6.10 | 0.830 | 0.890 | 0.860 | 0.770 | 0.898 | 0.834 | 0.822 | 0.844 | 0.83 | 0.692 | 0.765 | 0.729 | -0.15 | -2.83 | -2.11 | -3.98 | 0.82 | -1.85 | -1.14 | -3.00 | 0.35 | -2.32 | -1.61 | -3.47 | 1.47 | -1.20 | -0.49 | -2.35 | -0.98 | -0.51 | -1.62 | 0.51 | -0.47 | -1.12 |
| PbI2 | C2/m | 2dm-3464 | PbI2#C2__m#2dm-3464 | 4.88 | 3.34 | 4.00 | 2.59 | 5.15 | 4.44 | 4.91 | 3.91 | 0.856 | 0.883 | 0.869 | 0.812 | 0.861 | 0.836 | 0.794 | 0.786 | 0.79 | 0.703 | 0.705 | 0.704 | -0.26 | -1.80 | -1.14 | -2.56 | 0.44 | -1.10 | -0.44 | -1.85 | -0.03 | -1.57 | -0.91 | -2.32 | 0.97 | -0.57 | 0.09 | -1.32 | -0.70 | -0.23 | -1.23 | 0.23 | -0.47 | -1.00 |
| PBr3 | Pm | 2dm-2386 | PBr3#Pm#2dm-2386 | 2.41 | 0.86 | 1.46 | 0.45 | 2.78 | 2.09 | 2.12 | 1.57 | 0.809 | 0.816 | 0.812 | 0.712 | 0.753 | 0.732 | 0.841 | 0.873 | 0.86 | 0.710 | 0.693 | 0.702 | -0.37 | -1.92 | -1.32 | -2.33 | 0.32 | -1.23 | -0.63 | -1.64 | 0.29 | -1.26 | -0.66 | -1.67 | 0.83 | -0.72 | -0.11 | -1.12 | -0.69 | -0.66 | -1.20 | 0.66 | -0.03 | -0.54 |
| PbS | Cm | 2dm-3013 | PbS#Cm#2dm-3013 | 4.50 | 2.98 | 3.65 | 2.22 | 4.81 | 4.73 | 4.80 | 3.64 | 0.833 | 0.887 | 0.860 | 0.782 | 0.859 | 0.820 | 0.786 | 0.821 | 0.80 | 0.813 | 0.697 | 0.755 | -0.32 | -1.83 | -1.17 | -2.59 | -0.24 | -1.75 | -1.09 | -2.51 | -0.30 | -1.82 | -1.15 | -2.58 | 0.86 | -0.66 | 0.01 | -1.42 | -0.08 | -0.01 | -1.17 | 0.01 | -0.07 | -1.16 |
| PbS2 | C2/m | 2dm-773 | PbS2#C2__m#2dm-773 | 2.94 | 1.41 | 2.06 | 0.77 | 3.27 | 3.08 | 3.26 | 1.77 | 0.798 | 0.845 | 0.821 | 0.740 | 0.810 | 0.775 | 0.865 | 0.946 | 0.91 | 0.551 | 0.751 | 0.651 | -0.33 | -1.86 | -1.20 | -2.50 | -0.14 | -1.67 | -1.01 | -2.31 | -0.32 | -1.86 | -1.20 | -2.49 | 1.17 | -0.37 | 0.29 | -1.01 | -0.19 | -0.01 | -1.49 | 0.01 | -0.19 | -1.49 |
| PbSe | Cm | 2dm-458 | PbSe#Cm#2dm-458 | 4.16 | 2.69 | 3.35 | 2.01 | 4.47 | 4.39 | 4.44 | 3.11 | 0.837 | 0.890 | 0.863 | 0.793 | 0.859 | 0.826 | 0.800 | 0.775 | 0.79 | 0.867 | 0.736 | 0.801 | -0.31 | -1.78 | -1.12 | -2.46 | -0.23 | -1.70 | -1.04 | -2.38 | -0.27 | -1.74 | -1.08 | -2.43 | 1.05 | -0.42 | 0.24 | -1.10 | -0.08 | -0.04 | -1.36 | 0.04 | -0.04 | -1.32 |
| PbSe2 | Cm | 2dm-678 | PbSe2#Cm#2dm-678 | 2.16 | 0.65 | 1.29 | 0.18 | 2.35 | 1.94 | 1.98 | 1.52 | 0.802 | 0.839 | 0.821 | 0.737 | 0.800 | 0.769 | 0.848 | 0.880 | 0.86 | 0.779 | 0.751 | 0.765 | -0.19 | -1.69 | -1.06 | -2.17 | 0.22 | -1.28 | -0.65 | -1.76 | 0.18 | -1.33 | -0.70 | -1.80 | 0.64 | -0.87 | -0.23 | -1.34 | -0.41 | -0.36 | -0.83 | 0.36 | -0.05 | -0.46 |
| PCl3 | Amm2 | 2dm-2306 | PCl3#Amm2#2dm-2306 | 2.94 | 1.32 | 1.96 | 0.74 | 3.23 | 2.51 | 2.78 | 1.63 | 0.813 | 0.818 | 0.815 | 0.718 | 0.760 | 0.739 | 0.816 | 0.827 | 0.82 | 0.782 | 0.730 | 0.756 | -0.30 | -1.91 | -1.27 | -2.49 | 0.43 | -1.19 | -0.55 | -1.77 | 0.15 | -1.46 | -0.83 | -2.04 | 1.31 | -0.31 | 0.33 | -0.89 | -0.72 | -0.45 | -1.60 | 0.45 | -0.28 | -1.15 |
| Pd2Br4 | P1 | 2dm-2651 | Pd2Br4#P1#2dm-2651 | 2.36 | 0.83 | 1.50 | 0.19 | 2.73 | 2.21 | 2.69 | 1.66 | 0.784 | 0.801 | 0.793 | 0.705 | 0.745 | 0.725 | 0.809 | 0.869 | 0.84 | 0.729 | 0.710 | 0.719 | -0.36 | -1.90 | -1.22 | -2.53 | 0.15 | -1.38 | -0.71 | -2.02 | -0.33 | -1.86 | -1.19 | -2.50 | 0.70 | -0.83 | -0.16 | -1.47 | -0.52 | -0.04 | -1.06 | 0.04 | -0.48 | -1.03 |
| Pd2Cl4 | P1 | 2dm-2514 | Pd2Cl4#P1#2dm-2514 | 2.58 | 1.05 | 1.75 | 1.00 | 2.81 | 2.24 | 2.66 | 1.74 | 0.781 | 0.793 | 0.787 | 0.711 | 0.735 | 0.723 | 0.784 | 0.781 | 0.78 | 0.765 | 0.806 | 0.786 | -0.23 | -1.76 | -1.06 | -1.81 | 0.34 | -1.19 | -0.49 | -1.24 | -0.08 | -1.61 | -0.91 | -1.66 | 0.85 | -0.68 | 0.02 | -0.73 | -0.57 | -0.15 | -1.08 | 0.15 | -0.42 | -0.93 |
| Pd2S4 | P2_1/c | 2dm-3017 | Pd2S4#P2_1__c#2dm-3017 | 3.82 | 2.21 | 2.91 | 1.20 | 4.11 | 3.98 | 4.07 | 2.64 | 0.821 | 0.828 | 0.825 | 0.763 | 0.793 | 0.778 | 0.802 | 0.804 | 0.80 | 0.805 | 0.689 | 0.747 | -0.29 | -1.91 | -1.20 | -2.91 | -0.16 | -1.78 | -1.07 | -2.79 | -0.25 | -1.86 | -1.16 | -2.87 | 1.18 | -0.44 | 0.27 | -1.44 | -0.13 | -0.05 | -1.47 | 0.05 | -0.08 | -1.42 |
| Pd2Se4 | P2_1/c | 2dm-3426 | Pd2Se4#P2_1__c#2dm-3426 | 3.90 | 2.32 | 3.01 | 1.46 | 3.53 | 3.38 | 3.49 | 2.73 | 0.822 | 0.835 | 0.828 | 0.754 | 0.805 | 0.780 | 0.799 | 0.824 | 0.81 | 0.743 | 0.714 | 0.728 | 0.38 | -1.20 | -0.52 | -2.07 | 0.53 | -1.05 | -0.37 | -1.92 | 0.41 | -1.16 | -0.48 | -2.03 | 1.17 | -0.41 | 0.27 | -1.27 | -0.15 | -0.04 | -0.79 | 0.04 | -0.11 | -0.75 |
| PdN2 | P1 | 2dm-1990 | PdN2#P1#2dm-1990 | 2.85 | 1.04 | 1.69 | 0.31 | 3.22 | 2.67 | 2.91 | 2.10 | 0.754 | 0.812 | 0.783 | 0.650 | 0.812 | 0.731 | 0.786 | 0.770 | 0.78 | 0.820 | 0.749 | 0.785 | -0.37 | -2.18 | -1.52 | -2.90 | 0.18 | -1.63 | -0.98 | -2.36 | -0.07 | -1.87 | -1.22 | -2.60 | 0.75 | -1.06 | -0.40 | -1.78 | -0.55 | -0.30 | -1.12 | 0.30 | -0.25 | -0.82 |
| PdS2 | C2/m | 2dm-71 | PdS2#C2__m#2dm-71 | 3.40 | 1.87 | 2.51 | 1.30 | 3.70 | 3.52 | 3.65 | 2.10 | 0.814 | 0.818 | 0.816 | 0.743 | 0.772 | 0.757 | 0.815 | 0.805 | 0.81 | 0.751 | 0.733 | 0.742 | -0.31 | -1.83 | -1.19 | -2.41 | -0.12 | -1.65 | -1.01 | -2.22 | -0.25 | -1.77 | -1.13 | -2.35 | 1.30 | -0.23 | 0.42 | -0.80 | -0.18 | -0.06 | -1.61 | 0.06 | -0.13 | -1.55 |
| PdSe2 | C2/m | 2dm-384 | PdSe2#C2__m#2dm-384 | 2.72 | 1.24 | 1.86 | 0.80 | 2.97 | 2.52 | 2.58 | 1.36 | 0.820 | 0.823 | 0.821 | 0.747 | 0.780 | 0.764 | 0.827 | 0.821 | 0.82 | 0.731 | 0.760 | 0.746 | -0.25 | -1.73 | -1.11 | -2.17 | 0.20 | -1.28 | -0.66 | -1.72 | 0.14 | -1.34 | -0.72 | -1.78 | 1.36 | -0.12 | 0.50 | -0.56 | -0.45 | -0.39 | -1.61 | 0.39 | -0.06 | -1.22 |
| PF3 | Amm2 | 2dm-2119 | PF3#Amm2#2dm-2119 | 4.93 | 2.99 | 3.71 | 1.92 | 4.98 | 4.44 | 4.79 | 3.74 | 0.804 | 0.827 | 0.815 | 0.749 | 0.784 | 0.767 | 0.840 | 0.886 | 0.86 | 0.809 | 0.694 | 0.752 | -0.05 | -1.99 | -1.27 | -3.06 | 0.49 | -1.45 | -0.72 | -2.52 | 0.14 | -1.80 | -1.07 | -2.87 | 1.19 | -0.75 | -0.03 | -1.82 | -0.54 | -0.19 | -1.24 | 0.19 | -0.35 | -1.05 |
| PI3 | Amm2 | 2dm-2689 | PI3#Amm2#2dm-2689 | 2.36 | 0.88 | 1.47 | 0.57 | 2.76 | 2.17 | 2.58 | 1.75 | 0.804 | 0.817 | 0.810 | 0.708 | 0.772 | 0.740 | 0.821 | 0.853 | 0.84 | 0.793 | 0.693 | 0.743 | -0.39 | -1.88 | -1.29 | -2.18 | 0.19 | -1.30 | -0.71 | -1.60 | -0.22 | -1.71 | -1.12 | -2.01 | 0.61 | -0.88 | -0.29 | -1.18 | -0.58 | -0.17 | -1.00 | 0.17 | -0.41 | -0.83 |
| Pt2Br4 | P1 | 2dm-2236 | Pt2Br4#P1#2dm-2236 | 2.27 | 0.69 | 1.33 | 0.02 | 2.67 | 1.98 | 2.34 | 1.51 | 0.796 | 0.816 | 0.806 | 0.725 | 0.757 | 0.741 | 0.823 | 0.845 | 0.83 | 0.648 | 0.787 | 0.717 | -0.41 | -1.98 | -1.35 | -2.65 | 0.29 | -1.29 | -0.65 | -1.96 | -0.07 | -1.65 | -1.01 | -2.32 | 0.76 | -0.82 | -0.18 | -1.48 | -0.69 | -0.33 | -1.17 | 0.33 | -0.36 | -0.83 |
| Pt2Cl4 | P1 | 2dm-5213 | Pt2Cl4#P1#2dm-5213 | 2.51 | 0.91 | 1.57 | 0.05 | 2.92 | 2.29 | 2.72 | 1.74 | 0.788 | 0.810 | 0.799 | 0.725 | 0.756 | 0.740 | 0.791 | 0.800 | 0.80 | 0.731 | 0.749 | 0.740 | -0.41 | -2.01 | -1.36 | -2.88 | 0.22 | -1.37 | -0.72 | -2.24 | -0.21 | -1.81 | -1.15 | -2.67 | 0.78 | -0.82 | -0.17 | -1.69 | -0.64 | -0.20 | -1.19 | 0.20 | -0.43 | -0.98 |
| Pt2F4 | P1 | 2dm-2586 | Pt2F4#P1#2dm-2586 | 3.37 | 1.77 | 2.46 | 0.34 | 3.64 | 3.18 | 3.43 | 2.97 | 0.789 | 0.802 | 0.796 | 0.729 | 0.758 | 0.743 | 0.823 | 0.835 | 0.83 | 0.753 | 0.738 | 0.746 | -0.27 | -1.87 | -1.18 | -3.30 | 0.20 | -1.40 | -0.71 | -2.84 | -0.06 | -1.66 | -0.96 | -3.09 | 0.40 | -1.20 | -0.51 | -2.64 | -0.47 | -0.21 | -0.67 | 0.21 | -0.25 | -0.45 |
| Pt2S4 | P2_1/c | 2dm-1006 | Pt2S4#P2_1__c#2dm-1006 | 4.90 | 2.81 | 3.53 | 1.73 | 5.02 | 4.63 | 4.84 | 3.54 | 0.843 | 0.860 | 0.852 | 0.798 | 0.829 | 0.813 | 0.837 | 0.849 | 0.84 | 0.763 | 0.741 | 0.752 | -0.12 | -2.21 | -1.49 | -3.29 | 0.28 | -1.81 | -1.10 | -2.90 | 0.06 | -2.03 | -1.31 | -3.11 | 1.36 | -0.73 | -0.02 | -1.82 | -0.39 | -0.18 | -1.48 | 0.18 | -0.21 | -1.30 |
| Pt2Se4 | P1 | 2dm-523 | Pt2Se4#P1#2dm-523 | 2.32 | 0.74 | 1.38 | 0.10 | 2.70 | 2.46 | 2.65 | 2.45 | 0.826 | 0.836 | 0.831 | 0.779 | 0.806 | 0.792 | 0.769 | 0.792 | 0.78 | 0.792 | 0.719 | 0.755 | -0.38 | -1.96 | -1.33 | -2.60 | -0.14 | -1.72 | -1.08 | -2.35 | -0.33 | -1.91 | -1.28 | -2.55 | -0.13 | -1.71 | -1.08 | -2.35 | -0.24 | -0.05 | -0.25 | 0.05 | -0.20 | -0.20 |
| PtO2 | C2/m | 2dm-4178 | PtO2#C2__m#2dm-4178 | 4.89 | 3.30 | 4.02 | 1.67 | 4.77 | 4.36 | 4.60 | 3.57 | 0.806 | 0.817 | 0.812 | 0.753 | 0.785 | 0.769 | 0.839 | 0.853 | 0.85 | 0.790 | 0.818 | 0.804 | 0.12 | -1.47 | -0.76 | -3.10 | 0.53 | -1.06 | -0.34 | -2.68 | 0.29 | -1.29 | -0.58 | -2.92 | 1.32 | -0.27 | 0.45 | -1.90 | -0.42 | -0.18 | -1.20 | 0.18 | -0.24 | -1.03 |
| PtS2 | C2/m | 2dm-4136 | PtS2#C2__m#2dm-4136 | 4.35 | 2.76 | 3.45 | 1.89 | 4.68 | 4.57 | 4.65 | 2.96 | 0.830 | 0.841 | 0.836 | 0.772 | 0.809 | 0.790 | 0.797 | 0.836 | 0.82 | 0.762 | 0.742 | 0.752 | -0.32 | -1.92 | -1.23 | -2.79 | -0.22 | -1.81 | -1.13 | -2.68 | -0.29 | -1.89 | -1.20 | -2.75 | 1.39 | -0.20 | 0.48 | -1.07 | -0.11 | -0.03 | -1.71 | 0.03 | -0.08 | -1.68 |
| PtSe2 | C2/m | 2dm-3287 | PtSe2#C2__m#2dm-3287 | 3.57 | 2.05 | 2.70 | 1.48 | 3.84 | 3.69 | 3.80 | 2.20 | 0.841 | 0.842 | 0.841 | 0.773 | 0.806 | 0.789 | 0.838 | 0.836 | 0.84 | 0.752 | 0.709 | 0.730 | -0.27 | -1.78 | -1.14 | -2.35 | -0.12 | -1.63 | -0.99 | -2.21 | -0.23 | -1.75 | -1.10 | -2.32 | 1.37 | -0.15 | 0.50 | -0.72 | -0.15 | -0.04 | -1.64 | 0.04 | -0.11 | -1.60 |
| Sb2 | C2/m | 2dm-3356 | Sb2#C2__m#2dm-3356 | 3.28 | 1.71 | 2.39 | 1.22 | 3.64 | 3.52 | 3.59 | 1.96 | 0.858 | 0.866 | 0.862 | 0.769 | 0.838 | 0.803 | 0.814 | 0.847 | 0.83 | 0.735 | 0.820 | 0.778 | -0.35 | -1.92 | -1.24 | -2.41 | -0.24 | -1.81 | -1.13 | -2.30 | -0.31 | -1.88 | -1.20 | -2.37 | 1.32 | -0.25 | 0.43 | -0.74 | -0.12 | -0.04 | -1.68 | 0.04 | -0.07 | -1.63 |
| Sb2As2 | P1 | 2dm-2075 | Sb2As2#P1#2dm-2075 | 2.06 | 0.72 | 1.25 | 0.51 | 2.39 | 2.25 | 2.26 | 2.24 | 0.857 | 0.858 | 0.857 | 0.782 | 0.834 | 0.808 | 0.845 | 0.884 | 0.86 | 0.811 | 0.786 | 0.798 | -0.33 | -1.67 | -1.14 | -1.88 | -0.19 | -1.53 | -1.00 | -1.74 | -0.20 | -1.54 | -1.01 | -1.75 | -0.18 | -1.52 | -0.99 | -1.73 | -0.14 | -0.13 | -0.15 | 0.13 | -0.01 | -0.02 |
| Sb2N2 | P1 | 2dm-5216 | Sb2N2#P1#2dm-5216 | 4.41 | 2.67 | 3.37 | 1.86 | 4.54 | 3.88 | 4.34 | 3.06 | 0.819 | 0.869 | 0.844 | 0.788 | 0.850 | 0.819 | 0.736 | 0.760 | 0.75 | 0.729 | 0.740 | 0.735 | -0.13 | -1.86 | -1.16 | -2.68 | 0.53 | -1.21 | -0.51 | -2.02 | 0.07 | -1.67 | -0.97 | -2.48 | 1.35 | -0.39 | 0.31 | -1.20 | -0.65 | -0.19 | -1.48 | 0.19 | -0.46 | -1.28 |
| Sb2O3 | Pm | 2dm-443 | Sb2O3#Pm#2dm-443 | 4.67 | 2.95 | 3.66 | 2.07 | 4.81 | 4.10 | 4.62 | 4.04 | 0.836 | 0.878 | 0.857 | 0.793 | 0.857 | 0.825 | 0.791 | 0.825 | 0.81 | 0.723 | 0.713 | 0.718 | -0.14 | -1.86 | -1.15 | -2.74 | 0.58 | -1.15 | -0.43 | -2.02 | 0.06 | -1.67 | -0.95 | -2.54 | 0.63 | -1.09 | -0.38 | -1.96 | -0.71 | -0.19 | -0.77 | 0.19 | -0.52 | -0.58 |
| Sb2P2 | P1 | 2dm-2610 | Sb2P2#P1#2dm-2610 | 2.47 | 0.96 | 1.59 | 0.55 | 2.61 | 2.17 | 2.51 | 2.13 | 0.851 | 0.858 | 0.854 | 0.781 | 0.828 | 0.805 | 0.726 | 0.789 | 0.76 | 0.775 | 0.702 | 0.739 | -0.14 | -1.65 | -1.02 | -2.06 | 0.30 | -1.21 | -0.58 | -1.61 | -0.04 | -1.55 | -0.92 | -1.95 | 0.34 | -1.17 | -0.54 | -1.57 | -0.44 | -0.10 | -0.48 | 0.10 | -0.34 | -0.38 |
| Sb2Te2I2 | P-1 | 2dm-3398 | Sb2Te2I2#P-1#2dm-3398 | 3.39 | 1.79 | 2.43 | 1.28 | 3.73 | 3.54 | 3.58 | 2.22 | 0.852 | 0.872 | 0.862 | 0.788 | 0.838 | 0.813 | 0.783 | 0.776 | 0.78 | 0.698 | 0.698 | 0.698 | -0.34 | -1.94 | -1.30 | -2.45 | -0.15 | -1.75 | -1.11 | -2.26 | -0.19 | -1.79 | -1.15 | -2.30 | 1.17 | -0.44 | 0.21 | -0.95 | -0.19 | -0.15 | -1.50 | 0.15 | -0.04 | -1.36 |
| SbBr3 | Pm | 2dm-2728 | SbBr3#Pm#2dm-2728 | 3.29 | 1.68 | 2.33 | 1.18 | 3.63 | 2.95 | 3.59 | 2.13 | 0.832 | 0.850 | 0.841 | 0.748 | 0.801 | 0.774 | 0.792 | 0.802 | 0.80 | 0.738 | 0.728 | 0.733 | -0.34 | -1.94 | -1.30 | -2.45 | 0.34 | -1.26 | -0.62 | -1.77 | -0.30 | -1.91 | -1.27 | -2.42 | 1.16 | -0.45 | 0.19 | -0.96 | -0.68 | -0.03 | -1.49 | 0.03 | -0.64 | -1.46 |
| SbBr5 | P1 | 2dm-2745 | SbBr5#P1#2dm-2745 | 2.21 | 0.46 | 1.13 | 1.77 | 2.61 | 2.47 | 2.59 | 2.27 | 0.736 | 0.833 | 0.784 | 0.649 | 0.792 | 0.720 | 0.813 | 0.813 | 0.81 | 0.745 | 0.718 | 0.732 | -0.39 | -2.15 | -1.47 | -0.84 | -0.26 | -2.02 | -1.34 | -0.71 | -0.38 | -2.13 | -1.45 | -0.82 | -0.06 | -1.81 | -1.14 | -0.50 | -0.13 | -0.02 | -0.34 | 0.02 | -0.11 | -0.32 |
| SbCl3 | Pm | 2dm-2244 | SbCl3#Pm#2dm-2244 | 3.73 | 2.07 | 2.75 | 1.44 | 4.00 | 3.23 | 3.61 | 2.81 | 0.834 | 0.861 | 0.848 | 0.761 | 0.813 | 0.787 | 0.800 | 0.814 | 0.81 | 0.753 | 0.690 | 0.721 | -0.27 | -1.93 | -1.25 | -2.56 | 0.50 | -1.16 | -0.48 | -1.79 | 0.11 | -1.54 | -0.87 | -2.17 | 0.92 | -0.73 | -0.06 | -1.37 | -0.77 | -0.39 | -1.19 | 0.39 | -0.38 | -0.81 |
| SbI3 | Pm | 2dm-2129 | SbI3#Pm#2dm-2129 | 2.78 | 1.24 | 1.85 | 0.87 | 3.18 | 2.53 | 3.00 | 1.85 | 0.825 | 0.842 | 0.833 | 0.731 | 0.784 | 0.758 | 0.752 | 0.814 | 0.78 | 0.764 | 0.710 | 0.737 | -0.40 | -1.94 | -1.33 | -2.30 | 0.25 | -1.29 | -0.68 | -1.65 | -0.23 | -1.77 | -1.15 | -2.13 | 0.93 | -0.62 | 0.00 | -0.98 | -0.65 | -0.17 | -1.32 | 0.17 | -0.48 | -1.15 |
| SeO3 | Cm | 2dm-2357 | SeO3#Cm#2dm-2357 | 4.16 | 2.17 | 2.91 | 0.49 | 4.48 | 3.37 | 4.33 | 2.56 | 0.769 | 0.815 | 0.792 | 0.689 | 0.785 | 0.737 | 0.815 | 0.834 | 0.82 | 0.742 | 0.724 | 0.733 | -0.32 | -2.30 | -1.56 | -3.99 | 0.79 | -1.20 | -0.46 | -2.88 | -0.17 | -2.15 | -1.41 | -3.84 | 1.60 | -0.39 | 0.35 | -2.07 | -1.11 | -0.15 | -1.91 | 0.15 | -0.96 | -1.76 |
| Si2 | P-1 | 2dm-5982 | Si2#P-1#2dm-5982 | 2.87 | 1.51 | 2.05 | 1.25 | 3.18 | 2.98 | 3.13 | 2.51 | 0.816 | 0.819 | 0.818 | 0.703 | 0.789 | 0.746 | 0.725 | 0.727 | 0.73 | 0.588 | 0.707 | 0.647 | -0.31 | -1.66 | -1.13 | -1.93 | -0.11 | -1.46 | -0.92 | -1.72 | -0.26 | -1.61 | -1.08 | -1.88 | 0.36 | -1.00 | -0.46 | -1.26 | -0.20 | -0.05 | -0.67 | 0.05 | -0.15 | -0.62 |
| Si2 | P-1 | 2dm-5934 | Si2#P-1#2dm-5934 | 2.81 | 1.46 | 2.00 | 1.19 | 3.12 | 2.61 | 3.02 | 2.64 | 0.812 | 0.816 | 0.814 | 0.707 | 0.778 | 0.742 | 0.747 | 0.739 | 0.74 | 0.782 | 0.707 | 0.744 | -0.30 | -1.65 | -1.12 | -1.93 | 0.21 | -1.14 | -0.61 | -1.42 | -0.21 | -1.56 | -1.02 | -1.83 | 0.17 | -1.18 | -0.64 | -1.45 | -0.51 | -0.10 | -0.48 | 0.10 | -0.41 | -0.38 |
| Si2H2 | C2/m | 2dm-4712 | Si2H2#C2__m#2dm-4712 | 4.65 | 2.95 | 3.63 | 2.20 | 4.87 | 4.17 | 4.81 | 3.72 | 0.854 | 0.867 | 0.860 | 0.791 | 0.846 | 0.819 | 0.748 | 0.773 | 0.76 | 0.763 | 0.727 | 0.745 | -0.23 | -1.93 | -1.25 | -2.67 | 0.47 | -1.23 | -0.55 | -1.97 | -0.17 | -1.87 | -1.19 | -2.61 | 0.93 | -0.77 | -0.09 | -1.51 | -0.70 | -0.06 | -1.16 | 0.06 | -0.64 | -1.10 |
| Si2O2 | P4mm | 2dm-344 | Si2O2#P4mm#2dm-344 | 4.00 | 3.15 | 3.91 | 1.99 | 4.21 | 3.55 | 4.19 | 3.34 | 0.850 | 0.889 | 0.870 | 0.813 | 0.869 | 0.841 | 0.829 | 0.852 | 0.84 | 0.718 | 0.700 | 0.709 | -0.21 | -1.06 | -0.30 | -2.22 | 0.45 | -0.40 | 0.36 | -1.56 | -0.19 | -1.04 | -0.28 | -2.20 | 0.66 | -0.19 | 0.57 | -1.35 | -0.66 | -0.02 | -0.87 | 0.02 | -0.64 | -0.85 |
| Si2S2 | Pmmn | 2dm-373 | Si2S2#Pmmn#2dm-373 | 4.23 | 2.45 | 3.16 | 1.66 | 4.60 | 4.33 | 4.38 | 2.98 | 0.839 | 0.862 | 0.850 | 0.788 | 0.845 | 0.816 | 0.808 | 0.836 | 0.82 | 0.724 | 0.738 | 0.731 | -0.36 | -2.14 | -1.43 | -2.94 | -0.10 | -1.88 | -1.17 | -2.67 | -0.14 | -1.92 | -1.22 | -2.72 | 1.25 | -0.52 | 0.18 | -1.32 | -0.26 | -0.22 | -1.62 | 0.22 | -0.04 | -1.40 |
| Si2S2 | Pmn2_1 | 2dm-279 | Si2S2#Pmn2_1#2dm-279 | 3.91 | 2.15 | 2.85 | 1.40 | 4.25 | 4.06 | 4.22 | 2.82 | 0.848 | 0.871 | 0.860 | 0.795 | 0.839 | 0.817 | 0.814 | 0.844 | 0.83 | 0.742 | 0.756 | 0.749 | -0.33 | -2.10 | -1.40 | -2.85 | -0.14 | -1.91 | -1.21 | -2.66 | -0.30 | -2.07 | -1.37 | -2.82 | 1.09 | -0.68 | 0.03 | -1.42 | -0.19 | -0.03 | -1.43 | 0.03 | -0.16 | -1.39 |
| Si2Se2 | Pmm2 | 2dm-117 | Si2Se2#Pmm2#2dm-117 | 3.52 | 1.88 | 2.51 | 1.29 | 3.78 | 3.18 | 3.77 | 2.48 | 0.850 | 0.857 | 0.854 | 0.778 | 0.843 | 0.810 | 0.800 | 0.808 | 0.80 | 0.893 | 0.731 | 0.812 | -0.27 | -1.90 | -1.27 | -2.49 | 0.33 | -1.30 | -0.67 | -1.89 | -0.25 | -1.89 | -1.26 | -2.47 | 1.03 | -0.60 | 0.03 | -1.19 | -0.60 | -0.02 | -1.30 | 0.02 | -0.58 | -1.29 |
| Si2Te2 | Pmm2 | 2dm-629 | Si2Te2#Pmm2#2dm-629 | 2.72 | 1.22 | 1.78 | 0.83 | 3.04 | 2.53 | 2.81 | 1.75 | 0.848 | 0.854 | 0.851 | 0.775 | 0.828 | 0.802 | 0.804 | 0.799 | 0.80 | 0.703 | 0.728 | 0.715 | -0.31 | -1.82 | -1.26 | -2.21 | 0.20 | -1.31 | -0.75 | -1.70 | -0.08 | -1.58 | -1.03 | -1.98 | 0.97 | -0.53 | 0.03 | -0.93 | -0.51 | -0.23 | -1.28 | 0.23 | -0.28 | -1.05 |
| Si3N4 | Pm | 2dm-5499 | Si3N4#Pm#2dm-5499 | 4.82 | 3.33 | 4.04 | 1.99 | 4.95 | 4.18 | 4.70 | 4.13 | 0.837 | 0.914 | 0.875 | 0.803 | 0.893 | 0.848 | 0.816 | 0.818 | 0.82 | 0.737 | 0.731 | 0.734 | -0.13 | -1.62 | -0.90 | -2.96 | 0.63 | -0.85 | -0.14 | -2.19 | 0.11 | -1.38 | -0.66 | -2.71 | 0.68 | -0.81 | -0.09 | -2.14 | -0.77 | -0.24 | -0.81 | 0.24 | -0.52 | -0.57 |
| SiBr2 | Cm | 2dm-68 | SiBr2#Cm#2dm-68 | 4.38 | 2.72 | 3.38 | 1.89 | 4.71 | 3.99 | 4.33 | 3.26 | 0.842 | 0.863 | 0.852 | 0.802 | 0.837 | 0.820 | 0.824 | 0.854 | 0.84 | 0.770 | 0.696 | 0.733 | -0.32 | -1.99 | -1.32 | -2.81 | 0.39 | -1.27 | -0.61 | -2.10 | 0.05 | -1.61 | -0.95 | -2.44 | 1.12 | -0.54 | 0.12 | -1.37 | -0.72 | -0.37 | -1.45 | 0.37 | -0.34 | -1.07 |
| SiBr4 | P4/mmm | 2dm-10 | SiBr4#P4__mmm#2dm-10 | 2.91 | 1.34 | 2.04 | 0.41 | 3.26 | 2.57 | 2.98 | 2.55 | 0.853 | 0.863 | 0.858 | 0.800 | 0.829 | 0.814 | 0.828 | 0.852 | 0.84 | 0.771 | 0.714 | 0.742 | -0.35 | -1.92 | -1.22 | -2.85 | 0.35 | -1.22 | -0.53 | -2.16 | -0.06 | -1.63 | -0.93 | -2.56 | 0.36 | -1.21 | -0.51 | -2.14 | -0.69 | -0.28 | -0.71 | 0.28 | -0.41 | -0.42 |
| SiC | P1 | 2dm-2686 | SiC#P1#2dm-2686 | 5.12 | 3.52 | 4.22 | 2.69 | 5.37 | 5.26 | 5.37 | 3.71 | 0.839 | 0.907 | 0.873 | 0.783 | 0.877 | 0.830 | 0.766 | 0.776 | 0.77 | 0.747 | 0.686 | 0.717 | -0.24 | -1.85 | -1.15 | -2.68 | -0.14 | -1.74 | -1.04 | -2.57 | -0.24 | -1.85 | -1.15 | -2.67 | 1.41 | -0.19 | 0.51 | -1.02 | -0.11 | 0.00 | -1.66 | 0.00 | -0.11 | -1.65 |
| SiCl2 | Cm | 2dm-126 | SiCl2#Cm#2dm-126 | 4.86 | 3.15 | 3.84 | 2.22 | 5.16 | 5.00 | 5.14 | 3.89 | 0.844 | 0.868 | 0.856 | 0.807 | 0.841 | 0.824 | 0.843 | 0.861 | 0.85 | 0.761 | 0.789 | 0.775 | -0.31 | -2.02 | -1.33 | -2.94 | -0.14 | -1.85 | -1.16 | -2.78 | -0.28 | -1.99 | -1.30 | -2.92 | 0.96 | -0.75 | -0.05 | -1.67 | -0.16 | -0.02 | -1.27 | 0.02 | -0.14 | -1.25 |
| SiF2 | Amm2 | 2dm-2020 | SiF2#Amm2#2dm-2020 | 5.78 | 2.88 | 3.60 | 1.89 | 5.65 | 5.13 | 5.39 | 3.81 | 0.819 | 0.848 | 0.833 | 0.759 | 0.838 | 0.798 | 0.744 | 0.844 | 0.79 | 0.806 | 0.711 | 0.759 | 0.13 | -2.77 | -2.05 | -3.76 | 0.65 | -2.25 | -1.53 | -3.24 | 0.39 | -2.51 | -1.79 | -3.50 | 1.97 | -0.93 | -0.21 | -1.92 | -0.52 | -0.26 | -1.84 | 0.26 | -0.26 | -1.58 |
| SiF2 | Cm | 2dm-533 | SiF2#Cm#2dm-533 | 4.85 | 3.09 | 3.84 | 1.88 | 5.11 | 4.43 | 4.97 | 4.43 | 0.838 | 0.884 | 0.861 | 0.797 | 0.861 | 0.829 | 0.790 | 0.823 | 0.81 | 0.718 | 0.734 | 0.726 | -0.26 | -2.02 | -1.27 | -3.23 | 0.42 | -1.34 | -0.59 | -2.55 | -0.12 | -1.88 | -1.14 | -3.09 | 0.42 | -1.33 | -0.59 | -2.55 | -0.68 | -0.14 | -0.68 | 0.14 | -0.54 | -0.55 |
| SiGe | Cm | 2dm-2388 | SiGe#Cm#2dm-2388 | 2.75 | 1.42 | 1.95 | 1.17 | 3.06 | 2.92 | 3.01 | 2.94 | 0.819 | 0.820 | 0.819 | 0.757 | 0.805 | 0.781 | 0.783 | 0.828 | 0.81 | 0.705 | 0.748 | 0.726 | -0.31 | -1.64 | -1.11 | -1.89 | -0.16 | -1.50 | -0.97 | -1.75 | -0.26 | -1.59 | -1.06 | -1.84 | -0.18 | -1.52 | -0.99 | -1.76 | -0.14 | -0.05 | -0.13 | 0.05 | -0.09 | -0.07 |
| SiI2 | Amm2 | 2dm-2084 | SiI2#Amm2#2dm-2084 | 3.95 | 2.26 | 2.92 | 1.66 | 4.18 | 3.53 | 3.61 | 2.52 | 0.846 | 0.852 | 0.849 | 0.782 | 0.824 | 0.803 | 0.754 | 0.772 | 0.76 | 0.830 | 0.726 | 0.778 | -0.23 | -1.92 | -1.26 | -2.52 | 0.42 | -1.27 | -0.61 | -1.86 | 0.34 | -1.35 | -0.69 | -1.95 | 1.43 | -0.26 | 0.40 | -0.85 | -0.65 | -0.57 | -1.66 | 0.57 | -0.08 | -1.09 |
| SiI2 | Cm | 2dm-1657 | SiI2#Cm#2dm-1657 | 4.11 | 2.46 | 3.11 | 1.77 | 4.42 | 3.75 | 3.88 | 2.77 | 0.852 | 0.859 | 0.855 | 0.802 | 0.832 | 0.817 | 0.770 | 0.791 | 0.78 | 0.759 | 0.591 | 0.675 | -0.31 | -1.96 | -1.31 | -2.65 | 0.36 | -1.29 | -0.64 | -1.98 | 0.23 | -1.42 | -0.77 | -2.11 | 1.34 | -0.31 | 0.34 | -1.00 | -0.67 | -0.54 | -1.65 | 0.54 | -0.13 | -1.11 |
| SiPb | Cm | 2dm-2391 | SiPb#Cm#2dm-2391 | 1.48 | 0.31 | 0.79 | 0.18 | 1.89 | 1.80 | 1.86 | 1.65 | 0.816 | 0.837 | 0.826 | 0.741 | 0.809 | 0.775 | 0.817 | 0.831 | 0.82 | 0.834 | 0.737 | 0.786 | -0.41 | -1.58 | -1.10 | -1.70 | -0.32 | -1.50 | -1.01 | -1.62 | -0.38 | -1.56 | -1.08 | -1.68 | -0.17 | -1.34 | -0.86 | -1.46 | -0.08 | -0.02 | -0.24 | 0.02 | -0.06 | -0.22 |
| SiS | Cm | 2dm-776 | SiS#Cm#2dm-776 | 4.89 | 3.04 | 3.78 | 2.21 | 5.18 | 4.41 | 4.91 | 3.81 | 0.834 | 0.855 | 0.844 | 0.777 | 0.838 | 0.808 | 0.779 | 0.781 | 0.78 | 0.690 | 0.720 | 0.705 | -0.29 | -2.14 | -1.40 | -2.96 | 0.48 | -1.37 | -0.63 | -2.20 | -0.02 | -1.87 | -1.13 | -2.70 | 1.08 | -0.77 | -0.04 | -1.60 | -0.77 | -0.27 | -1.36 | 0.27 | -0.50 | -1.10 |
| SiS2 | P-1 | 2dm-345 | SiS2#P-1#2dm-345 | 3.93 | 2.27 | 2.95 | 1.40 | 4.26 | 4.17 | 4.26 | 2.94 | 0.849 | 0.870 | 0.860 | 0.807 | 0.853 | 0.830 | 0.821 | 0.827 | 0.82 | 0.732 | 0.762 | 0.747 | -0.33 | -2.00 | -1.31 | -2.86 | -0.24 | -1.90 | -1.22 | -2.77 | -0.33 | -1.99 | -1.31 | -2.86 | 0.99 | -0.67 | 0.01 | -1.54 | -0.09 | 0.00 | -1.32 | 0.00 | -0.09 | -1.32 |
| SiSe2 | C2/m | 2dm-309 | SiSe2#C2__m#2dm-309 | 2.90 | 1.26 | 1.94 | 0.59 | 3.23 | 3.03 | 3.21 | 1.61 | 0.851 | 0.866 | 0.858 | 0.808 | 0.850 | 0.829 | 0.777 | 0.852 | 0.81 | 0.730 | 0.749 | 0.740 | -0.33 | -1.96 | -1.29 | -2.63 | -0.13 | -1.76 | -1.09 | -2.43 | -0.31 | -1.95 | -1.27 | -2.62 | 1.28 | -0.35 | 0.32 | -1.02 | -0.20 | -0.01 | -1.61 | 0.01 | -0.19 | -1.60 |
| SnBr4 | P4/mmm | 2dm-1152 | SnBr4#P4__mmm#2dm-1152 | 4.57 | 1.56 | 3.26 | 0.59 | 4.67 | 4.79 | 4.76 | 2.79 | 0.867 | 0.883 | 0.875 | 0.812 | 0.835 | 0.824 | 0.792 | 0.817 | 0.80 | 0.747 | 0.781 | 0.764 | -0.10 | -3.11 | -1.42 | -4.08 | -0.22 | -3.23 | -1.54 | -4.20 | -0.19 | -3.20 | -1.51 | -4.17 | 1.79 | -1.23 | 0.47 | -2.19 | 0.12 | 0.09 | -1.89 | -0.09 | 0.03 | -1.98 |
| SnC | Cm | 2dm-2507 | SnC#Cm#2dm-2507 | 3.45 | 2.01 | 2.70 | 1.16 | 3.78 | 3.67 | 3.75 | 2.23 | 0.817 | 0.855 | 0.836 | 0.763 | 0.832 | 0.797 | 0.810 | 0.886 | 0.85 | 0.746 | 0.693 | 0.720 | -0.33 | -1.77 | -1.08 | -2.62 | -0.21 | -1.65 | -0.97 | -2.51 | -0.30 | -1.74 | -1.05 | -2.59 | 1.23 | -0.22 | 0.47 | -1.07 | -0.11 | -0.03 | -1.55 | 0.03 | -0.08 | -1.52 |
| SnGe | P1 | 2dm-4906 | SnGe#P1#2dm-4906 | 1.56 | 0.32 | 0.81 | 0.19 | 1.97 | 1.88 | 1.94 | 1.74 | 0.829 | 0.839 | 0.834 | 0.771 | 0.826 | 0.798 | 0.796 | 0.834 | 0.81 | 0.805 | 0.809 | 0.807 | -0.41 | -1.66 | -1.16 | -1.79 | -0.32 | -1.56 | -1.07 | -1.70 | -0.38 | -1.62 | -1.13 | -1.76 | -0.18 | -1.42 | -0.92 | -1.55 | -0.09 | -0.03 | -0.23 | 0.03 | -0.06 | -0.20 |
| SO3 | Cm | 2dm-2250 | SO3#Cm#2dm-2250 | 4.69 | 2.18 | 2.92 | 0.48 | 4.76 | 4.41 | 4.64 | 2.81 | 0.784 | 0.841 | 0.812 | 0.736 | 0.816 | 0.776 | 0.846 | 0.856 | 0.85 | 0.750 | 0.742 | 0.746 | -0.07 | -2.59 | -1.84 | -4.28 | 0.28 | -2.23 | -1.49 | -3.93 | 0.05 | -2.46 | -1.72 | -4.16 | 1.89 | -0.63 | 0.11 | -2.32 | -0.35 | -0.12 | -1.96 | 0.12 | -0.23 | -1.83 |
| SrO2 | P4/mmm | 2dm-11 | SrO2#P4__mmm#2dm-11 | 6.12 | 4.78 | 5.47 | 2.96 | 5.94 | 5.37 | 5.73 | 5.04 | 0.835 | 0.930 | 0.883 | 0.816 | 0.915 | 0.865 | 0.811 | 0.810 | 0.81 | 0.734 | 0.744 | 0.739 | 0.18 | -1.15 | -0.46 | -2.98 | 0.75 | -0.59 | 0.10 | -2.42 | 0.39 | -0.95 | -0.26 | -2.77 | 1.08 | -0.26 | 0.43 | -2.09 | -0.56 | -0.21 | -0.89 | 0.21 | -0.36 | -0.69 |
| Te2Au2 | P2/m | 2dm-1409 | Te2Au2#P2__m#2dm-1409 | 2.12 | 0.61 | 1.23 | 0.26 | 2.51 | 2.28 | 2.46 | 2.14 | 0.843 | 0.844 | 0.844 | 0.764 | 0.800 | 0.782 | 0.827 | 0.852 | 0.84 | 0.801 | 0.723 | 0.762 | -0.39 | -1.90 | -1.28 | -2.25 | -0.16 | -1.67 | -1.06 | -2.03 | -0.34 | -1.84 | -1.23 | -2.20 | -0.02 | -1.52 | -0.91 | -1.88 | -0.22 | -0.05 | -0.37 | 0.05 | -0.17 | -0.32 |
| Te2Au4 | P1 | 2dm-1349 | Te2Au4#P1#2dm-1349 | 2.47 | 0.94 | 1.56 | 0.40 | 2.86 | 2.63 | 2.86 | 2.14 | 0.867 | 0.887 | 0.877 | 0.801 | 0.872 | 0.837 | 0.814 | 0.830 | 0.82 | 0.721 | 0.726 | 0.724 | -0.39 | -1.92 | -1.30 | -2.46 | -0.16 | -1.69 | -1.07 | -2.22 | -0.39 | -1.92 | -1.30 | -2.46 | 0.33 | -1.20 | -0.58 | -1.74 | -0.23 | 0.00 | -0.72 | 0.00 | -0.23 | -0.72 |
| Te2Mo | Amm2 | 2dm-5370 | Te2Mo#Amm2#2dm-5370 | 3.47 | 1.83 | 2.50 | 1.37 | 3.57 | 3.09 | 3.54 | 2.65 | 0.818 | 0.821 | 0.820 | 0.783 | 0.832 | 0.807 | 0.809 | 0.836 | 0.82 | 0.751 | 0.708 | 0.730 | -0.10 | -1.73 | -1.07 | -2.20 | 0.37 | -1.26 | -0.59 | -1.72 | -0.07 | -1.71 | -1.04 | -2.17 | 0.81 | -0.82 | -0.16 | -1.28 | -0.48 | -0.03 | -0.91 | 0.03 | -0.45 | -0.89 |
| Te2O4 | Pmmn | 2dm-3164 | Te2O4#Pmmn#2dm-3164 | 4.20 | 2.46 | 3.22 | 1.77 | 4.43 | 3.82 | 4.13 | 2.61 | 0.800 | 0.858 | 0.829 | 0.710 | 0.834 | 0.772 | 0.781 | 0.817 | 0.80 | 0.727 | 0.719 | 0.723 | -0.23 | -1.97 | -1.20 | -2.66 | 0.38 | -1.36 | -0.60 | -2.05 | 0.07 | -1.67 | -0.91 | -2.36 | 1.59 | -0.15 | 0.61 | -0.84 | -0.61 | -0.29 | -1.82 | 0.29 | -0.31 | -1.52 |
| Te2Pb2 | Pmmn | 2dm-407 | Te2Pb2#Pmmn#2dm-407 | 3.71 | 2.28 | 2.89 | 1.72 | 3.98 | 3.40 | 3.66 | 2.33 | 0.855 | 0.884 | 0.870 | 0.798 | 0.852 | 0.825 | 0.818 | 0.848 | 0.83 | 0.737 | 0.736 | 0.736 | -0.27 | -1.69 | -1.08 | -2.25 | 0.31 | -1.12 | -0.50 | -1.67 | 0.05 | -1.38 | -0.77 | -1.94 | 1.38 | -0.05 | 0.57 | -0.60 | -0.58 | -0.31 | -1.65 | 0.31 | -0.27 | -1.33 |
| Te2Pb2 | Pmn2_1 | 2dm-3176 | Te2Pb2#Pmn2_1#2dm-3176 | 3.23 | 1.77 | 2.33 | 1.30 | 3.51 | 3.00 | 3.35 | 1.76 | 0.861 | 0.876 | 0.868 | 0.800 | 0.849 | 0.825 | 0.782 | 0.741 | 0.76 | 0.724 | 0.750 | 0.737 | -0.27 | -1.74 | -1.18 | -2.21 | 0.23 | -1.23 | -0.67 | -1.70 | -0.11 | -1.58 | -1.02 | -2.05 | 1.48 | 0.01 | 0.57 | -0.46 | -0.51 | -0.16 | -1.75 | 0.16 | -0.35 | -1.59 |
| Te2Pd | C2/m | 2dm-5247 | Te2Pd#C2__m#2dm-5247 | 2.06 | 0.59 | 1.18 | 0.27 | 2.17 | 1.78 | 2.12 | 1.46 | 0.820 | 0.834 | 0.827 | 0.757 | 0.791 | 0.774 | 0.815 | 0.823 | 0.82 | 0.880 | 0.753 | 0.817 | -0.11 | -1.58 | -0.99 | -1.90 | 0.29 | -1.19 | -0.59 | -1.51 | -0.06 | -1.53 | -0.94 | -1.85 | 0.60 | -0.87 | -0.28 | -1.20 | -0.39 | -0.05 | -0.71 | 0.05 | -0.34 | -0.66 |
| Te2Pt | C2/m | 2dm-4255 | Te2Pt#C2__m#2dm-4255 | 2.68 | 1.14 | 1.76 | 0.72 | 2.78 | 2.31 | 2.39 | 1.21 | 0.839 | 0.849 | 0.844 | 0.777 | 0.814 | 0.795 | 0.831 | 0.871 | 0.85 | 0.713 | 0.719 | 0.716 | -0.10 | -1.64 | -1.02 | -2.06 | 0.37 | -1.18 | -0.55 | -1.60 | 0.28 | -1.26 | -0.63 | -1.68 | 1.47 | -0.07 | 0.55 | -0.49 | -0.47 | -0.39 | -1.57 | 0.39 | -0.08 | -1.18 |
| Te2W | Pm | 2dm-3050 | Te2W#Pm#2dm-3050 | 3.57 | 1.98 | 2.63 | 1.53 | 3.78 | 3.72 | 3.78 | 2.00 | 0.836 | 0.839 | 0.837 | 0.808 | 0.848 | 0.828 | 0.796 | 0.846 | 0.82 | 0.767 | 0.702 | 0.734 | -0.21 | -1.80 | -1.15 | -2.25 | -0.15 | -1.74 | -1.09 | -2.19 | -0.21 | -1.80 | -1.15 | -2.24 | 1.57 | -0.02 | 0.63 | -0.47 | -0.06 | 0.00 | -1.78 | 0.00 | -0.05 | -1.77 |
| Te4Pd2 | P1 | 2dm-941 | Te4Pd2#P1#2dm-941 | 0.55 | 0.24 | 0.78 | 0.04 | 0.43 | 0.34 | 0.42 | -0.31 | 0.803 | 0.815 | 0.809 | 0.750 | 0.796 | 0.773 | 0.768 | 0.844 | 0.81 | 0.746 | 0.739 | 0.742 | 0.12 | -0.19 | 0.35 | -0.38 | 0.21 | -0.10 | 0.44 | -0.29 | 0.12 | -0.19 | 0.36 | -0.38 | 0.85 | 0.54 | 1.09 | 0.35 | -0.09 | 0.00 | -0.73 | 0.00 | -0.09 | -0.73 |
| Te4Pt2 | P1 | 2dm-1087 | Te4Pt2#P1#2dm-1087 | 0.14 | 0.12 | 0.13 | 0.13 | 0.14 | 0.14 | 0.14 | -0.30 | 0.815 | 0.828 | 0.822 | 0.777 | 0.805 | 0.791 | 0.789 | 0.819 | 0.80 | 0.716 | 0.711 | 0.713 | 0.00 | -0.02 | -0.01 | -0.01 | 0.01 | -0.02 | 0.00 | -0.01 | 0.00 | -0.02 | -0.01 | -0.01 | 0.45 | 0.42 | 0.44 | 0.43 | 0.00 | 0.00 | -0.44 | 0.00 | 0.00 | -0.44 |
| TeO3 | Amm2 | 2dm-4882 | TeO3#Amm2#2dm-4882 | 3.95 | 2.45 | 3.16 | 0.87 | 4.02 | 3.59 | 3.92 | 2.85 | 0.760 | 0.821 | 0.790 | 0.692 | 0.789 | 0.741 | 0.816 | 0.839 | 0.83 | 0.756 | 0.688 | 0.722 | -0.07 | -1.57 | -0.86 | -3.15 | 0.36 | -1.14 | -0.43 | -2.72 | 0.03 | -1.47 | -0.76 | -3.05 | 1.10 | -0.40 | 0.32 | -1.98 | -0.43 | -0.10 | -1.17 | 0.10 | -0.33 | -1.07 |
| ThCl2 | P1 | 2dm-182 | ThCl2#P1#2dm-182 | 2.66 | 1.04 | 1.69 | 0.28 | 3.03 | 2.82 | 2.97 | 2.47 | 0.827 | 0.835 | 0.831 | 0.788 | 0.815 | 0.801 | 0.821 | 0.828 | 0.82 | 0.707 | 0.727 | 0.717 | -0.37 | -1.99 | -1.34 | -2.76 | -0.15 | -1.78 | -1.12 | -2.54 | -0.31 | -1.93 | -1.28 | -2.70 | 0.19 | -1.43 | -0.77 | -2.19 | -0.22 | -0.06 | -0.56 | 0.06 | -0.16 | -0.50 |
| ThF2 | Cm | 2dm-1048 | ThF2#Cm#2dm-1048 | 3.01 | 1.13 | 1.83 | 0.39 | 2.92 | 3.18 | 3.15 | 2.67 | 0.861 | 0.894 | 0.878 | 0.822 | 0.860 | 0.841 | 0.836 | 0.869 | 0.85 | 0.770 | 0.729 | 0.749 | 0.09 | -1.80 | -1.10 | -2.54 | -0.17 | -2.05 | -1.35 | -2.80 | -0.13 | -2.02 | -1.32 | -2.76 | 0.34 | -1.55 | -0.85 | -2.29 | 0.26 | 0.23 | -0.25 | -0.23 | 0.03 | -0.48 |
| ThI2 | Amm2 | 2dm-3738 | ThI2#Amm2#2dm-3738 | 2.39 | 0.80 | 1.43 | 0.26 | 2.55 | 2.11 | 2.54 | 1.84 | 0.833 | 0.836 | 0.834 | 0.791 | 0.820 | 0.805 | 0.748 | 0.714 | 0.73 | 0.709 | 0.727 | 0.718 | -0.16 | -1.75 | -1.12 | -2.29 | 0.28 | -1.31 | -0.68 | -1.85 | -0.15 | -1.74 | -1.11 | -2.28 | 0.56 | -1.04 | -0.41 | -1.57 | -0.44 | -0.01 | -0.71 | 0.01 | -0.43 | -0.70 |
| Ti2F4 | Pm | 2dm-502 | Ti2F4#Pm#2dm-502 | 3.21 | 1.55 | 2.28 | 0.70 | 2.86 | 2.77 | 2.85 | 2.12 | 0.763 | 0.777 | 0.770 | 0.739 | 0.772 | 0.756 | 0.743 | 0.709 | 0.73 | 0.762 | 0.727 | 0.745 | 0.35 | -1.30 | -0.58 | -2.16 | 0.43 | -1.22 | -0.49 | -2.07 | 0.36 | -1.29 | -0.57 | -2.15 | 1.09 | -0.56 | 0.16 | -1.41 | -0.09 | -0.01 | -0.74 | 0.01 | -0.08 | -0.73 |
| Ti2I4 | Pm | 2dm-1432 | Ti2I4#Pm#2dm-1432 | 2.62 | 0.53 | 1.12 | 0.27 | 2.44 | 2.81 | 2.80 | 1.47 | 0.825 | 0.838 | 0.831 | 0.770 | 0.730 | 0.750 | 0.817 | 0.845 | 0.83 | 0.699 | 0.770 | 0.734 | 0.18 | -1.91 | -1.32 | -2.17 | -0.19 | -2.28 | -1.69 | -2.54 | -0.18 | -2.27 | -1.67 | -2.53 | 1.15 | -0.94 | -0.35 | -1.20 | 0.37 | 0.36 | -0.97 | -0.36 | 0.01 | -1.33 |
| Ti2I4 | Pm | 2dm-1456 | Ti2I4#Pm#2dm-1456 | 2.26 | 0.53 | 1.13 | 0.14 | 2.40 | 2.41 | 2.41 | 1.46 | 0.828 | 0.786 | 0.807 | 0.765 | 0.724 | 0.744 | 0.835 | 0.889 | 0.86 | 0.802 | 0.772 | 0.787 | -0.14 | -1.88 | -1.28 | -2.26 | -0.15 | -1.88 | -1.29 | -2.27 | -0.15 | -1.88 | -1.28 | -2.27 | 0.80 | -0.93 | -0.33 | -1.31 | 0.01 | 0.00 | -0.95 | 0.00 | 0.00 | -0.95 |
| Ti2N2F2 | Pmmn | 2dm-3400 | Ti2N2F2#Pmmn#2dm-3400 | 3.25 | 1.77 | 2.42 | 0.70 | 3.61 | 2.93 | 3.51 | 2.65 | 0.787 | 0.821 | 0.804 | 0.735 | 0.803 | 0.769 | 0.820 | 0.839 | 0.83 | 0.757 | 0.820 | 0.789 | -0.36 | -1.84 | -1.19 | -2.91 | 0.33 | -1.15 | -0.50 | -2.23 | -0.26 | -1.74 | -1.09 | -2.81 | 0.60 | -0.88 | -0.23 | -1.95 | -0.68 | -0.10 | -0.96 | 0.10 | -0.58 | -0.86 |
| TiBr4 | P4/mmm | 2dm-2049 | TiBr4#P4__mmm#2dm-2049 | 3.89 | 1.35 | 2.03 | 0.38 | 4.17 | 3.68 | 3.93 | 3.20 | 0.765 | 0.830 | 0.798 | 0.708 | 0.832 | 0.770 | 0.778 | 0.854 | 0.82 | 0.745 | 0.798 | 0.772 | -0.28 | -2.82 | -2.14 | -3.79 | 0.20 | -2.34 | -1.66 | -3.31 | -0.04 | -2.58 | -1.90 | -3.55 | 0.68 | -1.85 | -1.18 | -2.83 | -0.48 | -0.24 | -0.96 | 0.24 | -0.24 | -0.73 |
| TiF2 | Cm | 2dm-380 | TiF2#Cm#2dm-380 | 3.84 | 2.19 | 2.89 | 1.36 | 4.17 | 4.05 | 4.16 | 2.87 | 0.753 | 0.786 | 0.769 | 0.722 | 0.763 | 0.742 | 0.799 | 0.819 | 0.81 | 0.721 | 0.734 | 0.727 | -0.33 | -1.98 | -1.27 | -2.81 | -0.21 | -1.86 | -1.15 | -2.69 | -0.32 | -1.98 | -1.27 | -2.80 | 0.97 | -0.69 | 0.02 | -1.51 | -0.12 | 0.00 | -1.29 | 0.00 | -0.12 | -1.29 |
| TiI2 | Cm | 2dm-45 | TiI2#Cm#2dm-45 | 2.67 | 1.09 | 1.68 | 0.81 | 2.77 | 2.42 | 2.63 | 1.90 | 0.764 | 0.783 | 0.773 | 0.727 | 0.778 | 0.753 | 0.812 | 0.850 | 0.83 | 0.814 | 0.725 | 0.770 | -0.11 | -1.69 | -1.10 | -1.96 | 0.25 | -1.33 | -0.74 | -1.60 | 0.04 | -1.54 | -0.95 | -1.82 | 0.77 | -0.81 | -0.22 | -1.09 | -0.36 | -0.15 | -0.87 | 0.15 | -0.21 | -0.73 |
| TiO2 | C2/m | 2dm-3816 | TiO2#C2__m#2dm-3816 | 5.84 | 4.33 | 5.05 | 2.73 | 5.94 | 5.16 | 5.72 | 5.08 | 0.784 | 0.793 | 0.788 | 0.767 | 0.782 | 0.775 | 0.837 | 0.882 | 0.86 | 0.755 | 0.723 | 0.739 | -0.10 | -1.60 | -0.88 | -3.21 | 0.68 | -0.83 | -0.11 | -2.43 | 0.12 | -1.39 | -0.67 | -2.99 | 0.76 | -0.75 | -0.03 | -2.36 | -0.77 | -0.21 | -0.85 | 0.21 | -0.56 | -0.64 |
| TiPbO3 | P4mm | 2dm-5482 | TiPbO3#P4mm#2dm-5482 | 5.30 | 3.68 | 4.44 | 2.50 | 5.49 | 5.42 | 5.48 | 4.64 | 0.796 | 0.921 | 0.858 | 0.776 | 0.908 | 0.842 | 0.841 | 0.868 | 0.85 | 0.767 | 0.671 | 0.719 | -0.19 | -1.81 | -1.05 | -2.98 | -0.12 | -1.74 | -0.98 | -2.91 | -0.18 | -1.80 | -1.04 | -2.98 | 0.66 | -0.96 | -0.20 | -2.13 | -0.07 | -0.01 | -0.85 | 0.01 | -0.07 | -0.85 |
| TiS2 | C2/m | 2dm-3731 | TiS2#C2__m#2dm-3731 | 2.25 | 0.77 | 1.40 | 0.05 | 2.63 | 2.44 | 2.60 | 1.77 | 0.790 | 0.832 | 0.811 | 0.750 | 0.814 | 0.782 | 0.781 | 0.756 | 0.77 | 0.887 | 0.728 | 0.807 | -0.38 | -1.85 | -1.23 | -2.58 | -0.19 | -1.67 | -1.05 | -2.40 | -0.35 | -1.83 | -1.21 | -2.56 | 0.48 | -1.00 | -0.38 | -1.73 | -0.19 | -0.03 | -0.86 | 0.03 | -0.16 | -0.83 |
| Tl2Br2 | Pmmn | 2dm-5125 | Tl2Br2#Pmmn#2dm-5125 | 5.51 | 3.95 | 4.62 | 3.13 | 5.40 | 5.18 | 5.31 | 4.96 | 0.870 | 0.924 | 0.897 | 0.829 | 0.904 | 0.867 | 0.813 | 0.851 | 0.83 | 0.697 | 0.749 | 0.723 | 0.10 | -1.45 | -0.78 | -2.27 | 0.33 | -1.22 | -0.56 | -2.04 | 0.20 | -1.35 | -0.69 | -2.18 | 0.55 | -1.00 | -0.34 | -1.83 | -0.23 | -0.10 | -0.45 | 0.10 | -0.13 | -0.35 |
| Tl2Cl2 | P4/nmm | 2dm-1712 | Tl2Cl2#P4__nmm#2dm-1712 | 5.99 | 4.52 | 5.19 | 3.63 | 6.01 | 5.26 | 5.80 | 5.23 | 0.866 | 0.933 | 0.899 | 0.830 | 0.913 | 0.872 | 0.818 | 0.844 | 0.83 | 0.820 | 0.739 | 0.779 | -0.02 | -1.50 | -0.82 | -2.39 | 0.73 | -0.75 | -0.07 | -1.64 | 0.19 | -1.29 | -0.61 | -2.18 | 0.76 | -0.71 | -0.04 | -1.60 | -0.75 | -0.21 | -0.78 | 0.21 | -0.54 | -0.57 |
| Tl2F2 | P4/nmm | 2dm-3772 | Tl2F2#P4__nmm#2dm-3772 | 6.29 | 4.64 | 5.34 | 3.61 | 6.36 | 5.57 | 5.99 | 5.35 | 0.851 | 0.929 | 0.890 | 0.793 | 0.907 | 0.850 | 0.815 | 0.826 | 0.82 | 0.725 | 0.703 | 0.714 | -0.08 | -1.72 | -1.03 | -2.75 | 0.71 | -0.93 | -0.24 | -1.96 | 0.29 | -1.35 | -0.66 | -2.38 | 0.94 | -0.71 | -0.01 | -1.74 | -0.79 | -0.37 | -1.02 | 0.37 | -0.42 | -0.64 |
| Tl2I2 | Pmmn | 2dm-3762 | Tl2I2#Pmmn#2dm-3762 | 5.12 | 3.56 | 4.22 | 2.84 | 4.72 | 4.22 | 4.62 | 4.18 | 0.869 | 0.913 | 0.891 | 0.822 | 0.888 | 0.855 | 0.760 | 0.738 | 0.75 | 0.723 | 0.724 | 0.723 | 0.40 | -1.15 | -0.49 | -1.88 | 0.90 | -0.65 | 0.01 | -1.38 | 0.50 | -1.05 | -0.39 | -1.78 | 0.94 | -0.62 | 0.05 | -1.34 | -0.50 | -0.10 | -0.54 | 0.10 | -0.40 | -0.44 |
| Tl2O | Cm | 2dm-4083 | Tl2O#Cm#2dm-4083 | 3.76 | 2.09 | 2.72 | 1.49 | 3.88 | 3.24 | 3.26 | 2.02 | 0.829 | 0.894 | 0.861 | 0.738 | 0.860 | 0.799 | 0.805 | 0.801 | 0.80 | 0.767 | 0.749 | 0.758 | -0.12 | -1.79 | -1.17 | -2.39 | 0.53 | -1.15 | -0.52 | -1.74 | 0.50 | -1.17 | -0.54 | -1.77 | 1.74 | 0.07 | 0.69 | -0.53 | -0.65 | -0.62 | -1.86 | 0.62 | -0.02 | -1.24 |
| Tl2O3 | C2 | 2dm-836 | Tl2O3#C2#2dm-836 | 3.42 | 1.95 | 2.65 | 0.93 | 3.74 | 3.06 | 3.46 | 2.86 | 0.765 | 0.849 | 0.807 | 0.679 | 0.796 | 0.737 | 0.757 | 0.737 | 0.75 | 0.758 | 0.702 | 0.730 | -0.32 | -1.79 | -1.09 | -2.81 | 0.36 | -1.11 | -0.41 | -2.13 | -0.04 | -1.50 | -0.80 | -2.53 | 0.56 | -0.91 | -0.21 | -1.93 | -0.68 | -0.28 | -0.88 | 0.28 | -0.40 | -0.60 |
| Tl2O3 | P1 | 2dm-2066 | Tl2O3#P1#2dm-2066 | 3.91 | 1.77 | 2.48 | 0.80 | 4.23 | 4.02 | 4.22 | 2.56 | 0.767 | 0.857 | 0.812 | 0.683 | 0.829 | 0.756 | 0.850 | 0.872 | 0.86 | 0.763 | 0.743 | 0.753 | -0.32 | -2.45 | -1.75 | -3.43 | -0.11 | -2.24 | -1.54 | -3.22 | -0.31 | -2.45 | -1.74 | -3.42 | 1.35 | -0.78 | -0.08 | -1.76 | -0.21 | -0.01 | -1.67 | 0.01 | -0.20 | -1.66 |
| Tl2O3 | P1 | 2dm-3783 | Tl2O3#P1#2dm-3783 | 2.94 | 1.45 | 2.13 | 0.64 | 3.31 | 3.23 | 3.27 | 1.69 | 0.770 | 0.839 | 0.805 | 0.714 | 0.826 | 0.770 | 0.808 | 0.856 | 0.83 | 0.700 | 0.760 | 0.730 | -0.38 | -1.86 | -1.18 | -2.67 | -0.29 | -1.77 | -1.10 | -2.59 | -0.34 | -1.82 | -1.14 | -2.63 | 1.24 | -0.24 | 0.43 | -1.06 | -0.08 | -0.04 | -1.62 | 0.04 | -0.04 | -1.58 |
| Tl2S2 | P1 | 2dm-2963 | Tl2S2#P1#2dm-2963 | 3.32 | 1.73 | 2.38 | 1.09 | 3.68 | 3.05 | 3.26 | 1.99 | 0.847 | 0.869 | 0.858 | 0.780 | 0.859 | 0.820 | 0.736 | 0.780 | 0.76 | 0.748 | 0.739 | 0.743 | -0.36 | -1.95 | -1.30 | -2.59 | 0.26 | -1.32 | -0.67 | -1.96 | 0.06 | -1.53 | -0.88 | -2.16 | 1.33 | -0.26 | 0.39 | -0.90 | -0.63 | -0.43 | -1.69 | 0.43 | -0.20 | -1.27 |
| Tl2S2 | P1 | 2dm-1404 | Tl2S2#P1#2dm-1404 | 2.78 | 1.36 | 2.00 | 0.77 | 3.15 | 3.01 | 3.08 | 1.86 | 0.823 | 0.857 | 0.840 | 0.777 | 0.825 | 0.801 | 0.801 | 0.831 | 0.82 | 0.734 | 0.748 | 0.741 | -0.37 | -1.79 | -1.15 | -2.37 | -0.23 | -1.65 | -1.01 | -2.23 | -0.30 | -1.72 | -1.08 | -2.30 | 0.92 | -0.50 | 0.14 | -1.09 | -0.14 | -0.07 | -1.29 | 0.07 | -0.07 | -1.22 |
| Tl2S3 | P1 | 2dm-1880 | Tl2S3#P1#2dm-1880 | 3.83 | 1.47 | 2.11 | 0.73 | 3.99 | 3.34 | 3.88 | 3.10 | 0.838 | 0.871 | 0.854 | 0.756 | 0.835 | 0.795 | 0.836 | 0.852 | 0.84 | 0.759 | 0.804 | 0.781 | -0.16 | -2.53 | -1.88 | -3.26 | 0.50 | -1.87 | -1.23 | -2.61 | -0.04 | -2.41 | -1.77 | -3.15 | 0.73 | -1.64 | -0.99 | -2.37 | -0.66 | -0.12 | -0.89 | 0.12 | -0.54 | -0.77 |
| Tl2S5 | P1 | 2dm-665 | Tl2S5#P1#2dm-665 | 2.82 | 1.35 | 2.00 | 0.75 | 2.89 | 2.62 | 2.84 | 2.29 | 0.837 | 0.873 | 0.855 | 0.775 | 0.835 | 0.805 | 0.827 | 0.839 | 0.83 | 0.746 | 0.722 | 0.734 | -0.07 | -1.54 | -0.89 | -2.14 | 0.20 | -1.27 | -0.62 | -1.87 | -0.02 | -1.49 | -0.84 | -2.09 | 0.52 | -0.95 | -0.29 | -1.55 | -0.26 | -0.05 | -0.59 | 0.05 | -0.21 | -0.54 |
| Tl2Se2 | Cm | 2dm-729 | Tl2Se2#Cm#2dm-729 | 2.46 | 1.08 | 1.69 | 0.59 | 2.86 | 2.79 | 2.85 | 1.49 | 0.832 | 0.862 | 0.847 | 0.776 | 0.828 | 0.802 | 0.808 | 0.829 | 0.82 | 0.744 | 0.754 | 0.749 | -0.40 | -1.78 | -1.17 | -2.27 | -0.33 | -1.71 | -1.10 | -2.20 | -0.39 | -1.76 | -1.16 | -2.26 | 0.97 | -0.41 | 0.20 | -0.90 | -0.07 | -0.01 | -1.37 | 0.01 | -0.06 | -1.36 |
| Tl2Se2 | Pm | 2dm-2962 | Tl2Se2#Pm#2dm-2962 | 2.63 | 1.08 | 1.70 | 0.55 | 3.02 | 2.75 | 2.95 | 2.27 | 0.849 | 0.853 | 0.851 | 0.774 | 0.857 | 0.816 | 0.808 | 0.818 | 0.81 | 0.725 | 0.714 | 0.720 | -0.39 | -1.94 | -1.32 | -2.47 | -0.12 | -1.67 | -1.05 | -2.20 | -0.33 | -1.87 | -1.25 | -2.40 | 0.35 | -1.19 | -0.57 | -1.72 | -0.27 | -0.07 | -0.75 | 0.07 | -0.20 | -0.68 |
| Tl2Se3 | P1 | 2dm-1024 | Tl2Se3#P1#2dm-1024 | 2.11 | 0.82 | 1.40 | 0.23 | 2.50 | 2.39 | 2.50 | 2.17 | 0.844 | 0.866 | 0.855 | 0.759 | 0.838 | 0.799 | 0.776 | 0.778 | 0.78 | 0.804 | 0.702 | 0.753 | -0.40 | -1.69 | -1.10 | -2.27 | -0.28 | -1.57 | -0.99 | -2.16 | -0.39 | -1.68 | -1.10 | -2.27 | -0.06 | -1.35 | -0.77 | -1.94 | -0.12 | -0.01 | -0.33 | 0.01 | -0.11 | -0.33 |
| Tl2Se5 | P1 | 2dm-233 | Tl2Se5#P1#2dm-233 | 2.76 | 1.15 | 1.78 | 0.57 | 3.11 | 2.94 | 3.04 | 1.65 | 0.839 | 0.874 | 0.857 | 0.768 | 0.839 | 0.804 | 0.787 | 0.790 | 0.79 | 0.807 | 0.814 | 0.810 | -0.35 | -1.96 | -1.33 | -2.55 | -0.18 | -1.79 | -1.16 | -2.37 | -0.29 | -1.89 | -1.27 | -2.48 | 1.11 | -0.50 | 0.13 | -1.08 | -0.17 | -0.07 | -1.46 | 0.07 | -0.10 | -1.39 |
| Tl2Te2 | Cm | 2dm-1104 | Tl2Te2#Cm#2dm-1104 | 2.29 | 0.97 | 1.54 | 0.55 | 2.65 | 2.51 | 2.60 | 2.01 | 0.838 | 0.865 | 0.851 | 0.769 | 0.823 | 0.796 | 0.817 | 0.833 | 0.83 | 0.745 | 0.732 | 0.739 | -0.36 | -1.68 | -1.11 | -2.10 | -0.22 | -1.54 | -0.97 | -1.96 | -0.31 | -1.63 | -1.06 | -2.05 | 0.28 | -1.04 | -0.48 | -1.47 | -0.14 | -0.05 | -0.63 | 0.05 | -0.09 | -0.59 |
| Tl2Te3 | P1 | 2dm-1315 | Tl2Te3#P1#2dm-1315 | 0.59 | 0.10 | 0.61 | 0.09 | 0.43 | 0.36 | 0.43 | 0.28 | 0.850 | 0.865 | 0.857 | 0.758 | 0.827 | 0.792 | 0.817 | 0.837 | 0.83 | 0.715 | 0.814 | 0.764 | 0.16 | -0.33 | 0.18 | -0.35 | 0.23 | -0.26 | 0.25 | -0.28 | 0.16 | -0.33 | 0.18 | -0.34 | 0.32 | -0.18 | 0.33 | -0.19 | -0.07 | 0.00 | -0.15 | 0.00 | -0.07 | -0.15 |
| Tl2Te5 | P1 | 2dm-1601 | Tl2Te5#P1#2dm-1601 | 2.80 | 1.05 | 1.63 | 0.54 | 3.15 | 2.98 | 3.13 | 2.37 | 0.851 | 0.881 | 0.866 | 0.776 | 0.845 | 0.810 | 0.789 | 0.815 | 0.80 | 0.812 | 0.740 | 0.776 | -0.36 | -2.10 | -1.53 | -2.62 | -0.18 | -1.93 | -1.35 | -2.44 | -0.33 | -2.08 | -1.50 | -2.60 | 0.43 | -1.32 | -0.75 | -1.84 | -0.17 | -0.02 | -0.78 | 0.02 | -0.15 | -0.76 |
| TlInS2 | P1 | 2dm-3849 | TlInS2#P1#2dm-3849 | 4.24 | 2.61 | 3.27 | 1.90 | 4.53 | 3.82 | 4.49 | 3.25 | 0.833 | 0.888 | 0.860 | 0.791 | 0.867 | 0.829 | 0.813 | 0.807 | 0.81 | 0.701 | 0.816 | 0.758 | -0.29 | -1.92 | -1.26 | -2.63 | 0.42 | -1.21 | -0.55 | -1.92 | -0.25 | -1.88 | -1.22 | -2.59 | 0.99 | -0.64 | 0.02 | -1.35 | -0.71 | -0.04 | -1.28 | 0.04 | -0.67 | -1.24 |
| UO2F2 | P2/m | 2dm-5199 | UO2F2#P2__m#2dm-5199 | 5.13 | 3.56 | 4.30 | 2.09 | 5.30 | 4.82 | 5.09 | 4.72 | 0.755 | 0.774 | 0.764 | 0.705 | 0.763 | 0.734 | 0.822 | 0.862 | 0.84 | 0.772 | 0.689 | 0.731 | -0.17 | -1.74 | -1.00 | -3.21 | 0.32 | -1.26 | -0.52 | -2.73 | 0.05 | -1.53 | -0.79 | -3.00 | 0.41 | -1.16 | -0.42 | -2.63 | -0.48 | -0.21 | -0.58 | 0.21 | -0.27 | -0.37 |
| VF5 | Pmmm | 2dm-2503 | VF5#Pmmm#2dm-2503 | 2.91 | 1.44 | 2.11 | 0.28 | 3.28 | 2.54 | 3.19 | 2.54 | 0.719 | 0.741 | 0.730 | 0.704 | 0.724 | 0.714 | 0.798 | 0.869 | 0.83 | 0.796 | 0.727 | 0.761 | -0.37 | -1.84 | -1.16 | -3.00 | 0.37 | -1.11 | -0.43 | -2.27 | -0.28 | -1.75 | -1.08 | -2.91 | 0.37 | -1.10 | -0.43 | -2.26 | -0.73 | -0.09 | -0.74 | 0.09 | -0.65 | -0.65 |
| W2N2 | P-1 | 2dm-2568 | W2N2#P-1#2dm-2568 | 0.50 | 0.23 | 0.78 | 0.01 | 0.28 | 0.27 | 0.28 | 0.25 | 0.826 | 0.841 | 0.833 | 0.812 | 0.842 | 0.827 | 0.792 | 0.820 | 0.81 | 0.783 | 0.730 | 0.756 | 0.22 | -0.05 | 0.50 | -0.27 | 0.23 | -0.04 | 0.51 | -0.26 | 0.22 | -0.05 | 0.50 | -0.27 | 0.25 | -0.02 | 0.53 | -0.24 | -0.01 | 0.00 | -0.03 | 0.00 | -0.01 | -0.02 |
| W2Se4 | Pm | 2dm-191 | W2Se4#Pm#2dm-191 | 0.65 | 0.39 | 0.93 | 0.13 | 0.73 | 0.41 | 0.69 | 0.32 | 0.838 | 0.855 | 0.846 | 0.821 | 0.852 | 0.836 | 0.765 | 0.769 | 0.77 | 0.759 | 0.712 | 0.735 | -0.09 | -0.34 | 0.20 | -0.60 | 0.23 | -0.02 | 0.52 | -0.28 | -0.05 | -0.30 | 0.24 | -0.56 | 0.33 | 0.07 | 0.61 | -0.18 | -0.32 | -0.04 | -0.41 | 0.04 | -0.28 | -0.38 |
| WO2F2 | Amm2 | 2dm-4828 | WO2F2#Amm2#2dm-4828 | 4.06 | 2.44 | 3.14 | 1.35 | 4.13 | 3.58 | 4.09 | 3.56 | 0.788 | 0.859 | 0.823 | 0.770 | 0.841 | 0.806 | 0.780 | 0.808 | 0.79 | 0.763 | 0.734 | 0.748 | -0.07 | -1.69 | -0.99 | -2.78 | 0.48 | -1.14 | -0.44 | -2.22 | -0.03 | -1.66 | -0.95 | -2.74 | 0.50 | -1.12 | -0.42 | -2.21 | -0.55 | -0.04 | -0.57 | 0.04 | -0.52 | -0.53 |
| WO3 | Pm | 2dm-605 | WO3#Pm#2dm-605 | 3.58 | 1.97 | 2.68 | 0.80 | 3.44 | 3.15 | 3.32 | 3.14 | 0.789 | 0.862 | 0.825 | 0.779 | 0.847 | 0.813 | 0.779 | 0.817 | 0.80 | 0.768 | 0.688 | 0.728 | 0.14 | -1.47 | -0.77 | -2.64 | 0.43 | -1.18 | -0.47 | -2.35 | 0.26 | -1.35 | -0.64 | -2.52 | 0.44 | -1.18 | -0.47 | -2.34 | -0.30 | -0.13 | -0.30 | 0.13 | -0.17 | -0.17 |
| WS2 | Amm2 | 2dm-3749 | WS2#Amm2#2dm-3749 | 4.31 | 2.64 | 3.33 | 2.09 | 4.66 | 4.62 | 4.65 | 2.95 | 0.838 | 0.844 | 0.841 | 0.818 | 0.835 | 0.826 | 0.819 | 0.842 | 0.83 | 0.687 | 0.793 | 0.740 | -0.35 | -2.01 | -1.32 | -2.56 | -0.32 | -1.98 | -1.29 | -2.53 | -0.35 | -2.01 | -1.32 | -2.56 | 1.36 | -0.30 | 0.39 | -0.85 | -0.03 | 0.00 | -1.71 | 0.00 | -0.03 | -1.71 |
| WSe2 | Amm2 | 2dm-3594 | WSe2#Amm2#2dm-3594 | 4.00 | 2.40 | 3.06 | 1.90 | 4.16 | 3.68 | 4.00 | 2.56 | 0.841 | 0.843 | 0.842 | 0.811 | 0.838 | 0.824 | 0.777 | 0.953 | 0.86 | 0.817 | 0.801 | 0.809 | -0.17 | -1.76 | -1.10 | -2.27 | 0.32 | -1.28 | -0.62 | -1.78 | 0.00 | -1.60 | -0.94 | -2.10 | 1.44 | -0.16 | 0.50 | -0.66 | -0.48 | -0.16 | -1.60 | 0.16 | -0.32 | -1.44 |
| YbBi2 | Pmm2 | 2dm-5533 | YbBi2#Pmm2#2dm-5533 | 2.47 | 0.89 | 1.52 | 0.35 | 2.85 | 2.65 | 2.72 | 2.18 | 0.859 | 0.880 | 0.870 | 0.794 | 0.845 | 0.819 | 0.809 | 0.872 | 0.84 | 0.740 | 0.768 | 0.754 | -0.38 | -1.95 | -1.33 | -2.49 | -0.18 | -1.76 | -1.13 | -2.30 | -0.25 | -1.82 | -1.20 | -2.36 | 0.29 | -1.28 | -0.66 | -1.83 | -0.20 | -0.13 | -0.67 | 0.13 | -0.07 | -0.54 |
| Zr2Br4 | P2_1/m | 2dm-1550 | Zr2Br4#P2_1__m#2dm-1550 | 3.33 | 1.21 | 1.91 | 0.63 | 3.38 | 3.43 | 3.41 | 2.65 | 0.854 | 0.844 | 0.849 | 0.820 | 0.801 | 0.810 | 0.819 | 0.842 | 0.83 | 0.861 | 0.751 | 0.806 | -0.05 | -2.17 | -1.48 | -2.75 | -0.10 | -2.22 | -1.53 | -2.81 | -0.08 | -2.20 | -1.50 | -2.78 | 0.69 | -1.43 | -0.74 | -2.02 | 0.05 | 0.03 | -0.74 | -0.03 | 0.02 | -0.76 |
| Zr2Cl4 | Pm | 2dm-1578 | Zr2Cl4#Pm#2dm-1578 | 2.97 | 1.31 | 1.99 | 0.73 | 3.32 | 3.18 | 3.28 | 2.53 | 0.832 | 0.837 | 0.834 | 0.815 | 0.837 | 0.826 | 0.719 | 0.737 | 0.73 | 0.719 | 0.690 | 0.704 | -0.35 | -2.01 | -1.33 | -2.60 | -0.21 | -1.87 | -1.19 | -2.46 | -0.31 | -1.96 | -1.28 | -2.55 | 0.44 | -1.22 | -0.54 | -1.80 | -0.14 | -0.05 | -0.80 | 0.05 | -0.09 | -0.75 |
| Zr2F4 | P2_1/m | 2dm-1711 | Zr2F4#P2_1__m#2dm-1711 | 2.71 | 1.12 | 1.78 | 0.60 | 3.06 | 2.90 | 3.04 | 2.67 | 0.814 | 0.823 | 0.818 | 0.778 | 0.840 | 0.809 | 0.818 | 0.840 | 0.83 | 0.728 | 0.769 | 0.748 | -0.35 | -1.94 | -1.29 | -2.46 | -0.19 | -1.78 | -1.12 | -2.30 | -0.34 | -1.92 | -1.27 | -2.45 | 0.04 | -1.54 | -0.89 | -2.07 | -0.16 | -0.02 | -0.39 | 0.02 | -0.15 | -0.38 |
| Zr2F4 | Pm | 2dm-1478 | Zr2F4#Pm#2dm-1478 | 3.45 | 1.80 | 2.50 | 1.09 | 3.34 | 3.09 | 3.31 | 2.66 | 0.825 | 0.828 | 0.827 | 0.795 | 0.809 | 0.802 | 0.812 | 0.822 | 0.82 | 0.712 | 0.733 | 0.723 | 0.10 | -1.54 | -0.84 | -2.26 | 0.35 | -1.29 | -0.59 | -2.01 | 0.13 | -1.51 | -0.82 | -2.23 | 0.79 | -0.85 | -0.16 | -1.57 | -0.25 | -0.03 | -0.69 | 0.03 | -0.22 | -0.66 |
| ZrCl2 | Cm | 2dm-3706 | ZrCl2#Cm#2dm-3706 | 3.40 | 1.77 | 2.45 | 1.17 | 3.68 | 3.54 | 3.63 | 2.22 | 0.831 | 0.837 | 0.834 | 0.801 | 0.815 | 0.808 | 0.738 | 0.858 | 0.80 | 0.726 | 0.809 | 0.768 | -0.27 | -1.91 | -1.23 | -2.51 | -0.13 | -1.77 | -1.09 | -2.37 | -0.22 | -1.86 | -1.18 | -2.46 | 1.19 | -0.45 | 0.23 | -1.05 | -0.14 | -0.05 | -1.46 | 0.05 | -0.09 | -1.41 |
| ZrNCl | Cm | 2dm-3763 | ZrNCl#Cm#2dm-3763 | 6.20 | 4.53 | 5.26 | 3.41 | 6.31 | 5.62 | 6.14 | 5.36 | 0.824 | 0.858 | 0.841 | 0.797 | 0.842 | 0.820 | 0.799 | 0.871 | 0.83 | 0.770 | 0.739 | 0.755 | -0.11 | -1.78 | -1.05 | -2.90 | 0.57 | -1.10 | -0.36 | -2.21 | 0.06 | -1.61 | -0.88 | -2.73 | 0.83 | -0.84 | -0.10 | -1.95 | -0.68 | -0.17 | -0.95 | 0.17 | -0.51 | -0.78 |
| ZrS2 | C2/m | 2dm-3194 | ZrS2#C2__m#2dm-3194 | 3.65 | 2.06 | 2.72 | 1.22 | 4.02 | 3.90 | 3.97 | 2.90 | 0.838 | 0.845 | 0.841 | 0.801 | 0.825 | 0.813 | 0.835 | 0.844 | 0.84 | 0.769 | 0.698 | 0.733 | -0.37 | -1.96 | -1.30 | -2.80 | -0.25 | -1.84 | -1.18 | -2.68 | -0.32 | -1.91 | -1.25 | -2.75 | 0.75 | -0.84 | -0.17 | -1.68 | -0.12 | -0.05 | -1.12 | 0.05 | -0.07 | -1.07 |
| ZrSe2 | C2/m | 2dm-3542 | ZrSe2#C2__m#2dm-3542 | 2.80 | 1.23 | 1.88 | 0.53 | 3.12 | 2.99 | 3.12 | 1.92 | 0.839 | 0.851 | 0.845 | 0.802 | 0.831 | 0.816 | 0.768 | 0.752 | 0.76 | 0.764 | 0.757 | 0.760 | -0.32 | -1.89 | -1.25 | -2.59 | -0.18 | -1.76 | -1.11 | -2.45 | -0.31 | -1.88 | -1.24 | -2.58 | 0.88 | -0.69 | -0.04 | -1.38 | -0.14 | -0.01 | -1.20 | 0.01 | -0.13 | -1.20 |